\documentclass[aps,pra,twocolumn,nobibnotes]{revtex4}
\usepackage[utf8]{inputenc}
\usepackage[T1]{fontenc}
\usepackage{bm}
\usepackage{ulem}
\usepackage{epsfig}
\usepackage{graphicx}
\usepackage{amssymb,amsmath,amsbsy,amsgen,amsfonts}
\usepackage{dcolumn}
\usepackage{amsthm}
\usepackage{mathtools}
\usepackage{mathrsfs}
\usepackage{latexsym}
\usepackage{array}
\usepackage{color}
\usepackage{amstext}
\usepackage{multirow}
\usepackage{bbold}
\usepackage{color}
\usepackage{xcolor}
\usepackage{listings}
\usepackage{tikz}
\usepackage{lipsum}
\usepackage{textcomp}

\makeatletter
\allowdisplaybreaks[1]

\makeatletter
\let\old@makecaption=\@makecaption
\usepackage{caption}
\usepackage{subcaption}
\let\@makecaption=\old@makecaption
\makeatother

\usepackage{epstopdf} 

\DeclareSymbolFont{symbols}{OMS}{cmsy}{m}{n}

\newcommand\numberthis{\addtocounter{equation}{1}\tag{\theequation}}

\newcommand{\ket}[1]{\ensuremath{\left|#1\right>}}

\newcommand{\secref}[1]{Sec.~\ref{#1}}
\newcommand{\appendref}[1]{Appendix~\ref{#1}}
\newcommand{\equatref}[1]{Eq.~(\ref{#1})}
\newcommand{\figref}[1]{Fig.~\ref{#1}}

\begin{document}

\title{Investigating the effect of noise channels on the quality of unitary t-designs}

\author{Conrad Strydom}
\email{conradstryd@gmail.com}
\author{Mark Tame}
\affiliation{Department of Physics, Stellenbosch University, Matieland 7602, South Africa}

\begin{abstract}
Unitary $t$-designs have a wide variety of applications in quantum information theory, such as quantum data encryption and randomised benchmarking.  However, experimental realisations of $t$-designs are subject to noise.  Here we investigate the effect of noise channels on the quality of single-qubit $t$-designs.  The noise channels we study are bit flips, phase flips, bit and phase flips, phase damping, amplitude damping, and depolarising noise.  We consider two noise models: the first has noise applied before the $t$-design unitary operations, while the second has noise applied after the unitary operations.  We show that the single-qubit 1-design is affected only by amplitude damping, while numeric results obtained for the 2-, 3-, 4-, and 5-designs suggest that a $2t$-design is significantly more sensitive to noise than a $(2t-1)$-design and that, with the exception of amplitude damping, a $(2t+1)$-design is as sensitive to noise as a $2t$-design.  Numeric results also reveal substantial variations in sensitivity to noise throughout the Bloch sphere.  In particular, $t$-designs appear to be most sensitive to noise when acting on pure states and least sensitive to noise for the maximally mixed state.  For depolarising noise, we show that our two noise models are equivalent, and for the other noise channels, numeric results obtained for the model where noise is applied after the unitaries reflect the transformation of the noise channel into a depolarising channel, an effect exploited in randomised benchmarking with 2-designs.
\end{abstract}


\maketitle

\section{Introduction}\label{sec:introduction} 

Unitary operators chosen randomly with respect to the Haar measure on the unitary group play a fundamental role in quantum information theory.  Unfortunately, the resources required to sample from the uniform Haar ensemble grow exponentially with the number of qubits~\cite{inefficient}.  Unitary $t$-designs are therefore used as an efficient substitute in many important applications (the resources required to implement an approximate $t$-design scale polynomially with the number of qubits~\cite{RCC4, MBT2}).  In particular, 1-designs are used for encrypting quantum data~\cite{appl1-1, appl1-2}; 2-designs are used for randomised benchmarking~\cite{RB1, RB2, RB3, RB4, RB5, RB6, RB7, RB8, RB9}, characterising correlations within multipartite quantum systems~\cite{appl2-1}, and formulating quantum mechanical models of black holes~\cite{appl2-2}; 3-designs are used for detecting entanglement~\cite{appl3-1, appl3-2, appl3-3, appl3-4} and solving black-box problems~\cite{appl3-5}; and 4-designs are used for quantum state distinction~\cite{appl4-1} and estimating the self-adjointness of quantum noise~\cite{appl4-2}.  Higher-order $t$-designs also find applications in noise estimation for even $t$~\cite{applH}.

Two main techniques for generating exact and approximate unitary $t$-designs exist, namely, random circuit constructions~\cite{RCC1, RCC2, RCC3, RCC4} and measurement-based techniques~\cite{MBT1, MBT2}.  Random circuit constructions involve the application of non-deterministic sequences of gates from a universal set.  In contrast, measurement-based techniques involve performing deterministic sequences of single-qubit measurements on highly entangled cluster states.  Irrespective of the method used, experimental realisations of $t$-designs are subject to noise~\cite{experimental1, experimental2}.

In this paper we investigate the effect of noise on the quality of unitary $t$-designs for single qubits, similar to the way the effect of noise on randomised benchmarking~\cite{noiseRB} and the variational quantum eigensolver~\cite{noiseVQE} have been simulated.  Since the extent to which applications of $t$-designs are affected by noise in the underlying $t$-design is likely to differ, depending on the application, we study the effect of noise on the quality of $t$-designs without reference to any particular application.  We determine the effect of the bit flip channel, the phase flip channel, the bit and phase flip channel, the phase damping channel, the amplitude damping channel, and the depolarising noise channel on the quality $t$-designs for $t\in\{1,2,3,4,5\}$.  To this end, we consider two noise models, one in which noise is applied before the unitary operations of the $t$-design and one in which noise is applied after the unitary operations.  This is in line with the noise models used in randomised benchmarking~\cite{RB1, RB2, RB3, RB4, RB5, RB6, RB7, RB8, RB9}, one of the primary applications of $t$-designs.

For the model where noise is applied before the unitary operations, we are able to show analytically that the quality of the single-qubit 1-design is completely unaffected by an arbitrary noise channel, and for the model where noise is applied after the unitaries, we show that the 1-design is unaffected by noise, unless amplitude damping is applied.  We obtain numeric results for the 2-design, 3-design, 4-design, and 5-design.  These results suggest that a $2t$-design is significantly more sensitive to noise than a $(2t-1)$-design and that, with the exception of the amplitude damping channel, a $(2t+1)$-design is as sensitive to noise as a $2t$-design.  We also find large variations in sensitivity to noise throughout the state space, with $t$-designs generally being most sensitive to noise for pure states and least sensitive to noise for the maximally mixed state.  The findings presented in this paper will be helpful for researchers studying and developing applications using $t$-designs under realistic conditions.  While we hope that our work will encourage research into the effect of noise on the quality of multi-qubit $t$-designs, we also note that there are many protocols which exclusively use single-qubit $t$-designs~\cite{appl2-1, appl3-1, appl3-2, applsingle} for which our results may have direct consequences.

Our paper is structured as follows.  In \secref{sec:background} we give the definition of a unitary $t$-design as well as the definitions of the various noise channels.  In \secref{sec:noisemodelling} we describe the two noise models considered.  Our main numeric results for the 2-design, 3-design, 4-design, and 5-design are presented in \secref{sec:results}.  A summary of the results and concluding remarks are given in \secref{sec:conclusion}.  Supplementary appendices follow, in which we present analytic results for the 1-design and further numeric results for the higher-order $t$-designs, as well as some important proofs.  These appendices also include complementary numeric results which give a geometric picture of the state dependence of the effect of noise channels on the quality of single-qubit $t$-designs.

\section{Background}\label{sec:background}

\subsection{Unitary t-designs}\label{sec:tdbackground}

An ensemble of unitary operators is an exact unitary $t$-design if its statistical moments are equal to the corresponding statistical moments of the uniform Haar ensemble up to order $t$.  For any matrix $\rho\in\mathcal{B}(H^{\otimes t})$, with $H=\mathbb{C}^2$ for single qubits, the expectation of the uniform Haar ensemble is given by
\begin{equation}
\mathbb{E}^t_H(\rho)=\int U^{\otimes t}\rho\left(U^{\otimes t}\right)^{\dagger}\,dU.
\end{equation}
An ensemble of unitaries $\{p_i,U_i\}$ is an $\epsilon$-approximate $t$-design if there exists an $\epsilon$ such that for all $\rho\in\mathcal{B}(H^{\otimes t})$,
\begin{equation}
(1-\epsilon)\mathbb{E}^t_H(\rho)\leq\sum_{i}p_iU_i^{\otimes t}\rho\left(U_i^{\otimes t}\right)^{\dagger}\leq(1+\epsilon)\mathbb{E}^t_H(\rho),
\label{eq:def}
\end{equation}
where the matrix inequality $A\leq B$ holds if $B-A$ is positive semidefinite~\cite{MBT1, MBT2}.  An exact $t$-design can also be defined as an $\epsilon$-approximate $t$-design with $\epsilon=0$.

Since the positive semidefinite property defines a partial order (the Loewner order) on Hermitian matrices, inequality~(\ref{eq:def}) is a natural generalisation of an error bound inequality from scalars to Hermitian matrices.  However, an interpretation of $\epsilon$ in terms of defining an error range for the Haar ensemble expectation determined with an $\epsilon$-approximate $t$-design is unclear, as the Haar ensemble expectation $\mathbb{E}^t_H(\rho)$ is a matrix comprising many different scalar entries, not a single scalar expectation value.  This is further complicated by the fact that inequality~(\ref{eq:def}) need not be symmetric, that is, the $\epsilon$ required to satisfy the left inequality may differ from the $\epsilon$ required to satisfy the right inequality, and only the larger of these, which is the $\epsilon$ required to satisfy inequality~(\ref{eq:def}), is known.  It is also unclear how $\epsilon$ can be linked to a distance measure.  Nevertheless, it is clear that at a fundamental level, $\epsilon$ quantifies an $\epsilon$-approximate $t$-design's ability to replicate the moments of the uniform Haar ensemble.  The smallest possible $\epsilon$ is zero, for which we recover an exact $t$-design, and any larger value quantifies the deviation from an exact $t$-design, which is unbounded in theory.  In practice, an arbitrarily chosen bound, which depends on the application at hand, is typically enforced~\cite{experimental2}.

Our models for a noisy $t$-design (see \secref{sec:noisemodelling}) rely heavily on the definition of an $\epsilon$-approximate $t$-design, as given by inequality (\ref{eq:def}).  We note that while there are many state-independent quantifiers of the extent to which a given ensemble of unitary operators deviates from an exact unitary $t$-design, such as the frame potential~\cite{framepot1, framepot2}, it is unclear how these can be applied in the context of noise modelling, since noise channels act on states and cannot be applied to the unitary operators directly.  In the next section we introduce the noise channels that we use to study noisy $t$-designs.

\subsection{Noise channels}\label{sec:ncbackground}

The action of a noise channel on an input density matrix $\rho$ is described by a completely positive and trace-preserving map $\varepsilon$ and the output density matrix is denoted by $\varepsilon(\rho)$.  In what follows we consider four different types of single-qubit noise channels.  These types of noise channels occur in many different physical systems~\cite{noise}.

\subsubsection{Flip channels}\label{sec:flipncbackground}

We consider the bit flip channel and the phase flip channel, as well as the bit and phase flip channel.  The bit flip channel~\cite{textbook} is described by
\begin{equation}
\varepsilon(\rho)=pX\rho X + (1-p)\rho,
\label{eq:bflip}
\end{equation}
that is, a bit flip is applied to a state $\rho$ with probability $p$.  The phase flip channel~\cite{textbook} is described by
\begin{equation}
\varepsilon(\rho)=pZ\rho Z + (1-p)\rho,
\label{eq:pflip}
\end{equation}
that is, a phase flip is applied to a state $\rho$ with probability $p$.  The bit and phase flip channel~\cite{textbook} is described by
\begin{equation}
\varepsilon(\rho)=pY\rho Y + (1-p)\rho,
\label{eq:bpflip}
\end{equation}
that is, a bit and phase flip, in the form $Y=\text{i}XZ$, is applied to a state $\rho$ with probability $p$.

\subsubsection{Phase damping channel}\label{sec:phasencbackground}

Phase damping is information loss from a quantum system without energy loss.  The phase damping channel~\cite{textbook} is described by
\begin{equation}
\varepsilon(\rho)=E_0\rho E_0^{\dagger} + E_1\rho E_1^{\dagger},
\label{eq:phasedamp}
\end{equation}
where
\begin{align*}
E_0&=\begin{pmatrix}1&0\\0&\sqrt{1-\lambda}\end{pmatrix}, \\
E_1&=\begin{pmatrix}0&0\\0&\sqrt{\lambda}\end{pmatrix}, \\
\end{align*}
with $\lambda\in[0,1]$.  The advantage of this parameterisation is that it leads to a convenient description of maximal phase damping if we set $\lambda=1$.  This parameterisation is related to the conventional parameterisation of the phase damping channel by
\begin{equation}
e^{-\frac{t}{2T_2}}=\sqrt{1-\lambda},
\end{equation}
where $t$ is the time and $T_2$ is the phase damping time constant, so the phase damping rate is given by $\Gamma_{\text{PD}}=\frac{1}{2T_2}$.  The parameter $\lambda$ in the phase damping channel is related to the parameter $p$ in the phase flip channel by
\begin{equation}
p=\frac{1}{2}\left(1+\sqrt{1-\lambda}\right).
\label{eq:damptoflip}
\end{equation}

\subsubsection{Amplitude damping channel}\label{sec:ampncbackground}

Amplitude damping is energy loss from a quantum system.  Energy loss occurs when the computational basis state $\ket{1}$ (excited state) decays into the computational basis state $\ket{0}$ (ground state).  The amplitude damping channel~\cite{textbook} is described by
\begin{equation}
\varepsilon(\rho)=E_0\rho E_0^{\dagger} + E_1\rho E_1^{\dagger},
\label{eq:ampdamp}
\end{equation}
where
\begin{align*}
E_0&=\begin{pmatrix}1&0\\0&\sqrt{1-\lambda}\end{pmatrix}, \\
E_1&=\begin{pmatrix}0&\sqrt{\lambda}\\0&0\end{pmatrix}, \\
\end{align*}
with $\lambda\in[0,1]$.  This parameterisation once again has the advantage that we can describe maximal amplitude damping by setting $\lambda=1$.  The parameterisation is related to the conventional parameterisation of the amplitude damping channel by
\begin{equation}
e^{-\frac{t}{2T_1}}=\sqrt{1-\lambda},
\end{equation}
where $t$ is the time and $T_1$ is the amplitude damping time constant, so the amplitude damping rate (decay rate) is given by $\Gamma_{\text{AD}}=\frac{1}{2T_1}$.

\subsubsection{Depolarising noise channel}\label{sec:depncbackground}

Depolarising noise is another common type of noise.  It is the simplest noise model for incoherent gate errors on noisy intermediate-scale quantum computers such as the IBM quantum processors~\cite{experimental2, dep1}.  The depolarising channel~\cite{textbook} is described by
\begin{equation}
\varepsilon(\rho)=\frac{p}{2}I + (1-p)\rho,
\label{eq:depnoise}
\end{equation}
that is, a state $\rho$ is replaced by the maximally mixed state with probability $p$.  This channel can also be written as
\begin{equation}
\varepsilon(\rho)=\frac{p}{3}\left(X\rho X + Y\rho Y + Z\rho Z\right) + (1-p)\rho,
\end{equation}
that is, in the depolarising channel, a bit flip, a phase flip, and a bit and phase flip are each applied with probability $\frac{p}{3}$.  Specialised error mitigation techniques are available to reduce the effect of depolarising noise on quantum computers~\cite{dep2}.  However, these methods can only be applied in applications of $t$-designs where the final outcome is an expectation value.

\section{Noise modelling}\label{sec:noisemodelling}

Our models for a noisy $t$-design use an adapted form of inequality (\ref{eq:def}), the defining inequality for an approximate $t$-design.  Even though the definition applies to any density matrix in $\mathcal{B}((\mathbb{C}^2)^{\otimes t})$, we restrict our noise models to density matrices which are $t$ copies of an arbitrary single-qubit density matrix, as was done in our previous work~\cite{experimental2}.  This has two major benefits, namely, that numeric results can be obtained efficiently for all $t$, since the number of parameters that need to be varied when creating samples of density matrices remains constant with increasing $t$, and that numeric results can be analysed geometrically, since single-qubit states can be represented by points in the Bloch sphere.  We therefore quantify the effect of a noise channel $\varepsilon$ on the quality of an exact single-qubit $t$-design $\{p_i,U_i\}$ using the smallest possible $\epsilon$ such that the inequality
\begin{equation}
(1-\epsilon)\mathbb{E}^t_H(\rho^{\otimes t})\leq\widetilde{\mathbb{E}}^t_H(\rho)\leq(1+\epsilon)\mathbb{E}^t_H(\rho^{\otimes t})
\label{eq:model}
\end{equation}
holds for all single-qubit density matrices $\rho$.  This $\epsilon$ quantifies the noisy $t$-design's ability to replicate the moments of the uniform Haar ensemble and represents a lower bound in the more general definition of an approximate $t$-design where $\rho\in\mathcal{B}((\mathbb{C}^2)^{\otimes t})$.

The definition of $\widetilde{\mathbb{E}}^t_H(\rho)$ depends on the noise model.  Inspired by the noise models typically used in randomised benchmarking~\cite{RB1, RB2, RB3, RB4, RB5, RB6, RB7, RB8, RB9}, we consider a noise model in which noise is applied before the unitary operations, for which we define
\begin{equation}
\widetilde{\mathbb{E}}^t_H(\rho)=\sum_{i}p_i\left(U_i\varepsilon(\rho) U_i^{\dagger}\right)^{\otimes t},
\label{eq:before}
\end{equation}
as well as a noise model in which noise is applied after the unitary operations, for which we define
\begin{equation}
\widetilde{\mathbb{E}}^t_H(\rho)=\sum_{i}p_i\left(\varepsilon\left(U_i\rho U_i^{\dagger}\right)\right)^{\otimes t}.
\label{eq:after}
\end{equation}
In both models, the same noise channel $\varepsilon$ is applied to each single-qubit state in the $t$-fold tensor product.  Both noise models are well-defined in the sense that the value of $\epsilon$ obtained is independent of the choice of ensemble and a general property of the $t$-design for a given $t$.  This is proven in \appendref{append:well-definedness}.

Based on the fact that for $t \ge 2$ a $t$-design transforms any noise channel into a depolarising channel~\cite{RB1, RB2, RB3, RB7, RB8, RB9}, one might expect our two noise models to be equivalent.  However, since we are studying the effect of a noise channel on the quality of a $t$-design, not the effect of a $t$-design on a noise channel, equivalence of the noise models is a question of whether the noisy Haar ensemble expectations given by Eqs.~(\ref{eq:before}) and~(\ref{eq:after}) are equal, not a question of whether the resulting depolarising channels are equal for the two models.  Our noise models are therefore not generally equivalent.  Furthermore, the $t$-fold tensor product makes it difficult to find a relation between the two noisy Haar ensemble expectations by commuting noise through the unitary operations.

We note that a third noise model, in which noise is applied during the unitary operations, could also be considered due to the finite time duration for these operations in an experimental realisation.  However, this is dependent on the method used to implement the unitaries in an experiment (e.g.,~with control pulses) and so we focus on the former two models which are implementation independent.

\section{Results}\label{sec:results}

Analytic results for the 1-design are presented in \appendref{append:1-design}.  For the model where noise is applied before the unitary operations, we were able to show that the quality of the 1-design is completely unaffected by an arbitrary noise channel, and for the model where noise is applied after the unitaries, we showed that the quality of the 1-design is unaffected by noise, unless amplitude damping is applied.  Furthermore, we showed that $\epsilon=\lambda$ quantifies the effect of the amplitude damping channel on the quality of the 1-design for the model where noise is applied after the unitaries.

When $t>1$, $\mathbb{E}^t_H(\rho^{\otimes t})$ depends on the state $\rho$, which makes it very difficult to obtain results analytically, since inequality~(\ref{eq:model}) contains variables other than $\epsilon$ and the noise parameter ($p$ or $\lambda$) and so it is difficult to obtain an expression for $\epsilon$ in terms of the noise parameter.  Numeric results were therefore obtained for the 2-design, 3-design, 4-design, and 5-design.  Few exact single-qubit $t$-designs exist for $t>5$ and so it is hard to obtain results for $t>5$.  It is also of little interest at present, since there are only a few known applications of $t$-designs for $t>4$.

With the notable exception of the amplitude damping channel, numeric results obtained for the 3-design are identical to those obtained for the 2-design and numeric results obtained for the 5-design are identical to those obtained for the 4-design.  Hence, in the sections which follow and in the appendices referenced, we only present numeric results for the 2-design and the 4-design, unless amplitude damping was applied.

\subsection{Implementation}\label{sec:impresults}

To obtain numeric results we require samples of single-qubit density matrices.  In our previous work~\cite{experimental2}, in which we numerically investigated the effect of depolarising noise on the quality of the 2-design and the 3-design, we found that the $\epsilon$ needed to satisfy inequality (\ref{eq:model}) for pure states is very large even for small values of the depolarising noise parameter $p$ (see \equatref{eq:depnoise}).  For a comprehensive numerical investigation, we therefore do not simply consider a sample of single-qubit density matrices distributed over the entire Bloch sphere, but rather consider various samples of density matrices restricted to different regions of the Bloch sphere.  Opening the study of noise in regions of the Bloch sphere may provide useful information that could be used in specific applications of $t$-designs, for instance, those where the full state space is not required.

To generate a sample of density matrices, we first generate 11 evenly spaced values of $r\in [0, r_t]$, 11 evenly spaced values of $\theta\in [0, \theta_t]$, and 11 evenly spaced values of $\phi\in [0, \phi_t]$, where $r_t$, $\theta_t$, and $\phi_t$ are the points of truncation of the radial coordinate, the polar angle, and the azimuthal angle, respectively.  Together these truncation points define the region of the Bloch sphere being considered.  Unless stated otherwise, $r_t=1$, $\theta_t=\pi$, and $\phi_t=2\pi$ were used so that the entire Bloch sphere was considered.  We then convert all $11^3=1331$ possible combinations of the generated values of the spherical coordinates $(r, \theta, \phi)$ into the Cartesian coordinates $(x, y, z)$ using
\begin{align*}
x&=r\sin{\theta}\cos{\phi}, \\
y&=r\sin{\theta}\sin{\phi}, \\
z&=r\cos{\theta} \\
\end{align*}
and obtain a sample of 1331 density matrices using~\cite{rho}
\begin{equation}
\rho=\frac{1}{2}\begin{pmatrix}1+z&x-\text{i}y\\x+\text{i}y&1-z\\\end{pmatrix}.
\label{eq:rho}
\end{equation}

Given $t\in\{2,3,4,5\}$, a noise channel, a noise model, and a sample of density matrices, we obtain $\epsilon$ numerically as follows.  For each single-qubit density matrix $\rho$ in the sample, we calculate $\mathbb{E}^t_H(\rho^{\otimes t})$ and $\widetilde{\mathbb{E}}^t_H(\rho)$ using the icosahedral group~\cite{platonic} (an exact unitary 5-design and therefore also an exact unitary $t$-design for any $t \leq 5$) and determine the smallest possible $\epsilon$ such that inequality (\ref{eq:model}) is satisfied.  The largest $\epsilon$ found is the smallest possible $\epsilon$ such that inequality (\ref{eq:model}) holds for all density matrices in the sample and is therefore the value with which we quantify the effect of the given noise channel on the quality of the $t$-design.

\subsection{Numeric results}\label{sec:numresultsbefore}

This section covers numeric results obtained for the model where noise is applied before the unitary operations.  Numeric results obtained for the model where noise is applied after the unitary operations follow similar trends for a given noise channel and are presented in \appendref{append:numresultsafter}.  The primary difference is that the values of $\epsilon$ obtained for a given $t$ and noise parameter ($p$ or $\lambda$) are generally slightly smaller for the model where noise is applied after the unitary operations.  For the depolarising noise channel, we prove in \appendref{append:depolarising} that the values of $\epsilon$ obtained for the two noise models are equal, and for the other noise channels, numeric results obtained for the model where noise is applied after the unitary operations reflect the transformation of the noise channel into a depolarising noise channel, an effect exploited in randomised benchmarking with 2-designs~\cite{RB1, RB2, RB3, RB7, RB8, RB9}.

\begin{figure*}
    \centering
    \begin{subfigure}[b]{.42\textwidth}
        \centering
        \includegraphics[scale=0.55]{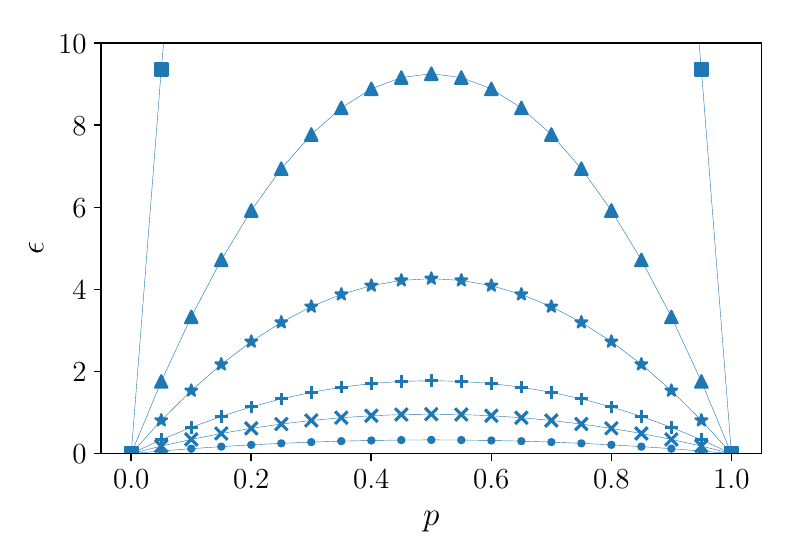}
        \caption{2-design}
        \label{fig:beforebflip2}
    \end{subfigure}
    \begin{subfigure}[b]{.42\textwidth}
        \centering
        \includegraphics[scale=0.55]{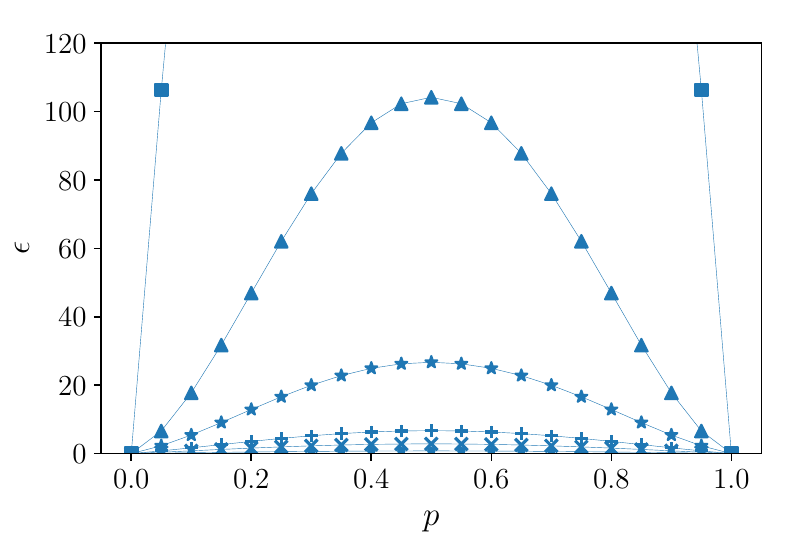}
        \caption{4-design}
        \label{fig:beforebflip4}
    \end{subfigure}
    \includegraphics[trim=0cm 3.8cm 0cm 7cm, clip, scale=0.52]{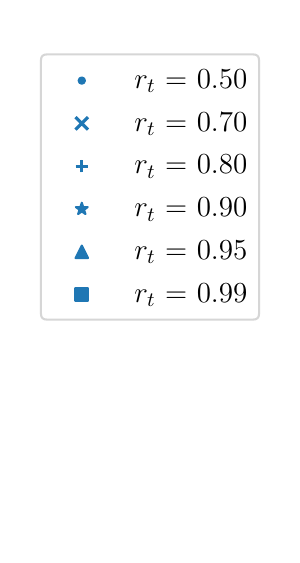}
    \caption{Effect of the bit flip channel (see \equatref{eq:bflip}) on the quality of the (a)~2-design and (b)~4-design for the model where noise is applied before the unitary operations, for different truncation radii $r_t$.}
\end{figure*}

\begin{figure}
    \centering
    \includegraphics[scale=0.58]{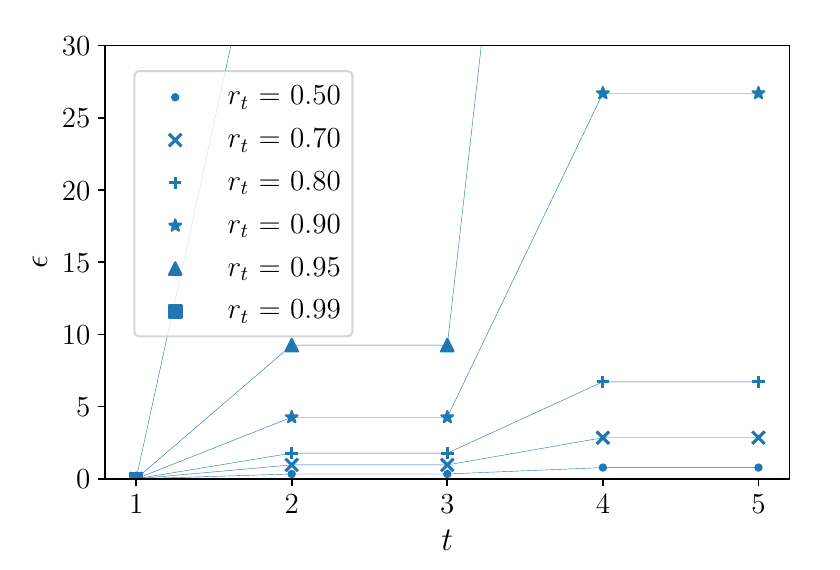}
    \caption{Plot of $\epsilon$ versus $t$ for the bit flip channel with $p=0.5$ (see \equatref{eq:bflip}) for the model where noise is applied before the unitary operations, for different truncation radii $r_t$.}
    \label{fig:beforebflip}
\end{figure}

\subsubsection{Flip channels}\label{sec:flipnumresultsbefore}

The effect of the bit flip channel (see \equatref{eq:bflip}) on the quality of the 2-design is shown in \figref{fig:beforebflip2}.  For each truncation radius considered (truncation angles fixed at $\theta_t=\pi$ and $\phi_t=2\pi$), $\epsilon$ versus $p$ is a parabola with maximum at $p=0.5$.  The maxima increase with increasing truncation radius, which shows that as the set of states considered is expanded to include states closer to the pure states at $r_t=1$, the 2-design becomes more sensitive to bit flips.  The symmetry of $\epsilon$ versus $p$ around $p=0.5$ can be explained as follows.  Note that states along the $x$-axis of the Bloch sphere, which are eigenstates of the Pauli $X$ operator, are unaffected by bit flips.  However, applying a bit flip to a state off the $x$-axis and its reflection in the $x$-axis with probability $p<0.5$ shifts both states by the same distance towards the $x$-axis of the Bloch sphere.  On the other hand, applying a bit flip to the state and its reflection with probability $p'=1-p>0.5$ shifts the state across the $x$-axis, to where its reflection was shifted when applying a bit flip with probability $p$, and shifts the state's reflection to where the state was shifted when applying a bit flip with probability $p$.  Therefore, applying a bit flip to all states in a sphere of radius $r_t$ with probability $p$ results in the same set of states as applying a bit flip to all states in that sphere with probability $1-p$.  Since $\epsilon$ must ensure that inequality (\ref{eq:model}) is satisfied for all states considered, $\epsilon$ depends only on the effect of the bit flip channel on the set of states in the sphere considered (not on the effect on individual states).  Hence the $\epsilon$ computed for a bit flip with probability $p$ is equal to the $\epsilon$ computed for a bit flip with probability $1-p$, so $\epsilon$ versus $p$ is symmetric about $p=0.5$.

The effect of the bit flip channel on the quality of the 4-design is shown in \figref{fig:beforebflip4}.  For each truncation radius, the maximum of $\epsilon$ versus $p$ still occurs at $p=0.5$, but $\epsilon$ versus $p$ now has a more sinusoidal shape.  The values of $\epsilon$ obtained for the 4-design are up to an order of magnitude larger than those obtained for the 2-design, for a fixed $p$ and $r_t$.  This shows that the 4-design is significantly more sensitive to bit flips than the 2-design.  To visualise the variation in sensitivity, we plot $\epsilon$ versus $t$ for $p=0.5$.  As can be seen in \figref{fig:beforebflip}, $\epsilon$ versus $t$ is a step function.  There is no increase in sensitivity to bit flips from $t=2$ to $t=3$ or from $t=4$ to $t=5$, but a significant increase in sensitivity from $t=3$ to $t=4$.

Numeric results obtained for the phase flip channel (see \equatref{eq:pflip}) and the bit and phase flip channel (see \equatref{eq:bpflip}) are identical to those obtained for the bit flip channel (shown in Figs.~\ref{fig:beforebflip2}, \ref{fig:beforebflip4}, and \ref{fig:beforebflip}).  To further investigate similarities and differences in the effect of these three channels on the quality of $t$-designs, we determine and visualise the region of the Bloch sphere (which we will refer to as the region of acceptable quality) for which a noisy $t$-design is able to replicate the moments of the uniform Haar ensemble, with a predefined accuracy, up to order $t$.  This investigation is presented in \appendref{append:regions}.  For each of the three flip channels, we find that the shape of the region of acceptable quality is similar to the shape into which the Bloch sphere is deformed by the relevant channel.  For example, the region of acceptable quality is an ellipsoid along the $x$-axis for the bit flip channel.  Since bit flips are performed by applying the Pauli $X$ operator to a state, states along the $x$-axis, which are closer to the eigenstates of the Pauli $X$ operator, are less affected by bit flips and so the quality remains acceptable for states along the $x$-axis even for a large bit flip probability.  The regions of acceptable quality for the three flip channels are thus identical up to a rotation, for a fixed $p$ and $t$, which explains why $\epsilon$ versus $p$ is the same for all three flip channels, for a fixed $r_t$ and $t$, as the full range of $\theta$ and $\phi$ is considered.

To further analyse the dependence of $\epsilon$ versus $p$ on the region of the Bloch sphere considered, we vary the truncation of the polar angle $\theta_t$ and the truncation of the azimuthal angle $\phi_t$.  These investigations are included in Appendices~\ref{append:polar} and \ref{append:azimuthal}, respectively.  We find that the phase flip channel is the only flip channel for which $\epsilon$ versus $p$ has a non-trivial dependence on $\theta_t$.  As $\theta_t$ is increased from 0 to $\frac{\pi}{2}$, the sample of density matrices is expanded to include states which are further from the eigenstates of the Pauli $Z$ operator and therefore more sensitive to phase flips, which results in a non-trivial dependence for the phase flip channel.  On the other hand, the states along the positive $z$-axis, which are among the furthest from the eigenstates of the Pauli $X$ and $Y$ operators, and therefore among the most sensitive to bit flips, and bit and phase flips, are included in the sample of density matrices for all $\theta_t$ and so the results for the bit flip channel and the bit and phase flip channel are independent of $\theta_t$.  The results are independent of $\phi_t$ for all three flip channels.  For the bit flip channel and the bit and phase flip channel, this can be attributed to the fact that the states along the positive $z$-axis are included in the sample of density matrices for all $\phi_t$.  For the phase flip channel, the independence of $\phi_t$ can be attributed to the fact that the smallest $\epsilon$ such that inequality (\ref{eq:model}) holds for a given state remains unchanged when that state is rotated about the $z$-axis.

\begin{figure*}
    \centering
    \begin{subfigure}[b]{.42\textwidth}
        \centering
        \includegraphics[scale=0.55]{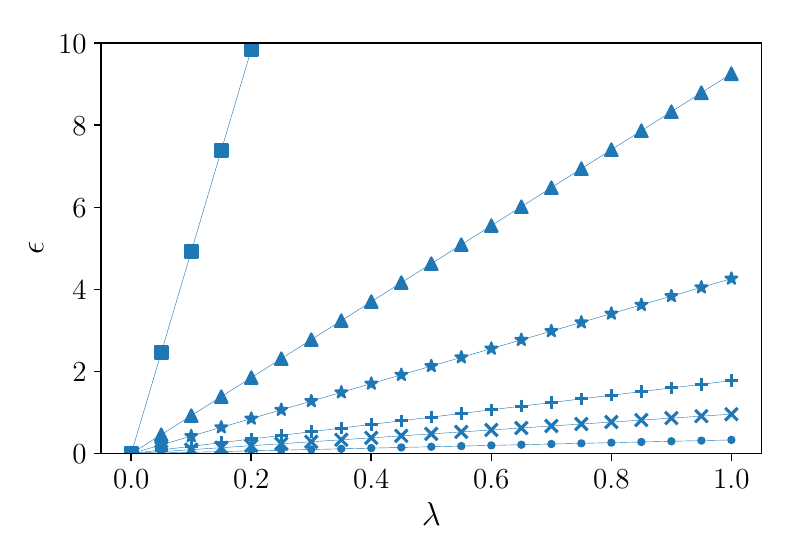}
        \caption{2-design}
        \label{fig:beforephasedamp2}
    \end{subfigure}
    \begin{subfigure}[b]{.42\textwidth}
        \centering
        \includegraphics[scale=0.55]{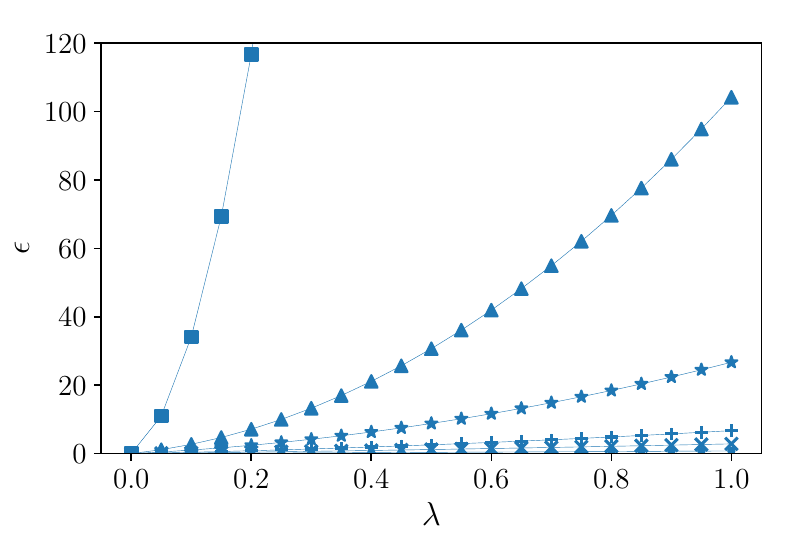}
        \caption{4-design}
        \label{fig:beforephasedamp4}
    \end{subfigure}
    \includegraphics[trim=0cm 3.8cm 0cm 7cm, clip, scale=0.52]{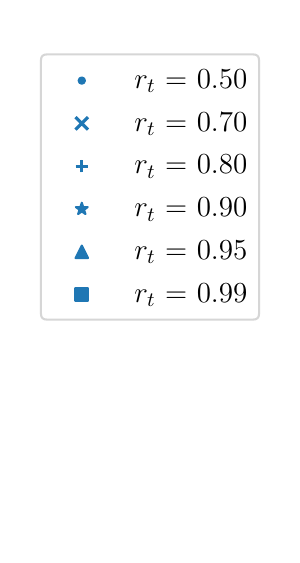}
    \caption{Effect of the phase damping channel (see \equatref{eq:phasedamp}) on the quality of the (a)~2-design and (b)~4-design for the model where noise is applied before the unitary operations, for different truncation radii $r_t$.}
\end{figure*}

\begin{figure}
    \centering
    \includegraphics[scale=0.58]{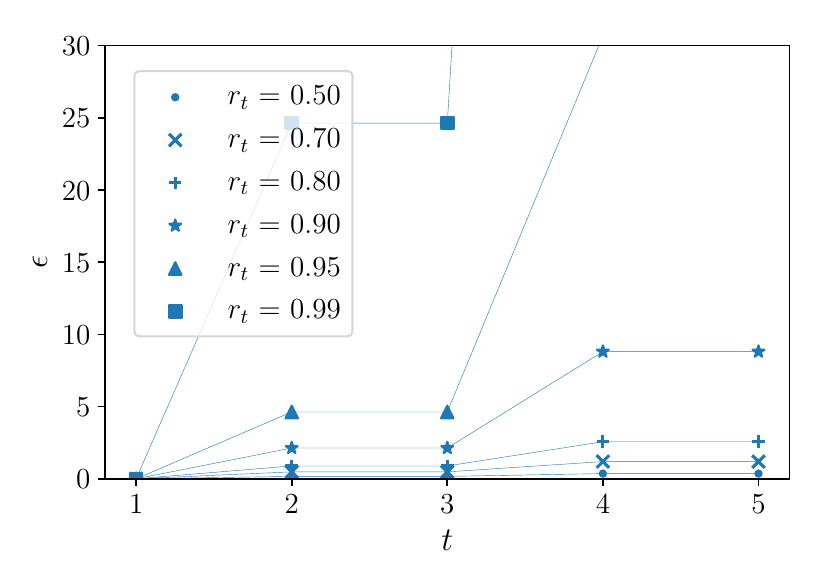}
    \caption{Plot of $\epsilon$ versus $t$ for the phase damping channel with $\lambda=0.5$ (see \equatref{eq:phasedamp}) for the model where noise is applied before the unitary operations, for different truncation radii $r_t$.}
    \label{fig:beforephasedamp}
\end{figure}

\subsubsection{Phase damping channel}\label{sec:phasenumresultsbefore}

The effect of the phase damping channel (see \equatref{eq:phasedamp}) on the quality of the 2-design is shown in \figref{fig:beforephasedamp2}.  For each truncation radius, $\epsilon$ increases linearly with $\lambda$.  The linear relation between $\epsilon$ and the parameter $\lambda$ in the phase damping channel can be attributed to the fact that both $\epsilon$ and $\lambda$ are quadratic functions of the parameter $p$ in the phase flip channel (see \figref{fig:beforebflip2} and \equatref{eq:damptoflip}, respectively).  The gradient of $\epsilon$ versus $\lambda$ increases with increasing truncation radius, similar to the way in which the maximum of $\epsilon$ versus $p$ increases with increasing truncation radius for the phase flip channel.

For the 4-design, $\epsilon$ versus $\lambda$ has a more exponential shape (see \figref{fig:beforephasedamp4}).  For a fixed $\lambda$, $\epsilon$ versus $t$ is a step function, as shown in \figref{fig:beforephasedamp}, just as $\epsilon$ versus $t$ is a step function for a fixed phase flip probability $p$.  We also investigate variations in sensitivity to phase damping throughout the Bloch sphere.  We find that both the shape of the region of acceptable quality and the dependence of $\epsilon$ versus $\lambda$ on $\theta_t$ and $\phi_t$ are similar to those of the phase flip channel (see Appendices~\ref{append:regions}, \ref{append:polar}, and \ref{append:azimuthal}, respectively).

\begin{figure*}
    \centering
    \includegraphics[scale=0.55]{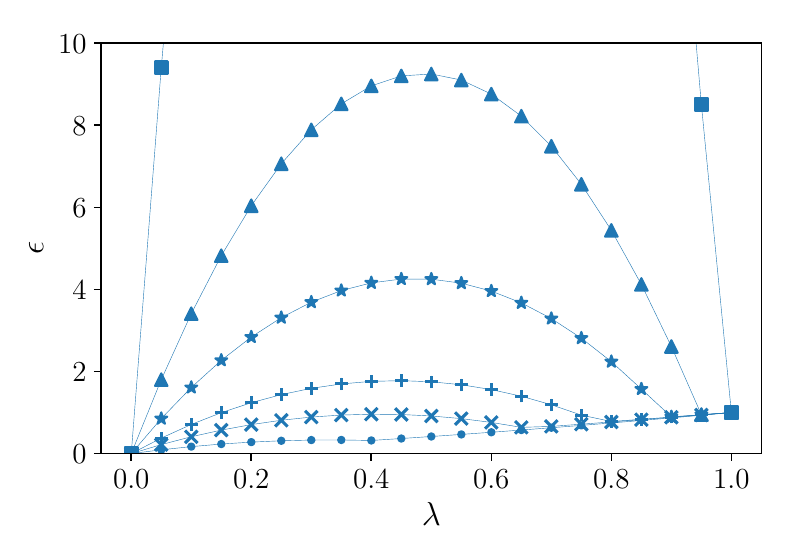}
    \includegraphics[scale=0.55]{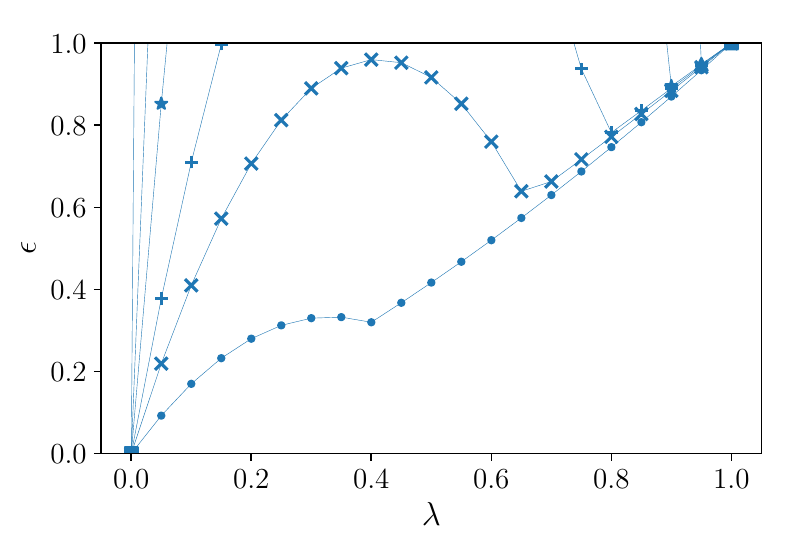}
    \includegraphics[trim=0cm 5cm 0cm 7cm, clip, scale=0.52]{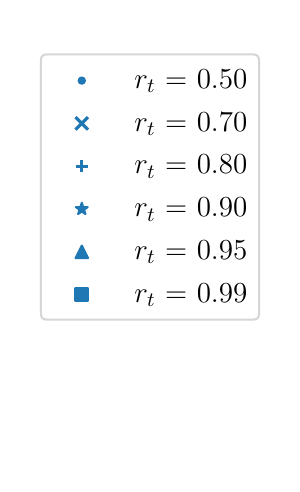}
    \caption{Effect of the amplitude damping channel (see \equatref{eq:ampdamp}) on the quality of the 2-design for the model where noise is applied before the unitary operations, for different truncation radii $r_t$.  The full set of results is shown on the left and the region in which the anomaly occurs is shown enlarged on the right.}
    \label{fig:beforeampdamp2}
\end{figure*}

\begin{figure*}
    \centering
    \includegraphics[scale=0.55]{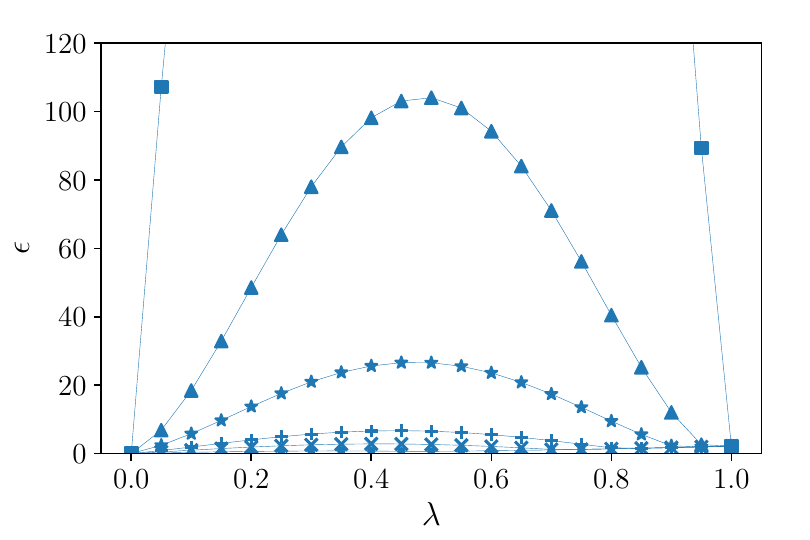}
    \includegraphics[scale=0.55]{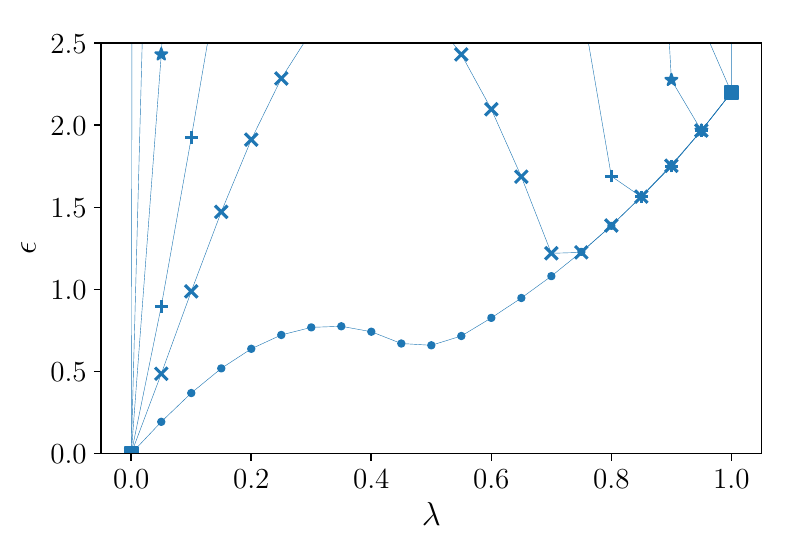}
    \includegraphics[trim=0cm 5cm 0cm 7cm, clip, scale=0.52]{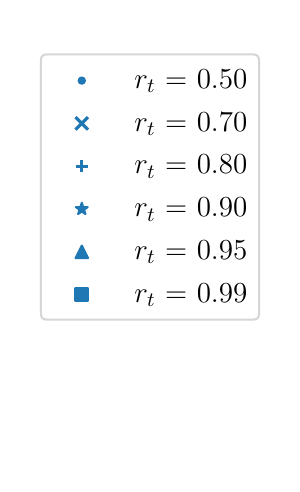}
    \caption{Effect of the amplitude damping channel (see \equatref{eq:ampdamp}) on the quality of the 4-design for the model where noise is applied before the unitary operations, for different truncation radii $r_t$.  The full set of results is shown on the left and the region in which the anomaly occurs is shown enlarged on the right.}
    \label{fig:beforeampdamp4}
\end{figure*}

\begin{figure}
    \centering
    \includegraphics[scale=0.58]{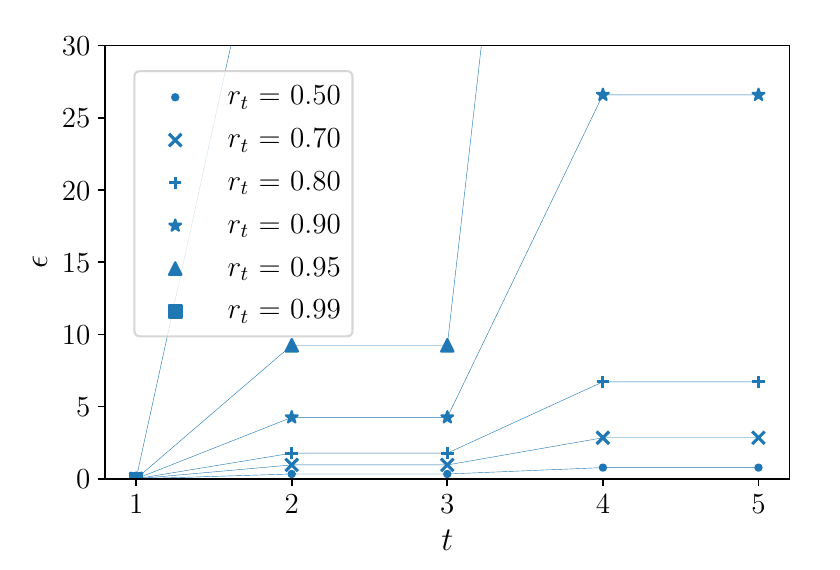}
    \caption{Plot of $\epsilon$ versus $t$ for the amplitude damping channel, with the parameter $\lambda$ (see \equatref{eq:ampdamp}) taken to be the turning point of $\epsilon$ versus $\lambda$, for the model where noise is applied before the unitary operations, for different truncation radii $r_t$.}
    \label{fig:beforeampdamp}
\end{figure}

\subsubsection{Amplitude damping channel}\label{sec:ampnumresultsbefore}

The effect of the amplitude damping channel (see \equatref{eq:ampdamp}) on the quality of the 2-design is shown in \figref{fig:beforeampdamp2}.  For the most part, $\epsilon$ versus $\lambda$ is a parabola with maximum either at or close to $\lambda=0.5$, but an anomaly occurs for large $\lambda$ and small $r_t$, where at a given $\lambda$, the trend spontaneously changes to strictly increasing.  Just as for the bit flip channel, the maxima of $\epsilon$ versus $\lambda$ increase with increasing truncation radius.  The similarities to the bit flip channel are to be expected, considering that the amplitude damping channel actually performs a bit flip on the state \ket{1} with a given probability (the difference being that the state \ket{0} is never flipped by the amplitude damping channel).  Bearing in mind that the amplitude damping channel shrinks and shifts any subsphere of states in the Bloch sphere up to the state \ket{0}, the anomaly can be interpreted as follows for a given $r_t$.  At the turning point of $\epsilon$ versus $\lambda$, the south pole of the shifted sphere crosses that sphere's initial equator.  The anomaly, where the trend changes to strictly increasing, occurs at the point where the south pole of the shifted sphere crosses its initial north pole.  In the limit $\lambda\to 1$ (maximal amplitude damping), all spheres are reduced to the state \ket{0}, which explains why $\epsilon=1$ for all $r_t$ at $\lambda=1$.

Just as for the other noise channels, numeric results obtained for the 3-design are identical to those obtained for the 2-design.  The effect of the amplitude damping channel on the quality of the 4-design is shown in \figref{fig:beforeampdamp4}.  The maxima of $\epsilon$ versus $\lambda$ occur at the same values of $\lambda$ as for the 2-design, and the anomaly still occurs for large $\lambda$ and small $r_t$, but $\epsilon$ versus $\lambda$ has a more sinusoidal shape.  Numeric results obtained for the 5-design are almost identical to those obtained for the 4-design.  The only difference is that for the 4-design $\epsilon$ increases to $2.20$ as $\lambda\to 1$, whereas for the 5-design $\epsilon$ increases to $4.33$ as $\lambda\to 1$.  We compare the values of $\epsilon$ obtained for different $t$-designs, for a fixed $\lambda$ and $r_t$, by plotting $\epsilon$ versus $t$ for different truncation radii, each time using the turning point of $\epsilon$ versus $\lambda$ as our fixed value of $\lambda$ (see \figref{fig:beforeampdamp}).  We find that $\epsilon$ versus $t$ is once again a step function and see a significant increase in sensitivity to amplitude damping from $t=3$ to $t=4$.  However, we note that for larger $\lambda$ and smaller $r_t$ there is also a slight increase in sensitivity to amplitude damping from $t=4$ to $t=5$.  The region of acceptable quality for the amplitude damping channel and the dependence of $\epsilon$ versus $\lambda$ on $\theta_t$ and $\phi_t$ are analysed in Appendices~\ref{append:regions}, \ref{append:polar}, and \ref{append:azimuthal}, respectively.

\begin{figure*}
    \centering
    \begin{subfigure}[b]{.42\textwidth}
        \centering
        \includegraphics[scale=0.55]{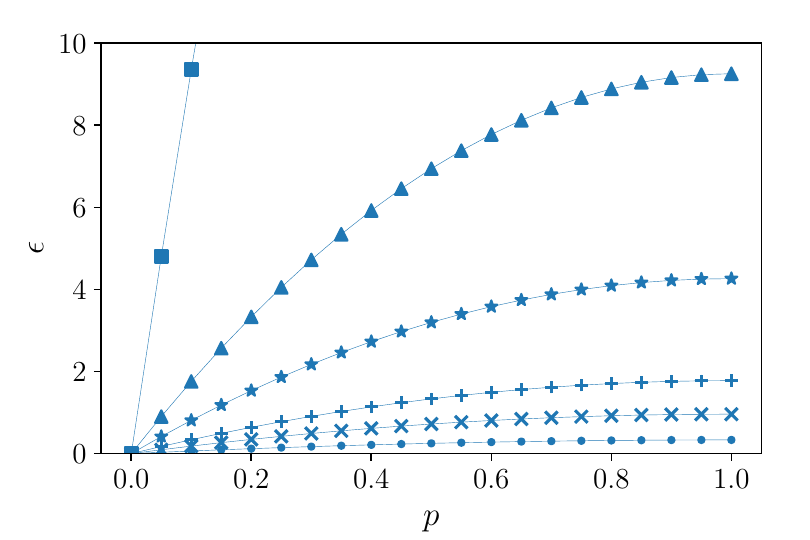}
        \caption{2-design}
        \label{fig:beforedepnoise2}
    \end{subfigure}
    \begin{subfigure}[b]{.42\textwidth}
        \centering
        \includegraphics[scale=0.55]{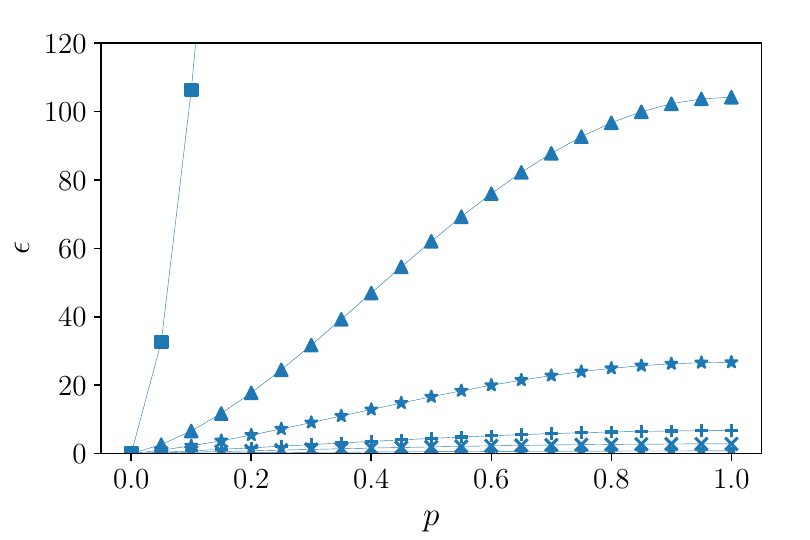}
        \caption{4-design}
        \label{fig:beforedepnoise4}
    \end{subfigure}
    \includegraphics[trim=0cm 3.8cm 0cm 7cm, clip, scale=0.52]{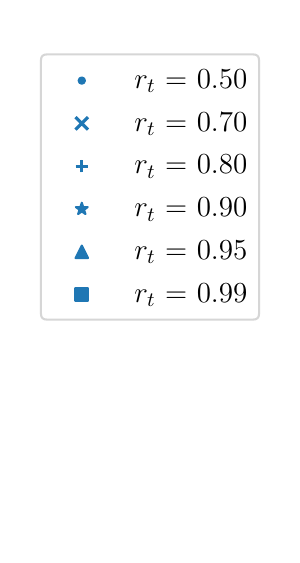}
    \caption{Effect of the depolarising noise channel (see \equatref{eq:depnoise}) on the quality of the (a)~2-design and (b)~4-design for the model where noise is applied before the unitary operations, for different truncation radii $r_t$.}
\end{figure*}

\begin{figure}
    \centering
    \includegraphics[scale=0.58]{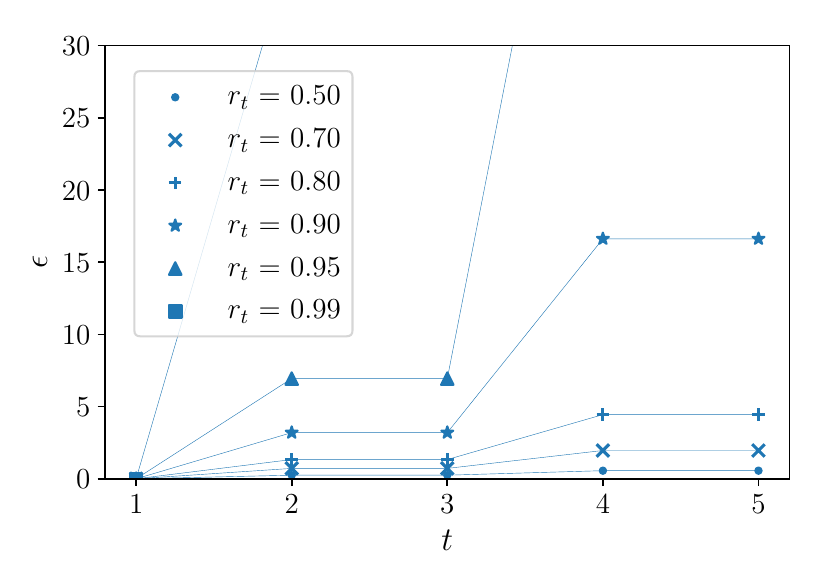}
    \caption{Plot of $\epsilon$ versus $t$ for the depolarising noise channel with $p=0.5$ (see \equatref{eq:depnoise}) for the model where noise is applied before the unitary operations, for different truncation radii $r_t$.}
    \label{fig:beforedepnoise}
\end{figure}

\subsubsection{Depolarising noise channel}\label{sec:depnumresultsbefore}

Finally, the effect of the depolarising channel (see \equatref{eq:depnoise}) on the quality of the 2-design is shown in \figref{fig:beforedepnoise2}.  These results have previously been published as part of the supplementary information for Ref.~\cite{experimental2}.  For each truncation radius considered, $\epsilon$ increases linearly with $p$, up to about $p=0.4$, after which the increase becomes more gradual.  In \figref{fig:beforedepnoise4} we see that for the 4-design, $\epsilon$ first increases rapidly with $p$ and then increases linearly with $p$ for $p\in[0.3, 0.6]$, after which the increase becomes more gradual.  Just as for all the other noise channels, the values of $\epsilon$ obtained for the 4-design are up to an order of magnitude larger than those obtained for the 2-design, for a fixed $p$ and $r_t$, once again confirming that the 4-design is significantly more sensitive to noise than the 2-design.  As expected, $\epsilon$ versus $t$ is a step function for a fixed $p$ and $r_t$ (see \figref{fig:beforedepnoise}).

As illustrated in \appendref{append:regions}, the region of acceptable quality for the depolarising noise channel has a spherical shape.  As such, the numeric results obtained for $\epsilon$ versus $p$ are independent of $\theta_t$ and $\phi_t$ (see Appendices~\ref{append:polar} and \ref{append:azimuthal}, respectively).

\section{Conclusion}\label{sec:conclusion} 

We studied the effect of different types of noise on the quality of single-qubit $t$-designs.  While we hope that our study will encourage research into the effect of noise on the quality of multi-qubit $t$-designs, we also note that there are many protocols which exclusively use single-qubit $t$-designs~\cite{appl2-1, appl3-1, appl3-2, applsingle} for which our work may have direct consequences.  The noise channels we investigated are the bit flip channel, the phase flip channel, the bit and phase flip channel, the phase damping channel, the amplitude damping channel, and the depolarising noise channel.  We quantified the effect of a noise channel on the quality of a $t$-design using the smallest possible $\epsilon$ such that a test inequality, adapted from the defining inequality for an $\epsilon$-approximate $t$-design, holds for all density matrices in a given sample.  Two noise models were considered, namely, a noise model in which noise is applied before the unitary operations and a noise model in which noise is applied after the unitary operations, in line with the noise models used in randomised benchmarking~\cite{RB1, RB2, RB3, RB4, RB5, RB6, RB7, RB8, RB9}.

We showed analytically that for the model where noise is applied before the unitary operations, the quality of the 1-design is completely unaffected by an arbitrary noise channel, and for the model where noise is applied after the unitaries, the quality of the 1-design is unaffected by noise, unless amplitude damping is applied (see \appendref{append:1-design}).  For the 2-design, 3-design, 4-design, and 5-design, results were obtained numerically using the icosahedral group~\cite{platonic}.  With the exception of the amplitude damping channel, $\epsilon$ versus $t$ is a step function for a fixed $p$ or $\lambda$.  We saw a significant increase in sensitivity to noise from $t=1$ to $t=2$ and an even larger increase in sensitivity to noise from $t=3$ to $t=4$, but no increase in sensitivity to noise from $t=2$ to $t=3$ or from $t=4$ to $t=5$, unless amplitude damping was applied.  Based on these results, we conjecture that for any $t$, a $(2t+1)$-design is as sensitive to noise as a $2t$-design, for any noise channel which deforms the Bloch sphere, but does not shift the Bloch sphere.  While it may be possible to prove this with induction on $t$ using recently discovered random circuit constructions for exact $t$-designs~\cite{appl4-2}, such a proof evaded the authors.  Further work in this direction is needed.  Developing and studying noise models which use the definition of a $t$-design in terms of a polynomial function~\cite{RB1} may also help to uncover this even versus odd behaviour.

For all the noise channels considered and for both noise models, $\epsilon$ increases with increasing truncation radius, for a fixed $t$ and noise parameter ($p$ or $\lambda$).  Hence $t$-designs become increasingly sensitive to noise as the set of states considered is expanded to include states further from the maximally mixed state at $r_t=0$ (for which the sensitivity to noise is least) and closer to the pure states at $r_t=1$ (for which the sensitivity to noise is greatest).  To further investigate variations in sensitivity to noise throughout the Bloch sphere, we determined the region of acceptable quality (region of the Bloch sphere for which a noisy $t$-design is able to replicate the moments of the uniform Haar ensemble, with a predefined accuracy, up to order $t$) for each of the noise channels (see \appendref{append:regions}).  For the model where noise is applied before the unitary operations, the shape of the region of acceptable quality for each noise channel is similar to the shape into which the Bloch sphere is deformed by the relevant noise channel.

For the model where noise is applied after the unitary operations, the region of acceptable quality has a spherical shape for all the noise channels considered.  Hence our numeric results reflect the transformation of a noise channel into a depolarising channel, an effect exploited in randomised benchmarking with 2-designs~\cite{RB1, RB2, RB3, RB7, RB8, RB9}, when the noise is applied after the unitary operations.  For the depolarising noise channel, our two noise models are equivalent (proven in \appendref{append:depolarising}).  For the other noise channels, $t$-designs generally show reduced sensitivity to noise for the model where noise is applied after the unitary operations (see \appendref{append:numresultsafter}), which seems to suggest that the process by which a noise channel is transformed into a depolarising channel (so the quality of $t$-designs is affected equally for all states at a given radial distance from the maximally mixed state) mitigates the effect of the noise channel on the quality of $t$-designs.  Future work going beyond states that are $t$-fold tensor products of single-qubit states and their geometric interpretation, as well as investigations into the effect of noise on the quality of multi-qubit $t$-designs, will help to elucidate further behaviour of $t$-designs under the effects of noise.  This kind of work will be helpful for researchers studying and developing applications using $t$-designs under realistic conditions.

\begin{acknowledgements}
The authors thank Dario Trinchero for valuable comments and insights.  This research was supported by the South African National Research Foundation, the University of Stellenbosch, the South African Research Chair Initiative of the Department of Science and Technology and National Research Foundation, and the South African Quantum Technology Initiative of the Department of Science and Technology.
\end{acknowledgements}


\appendix

\onecolumngrid

\newpage

\section{Proof of well-definedness of noise models}\label{append:well-definedness} 

Let $\{p_i, U_i\}$ and $\{q_i, V_i\}$ be exact unitary $t$-designs and let
\begin{equation}
\varepsilon(\rho)=\sum_{k}E_k\rho E_k^{\dagger}
\label{eq:noisedef}
\end{equation}
be a noise channel.  We note that
\begin{equation}
\mathbb{E}^t_H(\rho^{\otimes t})=\sum_{i}p_i\left(U_i\rho U_i^{\dagger}\right)^{\otimes t}=\sum_{i}q_i\left(V_i\rho V_i^{\dagger}\right)^{\otimes t}
\label{eq:designdef}
\end{equation}
by the definition of an exact unitary $t$-design.  Let $\widetilde{\mathbb{E}}^t_{H,U}(\rho)$ and $\widetilde{\mathbb{E}}^t_{H,V}(\rho)$ denote $\widetilde{\mathbb{E}}^t_H(\rho)$ determined using $\{p_i, U_i\}$ and $\{q_i, V_i\}$, respectively, each with noise applied.  For both noise models, we will show that $\widetilde{\mathbb{E}}^t_{H,U}(\rho)=\widetilde{\mathbb{E}}^t_{H,V}(\rho)$ for all $\rho$, from which it follows that the smallest $\epsilon$ such that inequality (\ref{eq:model}) holds for all density matrices is the same for the two ensembles.

\subsection{Proof for the model where noise is applied before the unitary operations}\label{append:well-definednessbefore} 

Let $\rho$ be a density matrix.  Since $\varepsilon(\rho)$ is also a density matrix, it follows from \equatref{eq:before} that $\widetilde{\mathbb{E}}^t_{H,U}(\rho)=\mathbb{E}^t_H((\varepsilon(\rho))^{\otimes t})$ and $\widetilde{\mathbb{E}}^t_{H,V}(\rho)=\mathbb{E}^t_H((\varepsilon(\rho))^{\otimes t})$, so that $\widetilde{\mathbb{E}}^t_{H,U}(\rho)=\widetilde{\mathbb{E}}^t_{H,V}(\rho)$.

\subsection{Proof for the model where noise is applied after the unitary operations}\label{append:well-definednessafter} 

Let $\rho$ be a density matrix.  Substituting \equatref{eq:noisedef} into \equatref{eq:after} and using the algebraic properties of the tensor product, we obtain
\begin{align*}
\widetilde{\mathbb{E}}^t_{H,U}(\rho)&=\sum_{i}p_i\left(\sum_kE_kU_i\rho U_i^{\dagger}E_k^{\dagger}\right)^{\otimes t} \\
&=\sum_{i}p_i\sum_{k_1}\sum_{k_2}\cdots\sum_{k_t}E_{k_1}U_i\rho U_i^{\dagger}E_{k_1}^{\dagger}\otimes E_{k_2}U_i\rho U_i^{\dagger}E_{k_2}^{\dagger}\otimes\cdots\otimes E_{k_t}U_i\rho U_i^{\dagger}E_{k_t}^{\dagger} \\
&=\sum_{k_1}\sum_{k_2}\cdots\sum_{k_t}\sum_{i}p_i\left(\bigotimes_{j=1}^{t}E_{k_j}\right)\left(U_i\rho U_i^{\dagger}\right)^{\otimes t}\left(\bigotimes_{j=1}^{t}E_{k_j}^{\dagger}\right).
\numberthis
\label{eq:calculation}
\end{align*}
Multiplying both sides of \equatref{eq:designdef} by $\bigotimes_{j=1}^{t}E_{k_j}$ from the left and by $\bigotimes_{j=1}^{t}E_{k_j}^{\dagger}$ from the right, we get
\begin{equation}
\sum_{i}p_i\left(\bigotimes_{j=1}^{t}E_{k_j}\right)\left(U_i\rho U_i^{\dagger}\right)^{\otimes t}\left(\bigotimes_{j=1}^{t}E_{k_j}^{\dagger}\right)=\sum_{i}q_i\left(\bigotimes_{j=1}^{t}E_{k_j}\right)\left(V_i\rho V_i^{\dagger}\right)^{\otimes t}\left(\bigotimes_{j=1}^{t}E_{k_j}^{\dagger}\right).
\label{eq:term}
\end{equation}
Performing a term-by-term replacement in the $t$-fold summation of \equatref{eq:calculation} by substituting in \equatref{eq:term} yields
\begin{equation}
\widetilde{\mathbb{E}}^t_{H,U}(\rho)=\sum_{k_1}\sum_{k_2}\cdots\sum_{k_t}\sum_{i}q_i\left(\bigotimes_{j=1}^{t}E_{k_j}\right)\left(V_i\rho V_i^{\dagger}\right)^{\otimes t}\left(\bigotimes_{j=1}^{t}E_{k_j}^{\dagger}\right),
\end{equation}
from which it follows that $\widetilde{\mathbb{E}}^t_{H,U}(\rho)=\widetilde{\mathbb{E}}^t_{H,V}(\rho)$ if we then perform our original calculation in \equatref{eq:calculation} in reverse.

\section{Analytic results for the 1-design}\label{append:1-design} 

Using the Pauli 1-design, one can show that $\mathbb{E}^1_H(\rho)=\frac{1}{2}I$ for all density matrices $\rho$.  Let
\begin{equation}
\varepsilon(\rho)=\sum_{k}E_k\rho E_k^{\dagger}
\label{eq:noisechan}
\end{equation}
be an arbitrary noise channel.  For the model where noise is applied before the unitary operations, we will show that the quality of the 1-design is completely unaffected by an arbitrary noise channel, and for the model where noise is applied after the unitary operations, we will show that the quality of the 1-design is unaffected by noise, unless amplitude damping is applied.

\subsection{Analytic results for the model where noise is applied before the unitary operations}\label{append:1-designbefore} 

For any density matrix $\rho$, $\varepsilon(\rho)$ is a density matrix and so it follows from \equatref{eq:before} that $\widetilde{\mathbb{E}}^1_H(\rho)=\mathbb{E}^1_H(\varepsilon(\rho))=\frac{1}{2}I$.  It therefore follows that inequality (\ref{eq:model}) can be satisfied with $\epsilon=0$ and so the quality of the 1-design is completely unaffected by an arbitrary noise channel.

\subsection{Analytic results for the model where noise is applied after the unitary operations}\label{append:1-designafter} 

Let $\rho$ be a density matrix and let $\{p_i, U_i\}$ be an exact 1-design.  Substituting \equatref{eq:noisechan} into \equatref{eq:after}, we have
\begin{align*}
\widetilde{\mathbb{E}}^1_H(\rho)&=\sum_ip_i\sum_kE_kU_i\rho U_i^{\dagger}E_k^{\dagger} \\
&=\sum_kE_k \left(\sum_ip_iU_i\rho U_i^{\dagger}\right)E_k^{\dagger} \\
&=\sum_kE_k \left(\frac{1}{2}I\right)E_k^{\dagger} \\
&=\frac{1}{2}\sum_kE_kE_k^{\dagger} \\
&=\frac{1}{2}I,
\numberthis
\end{align*}
where we have recognised $\mathbb{E}^1_H(\rho)=\frac{1}{2}I$ and assumed that $E_k$ are Hermitian matrices, which is the case for all the noise channels defined in \secref{sec:ncbackground} except the amplitude damping channel.  Hence, inequality (\ref{eq:model}) can again be satisfied with $\epsilon=0$ for all the noise channels defined in \secref{sec:ncbackground} except the amplitude damping channel.

For the amplitude damping channel, we have
\begin{equation}
\widetilde{\mathbb{E}}^1_H(\rho)=\frac{1}{2}\sum_kE_kE_k^{\dagger}=\frac{1}{2}\begin{pmatrix}1+\lambda&0\\0&1-\lambda\end{pmatrix},
\end{equation}
and so
\begin{equation}
\widetilde{\mathbb{E}}^1_H(\rho)-(1-\epsilon)\mathbb{E}^1_H(\rho)=\frac{1}{2}\begin{pmatrix}\epsilon+\lambda&0\\0&\epsilon-\lambda\end{pmatrix}.
\end{equation}
It follows that for $(1-\epsilon)\mathbb{E}^1_H(\rho)\leq\widetilde{\mathbb{E}}^1_H(\rho)$ to hold, we must have $\epsilon\geq\lambda$, so all the eigenvalues of $\widetilde{\mathbb{E}}^1_H(\rho)-(1-\epsilon)\mathbb{E}^1_H(\rho)$ are non-negative, which ensures that $\widetilde{\mathbb{E}}^1_H(\rho)-(1-\epsilon)\mathbb{E}^1_H(\rho)$ is positive semidefinite.  Similarly, $\epsilon\geq\lambda$ ensures that $\widetilde{\mathbb{E}}^1_H(\rho)\leq(1+\epsilon)\mathbb{E}^1_H(\rho)$ holds.  Hence, inequality (\ref{eq:model}) can be satisfied with $\epsilon=\lambda$ for the amplitude damping channel and so $\epsilon=\lambda$ quantifies the effect of the amplitude damping channel on the quality of the 1-design for the model where noise is applied after the unitary operations.

\section{Numeric results for the model where noise is applied after the unitary operations}\label{append:numresultsafter}

\subsection{Flip channels}\label{append:flipnumresultsafter}

\begin{figure*}
    \centering
    \begin{subfigure}[b]{.42\textwidth}
        \centering
        \includegraphics[scale=0.55]{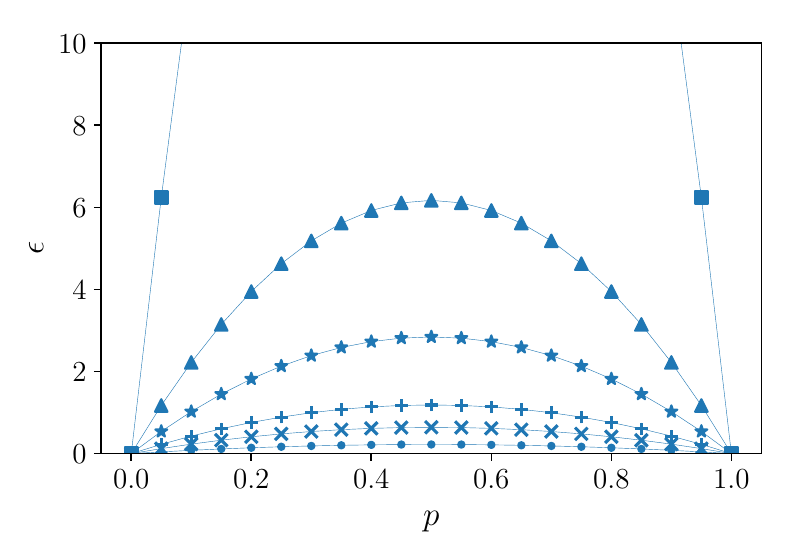}
        \caption{bit flip channel}
        \label{fig:afterbflip2}
    \end{subfigure}
    \begin{subfigure}[b]{.42\textwidth}
        \centering
        \includegraphics[scale=0.55]{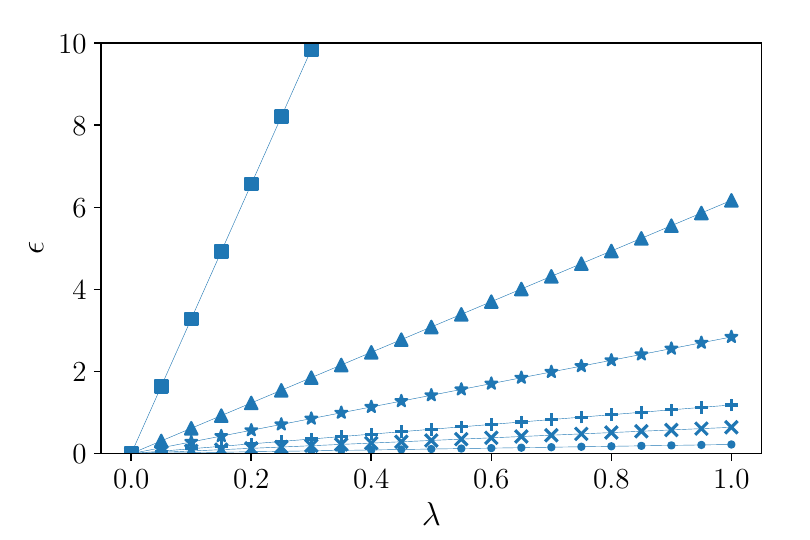}
        \caption{phase damping channel}
        \label{fig:afterphasedamp2}
    \end{subfigure}
    \includegraphics[trim=0cm 3.8cm 0cm 7cm, clip, scale=0.52]{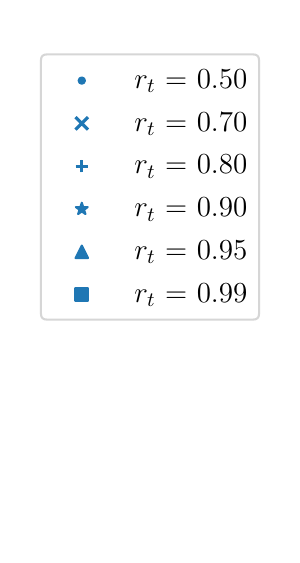}
    \caption{Effect of the (a) bit flip channel (see \equatref{eq:bflip}) and (b) phase damping channel (see \equatref{eq:phasedamp}) on the quality of the 2-design for the model where noise is applied after the unitary operations, for different truncation radii $r_t$.}
\end{figure*}

The effect of the bit flip channel (see \equatref{eq:bflip}) on the quality of the 2-design is shown in \figref{fig:afterbflip2}.  For the model where noise is applied after the unitary operations, $\epsilon$ versus $p$ is a parabola with maximum at $p=0.5$, just as for the model where noise is applied before the unitary operations (see \figref{fig:beforebflip2}).  However, the values of $\epsilon$ obtained for a given $p$ and $r_t$ are slightly smaller than those obtained for the model where noise is applied before the unitaries.  For the 4-design, $\epsilon$ versus $p$ for the model where noise is applied after the unitary operations has the same sinusoidal shape as for the model where noise is applied before the unitaries (see \figref{fig:beforebflip4}), but the values of $\epsilon$ are slightly smaller, just as for the 2-design.  Hence $t$-designs show reduced sensitivity to bit flips when applied after the unitary operations.

Numeric results obtained for the phase flip channel (see \equatref{eq:pflip}) and the bit and phase flip channel (see \equatref{eq:bpflip}) are identical to those obtained for the bit flip channel (shown in \figref{fig:afterbflip2}).  We again investigate similarities and differences by determining the regions of acceptable quality (see \appendref{append:regions}).  For all three flip channels, the region of acceptable quality has a spherical shape, that is, it is similar in shape to the region of acceptable quality for the depolarising noise channel for the model where noise is applied before the unitary operations.  Hence, we observe the transformation of each of the three flip channels into a depolarising channel when the flip channels are applied to states which have been randomised by the unitary operators, which explains why the results are the same for all three flip channels.  A 2-design's ability to transform an arbitrary noise channel into a depolarising noise channel is exploited in randomised benchmarking~\cite{RB1, RB2, RB3, RB7, RB8, RB9}.

\begin{figure*}
    \centering
    \begin{subfigure}[b]{.42\textwidth}
        \centering
        \includegraphics[scale=0.55]{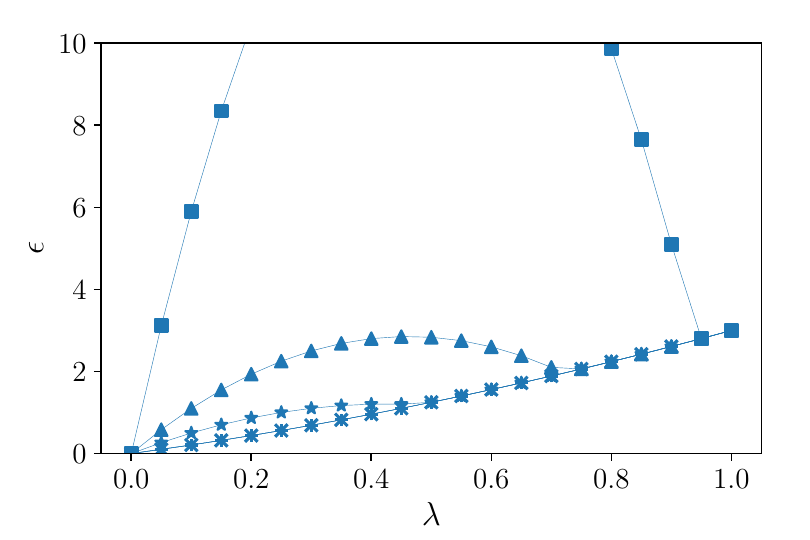}
        \caption{2-design}
        \label{fig:afterampdamp2}
    \end{subfigure}
    \begin{subfigure}[b]{.42\textwidth}
        \centering
        \includegraphics[scale=0.55]{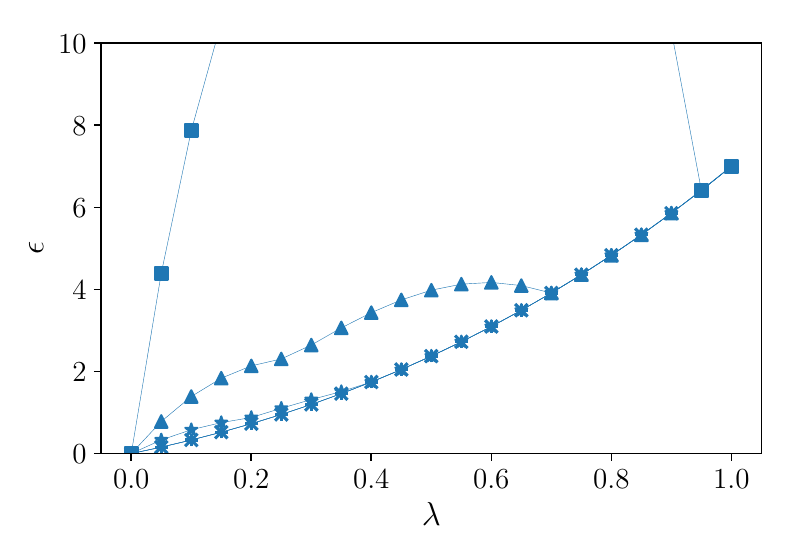}
        \caption{3-design}
        \label{fig:afterampdamp3}
    \end{subfigure}
    \includegraphics[trim=0cm 3.8cm 0cm 7cm, clip, scale=0.52]{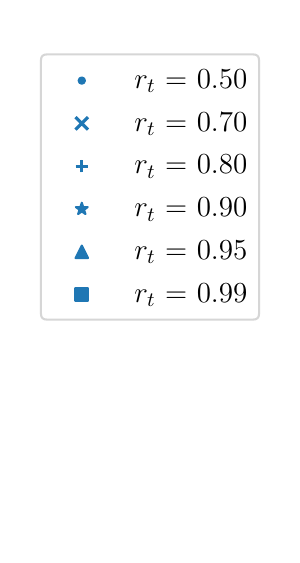}
    \caption{Effect of the amplitude damping channel (see \equatref{eq:ampdamp}) on the quality of the (a)~2-design and (b)~3-design for the model where noise is applied after the unitary operations, for different truncation radii $r_t$.}
\end{figure*}

\subsection{Phase damping channel}\label{append:phasenumresultsafter}

For the phase damping channel (see \equatref{eq:phasedamp}), $\epsilon$ versus $\lambda$ for the model where noise is applied after the unitary operations has the same shape as for the model where noise is applied before the unitary operations, for both the 2-design (shown in \figref{fig:afterphasedamp2}) and the 4-design, but the values of $\epsilon$ are slightly smaller for the model where noise is applied after the unitaries.  Just as for the phase flip channel, the region of acceptable quality for the phase damping channel has a spherical shape for the model where noise is applied after the unitaries (see \appendref{append:regions}).

\subsection{Amplitude damping channel}\label{append:ampnumresultsafter}

For the amplitude damping channel (see \equatref{eq:ampdamp}), $\epsilon$ versus $\lambda$ for the model where noise is applied after the unitary operations is once again similar to $\epsilon$ versus $\lambda$ for the model where noise is applied before the unitary operations, for the 2-design (shown in \figref{fig:afterampdamp2}), the 3-design (shown in \figref{fig:afterampdamp3}), the 4-design, and the 5-design.  The most notable differences are that the values of $\epsilon$ at the maxima are much smaller, the value of $\epsilon$ attained for $\lambda=1$ is much larger, and the anomaly occurs for much smaller $\lambda$ and much larger $r_t$ for the model where noise is applied after the unitary operations.  It is also worth noting that numeric results obtained for the 3-design differ significantly from those obtained for the 2-design (see Figs.~\ref{fig:afterampdamp2} and \ref{fig:afterampdamp3}) and that numeric results obtained for the 5-design differ significantly from those obtained for the 4-design.  The regions of acceptable quality are once again discussed in \appendref{append:regions}.

\subsection{Depolarising noise channel}\label{append:depnumresultsafter}

For the depolarising noise channel, we were able to show that the values of $\epsilon$ obtained for the two noise models are equal.  The proof is given in \appendref{append:depolarising}.

\section{Proof of equivalence of noise models for the depolarising noise channel}\label{append:depolarising} 

Let $\{p_i, U_i\}$ be an exact unitary $t$-design.  For the model where noise is applied before the unitary operations, we substitute \equatref{eq:depnoise} into \equatref{eq:before} to obtain
\begin{equation}
\widetilde{\mathbb{E}}^t_H(\rho)=\sum_{i}p_i\left(U_i\left(\frac{p}{2}I + (1-p)\rho\right) U_i^{\dagger}\right)^{\otimes t}=\sum_{i}p_i\left(\frac{p}{2}I + (1-p)U_i\rho U_i^{\dagger} \right)^{\otimes t},
\end{equation}
and for the model where noise is applied after the unitary operations, we substitute \equatref{eq:depnoise} into \equatref{eq:after} to obtain
\begin{equation}
\widetilde{\mathbb{E}}^t_H(\rho)=\sum_{i}p_i\left(\frac{p}{2}I + (1-p)U_i\rho U_i^{\dagger} \right)^{\otimes t}.
\end{equation}
Hence, for the depolarising noise channel, $\widetilde{\mathbb{E}}^t_H(\rho)$ is the same for the two noise models, so the smallest $\epsilon$ such that inequality (\ref{eq:model}) holds for all density matrices is the same for the two noise models.

\section{Visualisation of regions of acceptable quality}\label{append:regions} 

We determine and visualise regions of the Bloch sphere for which noisy $t$-designs are able to replicate the moments of the uniform Haar ensemble, with a predefined accuracy, up to order $t$.  In the sections which follow, we will refer to these regions as regions of acceptable quality.

\subsection{Implementation}\label{append:impregions} 

To determine regions of acceptable quality, we first generate 20 evenly spaced values of $x\in [-1,1]$, 20 evenly spaced values of $y\in [-1,1]$, and 20 evenly spaced values of $z\in [-1,1]$, thereby obtaining a sample of $20^3=8000$ coordinates $(x,y,z)$ in a cube which encloses the Bloch sphere.  Points which lie in the Bloch sphere, that is points such that $\sqrt{x^2+y^2+z^2}\leq 1$, correspond to valid states for which we obtain density matrices using \equatref{eq:rho}.  We now visualise the region of acceptable quality for a given $t\in\{2,3,4,5\}$, a given noise channel, and a given noise model as follows.  For each density matrix $\rho$, we calculate $\mathbb{E}^t_H(\rho^{\otimes t})$ and $\widetilde{\mathbb{E}}^t_H(\rho)$ using the icosahedral group~\cite{platonic} and plot the point $(x,y,z)$ corresponding to $\rho$ if and only if inequality (\ref{eq:model}) can be satisfied with $\epsilon\leq 0.5$.

Note that we have arbitrarily chosen $\epsilon=0.5$ as a threshold for acceptable quality, merely for the purposes of visualisation.  In practice, the value of $\epsilon$ required for the quality of a noisy $t$-design to be acceptable really depends on the application.  Furthermore, we note that it is not possible to use the $\epsilon$ for which an exact $(t-1)$-design is an approximate $t$-design as a threshold value for a noisy $t$-design, since this $\epsilon$ depends on the ensemble of unitaries that make up the exact $(t-1)$-design.  In fact, there are exact $(t-1)$-designs for which $\epsilon > 0.5$ when used as approximate $t$-designs.

\subsection{Numeric results for the model where noise is applied before the unitary operations}\label{append:numregionsbefore} 

\subsubsection{Bit flip channel}\label{append:bflipnumregionsbefore}

\begin{figure}
    \centering
    \begin{subfigure}[b]{.32\textwidth}
        \centering
        \includegraphics[trim=3.5cm 1cm 1.5cm 1.4cm, clip, scale=0.5]{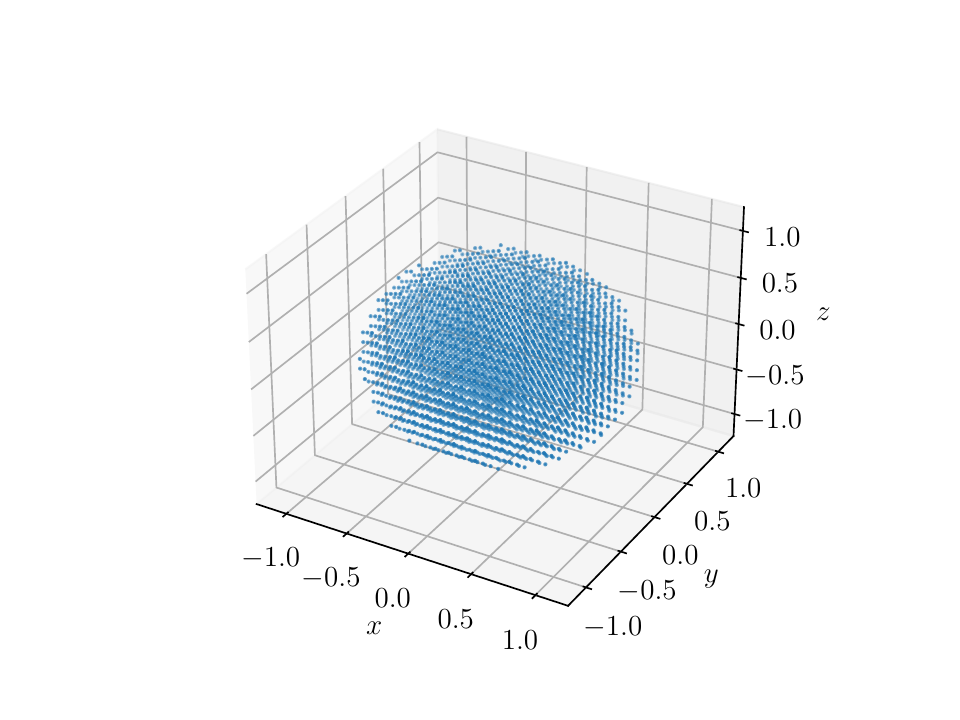}
        \caption{$p=0.01$}
    \end{subfigure}
    \begin{subfigure}[b]{.32\textwidth}
        \centering
        \includegraphics[trim=3.5cm 1cm 1.5cm 1.4cm, clip, scale=0.5]{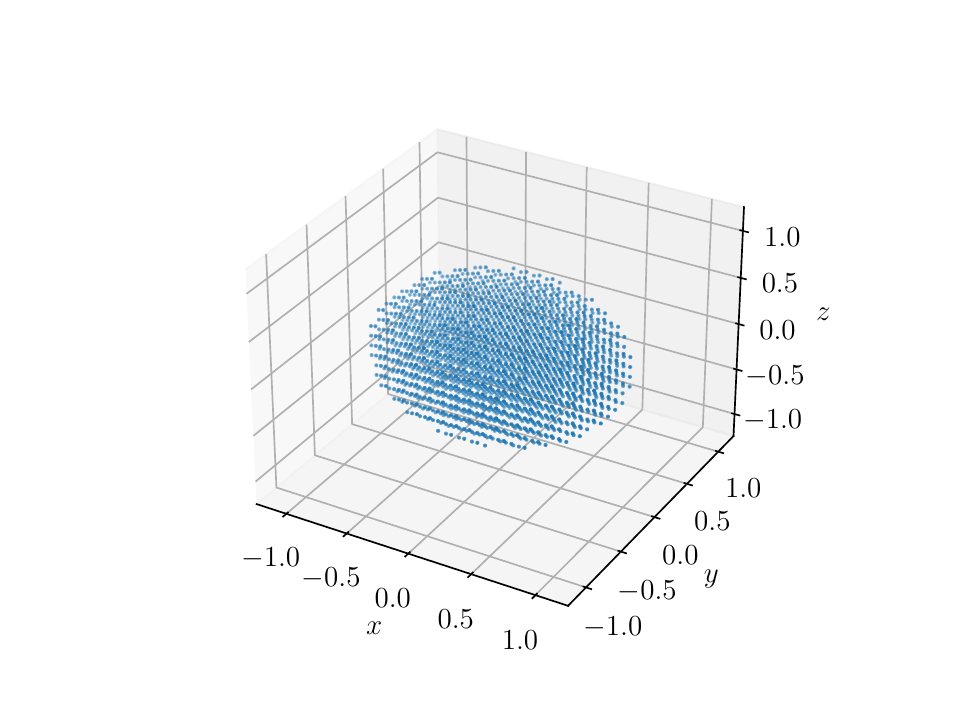}
        \caption{$p=0.1$}
    \end{subfigure}
    \begin{subfigure}[b]{.32\textwidth}
        \centering
        \includegraphics[trim=3.5cm 1cm 1.5cm 1.4cm, clip, scale=0.5]{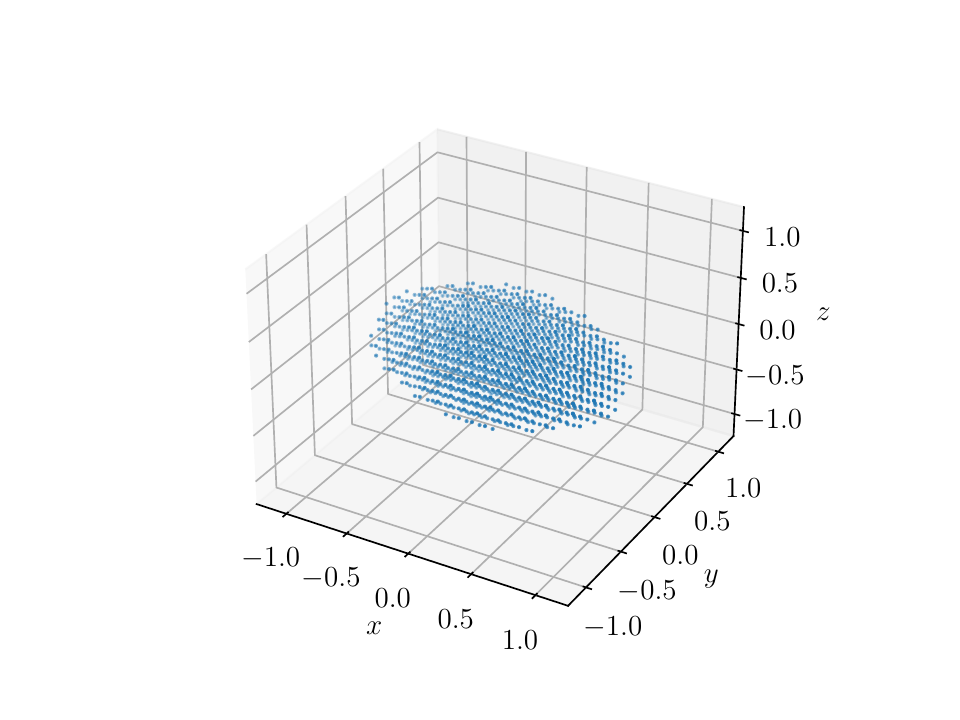}
        \caption{$p=0.3$}
    \end{subfigure}
    \caption{Regions of acceptable quality for the 2-design affected by the bit flip channel (see \equatref{eq:bflip}) for the model where noise is applied before the unitary operations, for different $p$.}
    \label{fig:regionsbeforebflip2}
\end{figure}

\begin{figure}
    \centering
    \begin{subfigure}[b]{.32\textwidth}
        \centering
        \includegraphics[trim=3.5cm 1cm 1.5cm 1.4cm, clip, scale=0.5]{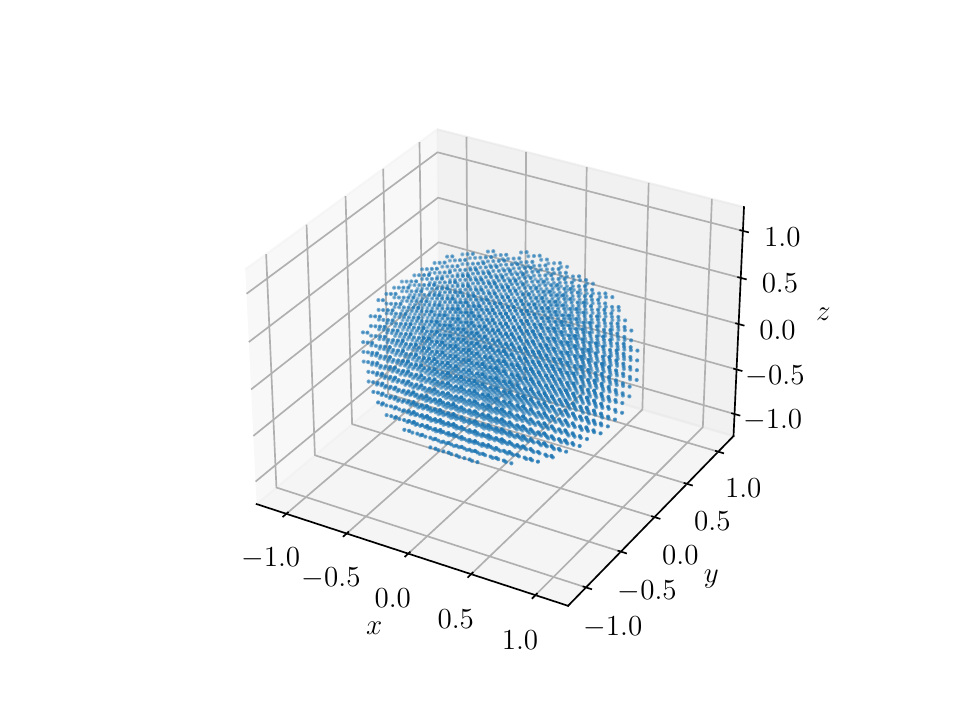}
        \caption{$p=0.01$}
    \end{subfigure}
    \begin{subfigure}[b]{.32\textwidth}
        \centering
        \includegraphics[trim=3.5cm 1cm 1.5cm 1.4cm, clip, scale=0.5]{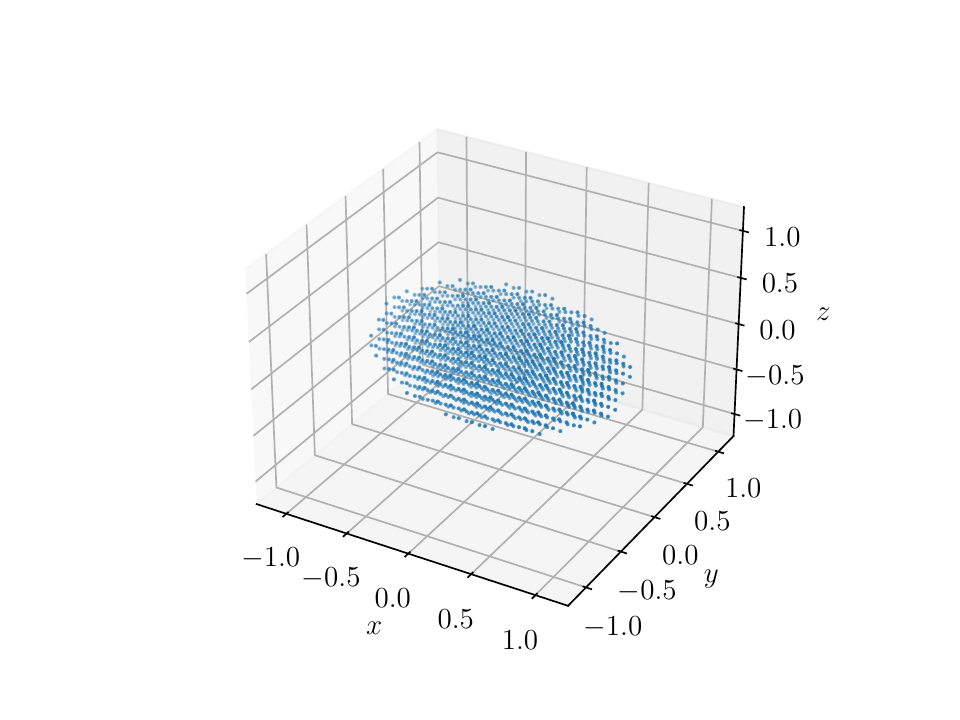}
        \caption{$p=0.1$}
    \end{subfigure}
    \begin{subfigure}[b]{.32\textwidth}
        \centering
        \includegraphics[trim=3.5cm 1cm 1.5cm 1.4cm, clip, scale=0.5]{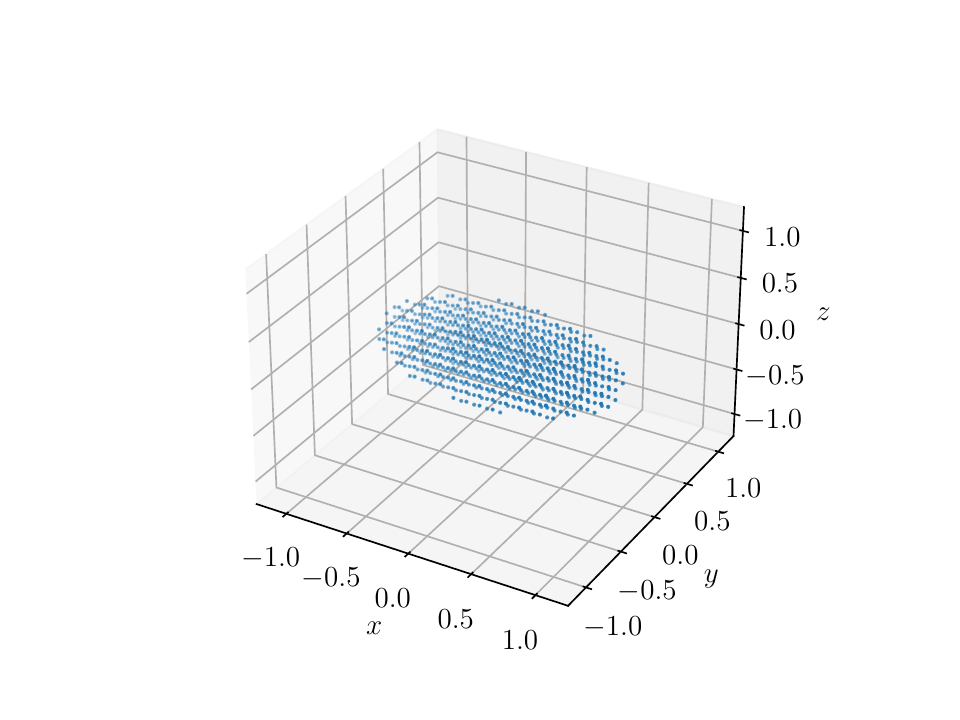}
        \caption{$p=0.3$}
    \end{subfigure}
    \caption{Regions of acceptable quality for the 4-design affected by the bit flip channel (see \equatref{eq:bflip}) for the model where noise is applied before the unitary operations, for different $p$.}
    \label{fig:regionsbeforebflip4}
\end{figure}

Regions of acceptable quality for the 2-design affected by the bit flip channel (see \equatref{eq:bflip}) are shown in \figref{fig:regionsbeforebflip2}, for various different $p$.  As expected, the size of the region of acceptable quality decreases, and its shape becomes more pronounced, as the bit flip probability $p$ increases.  For all $p$, the region of acceptable quality is an ellipsoid along the $x$-axis, that is, the region of acceptable quality is similar in shape and orientation to the region into which the Bloch sphere is deformed by the bit flip channel.  This can be understood as follows.  Since bit flips are performed by applying the Pauli $X$ operator to a state, states along the $x$-axis, which are closer to the eigenstates of the Pauli $X$ operator, are less affected by bit flips.  Hence the quality remains acceptable for states along the $x$-axis even for a large bit flip probability.  For the 4-design, the region of acceptable quality is similar in shape and orientation to the region of acceptable quality for the 2-design, but smaller in size for a fixed $p$ (see \figref{fig:regionsbeforebflip4}).  This simply confirms that the 4-design is more sensitive to bit flips than the 2-design.

\subsubsection{Phase flip channel}\label{append:pflipnumregionsbefore} 

\begin{figure}
    \centering
    \begin{subfigure}[b]{.32\textwidth}
        \centering
        \includegraphics[trim=3.5cm 1cm 1.5cm 1.4cm, clip, scale=0.5]{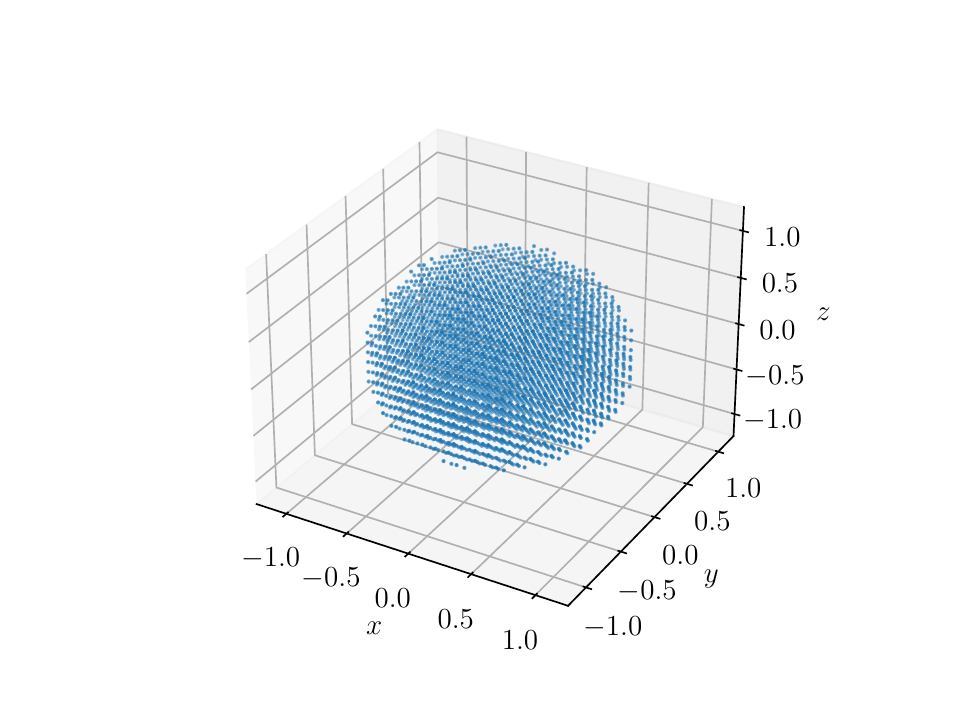}
        \caption{$p=0.01$}
    \end{subfigure}
    \begin{subfigure}[b]{.32\textwidth}
        \centering
        \includegraphics[trim=3.5cm 1cm 1.5cm 1.4cm, clip, scale=0.5]{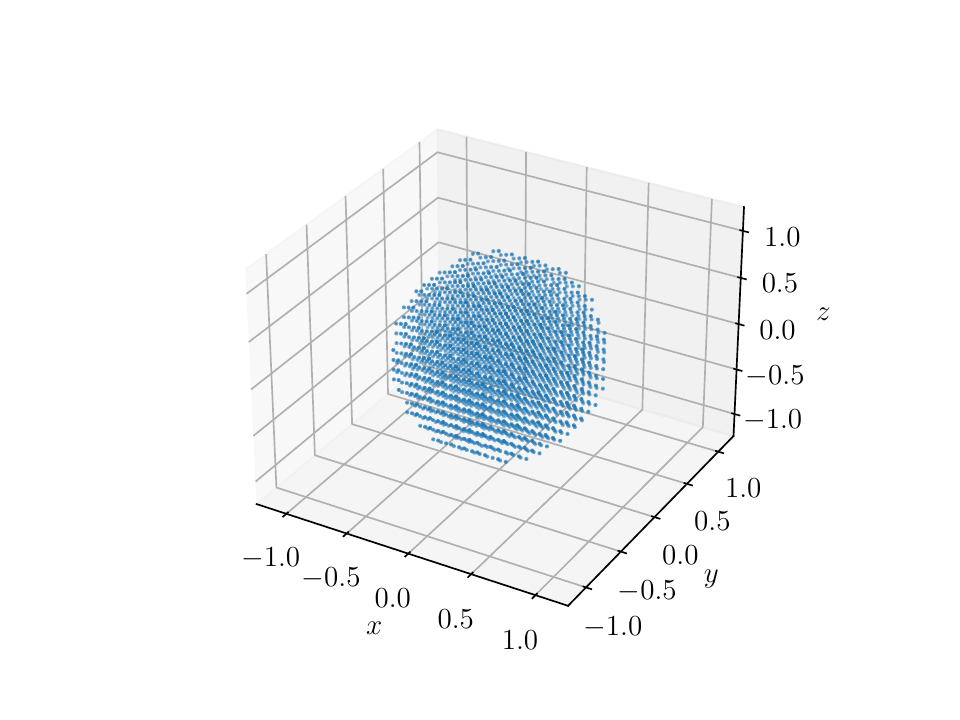}
        \caption{$p=0.1$}
    \end{subfigure}
    \begin{subfigure}[b]{.32\textwidth}
        \centering
        \includegraphics[trim=3.5cm 1cm 1.5cm 1.4cm, clip, scale=0.5]{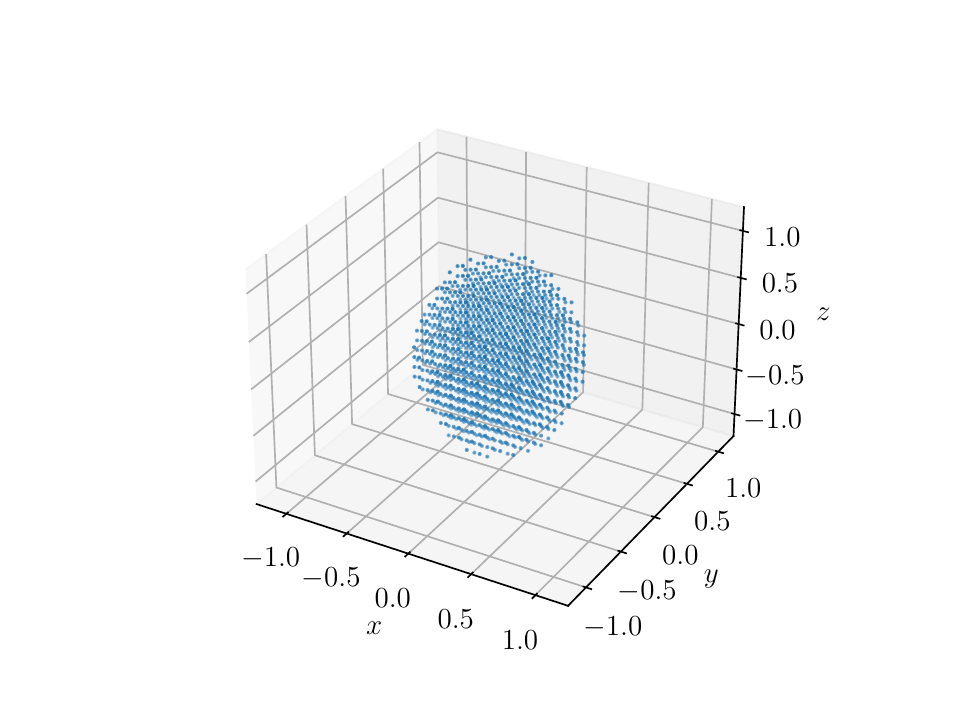}
        \caption{$p=0.3$}
    \end{subfigure}
    \caption{Regions of acceptable quality for the 2-design affected by the phase flip channel (see \equatref{eq:pflip}) for the model where noise is applied before the unitary operations, for different $p$.}
    \label{fig:regionsbeforepflip2}
\end{figure}

\begin{figure}
    \centering
    \begin{subfigure}[b]{.32\textwidth}
        \centering
        \includegraphics[trim=3.5cm 1cm 1.5cm 1.4cm, clip, scale=0.5]{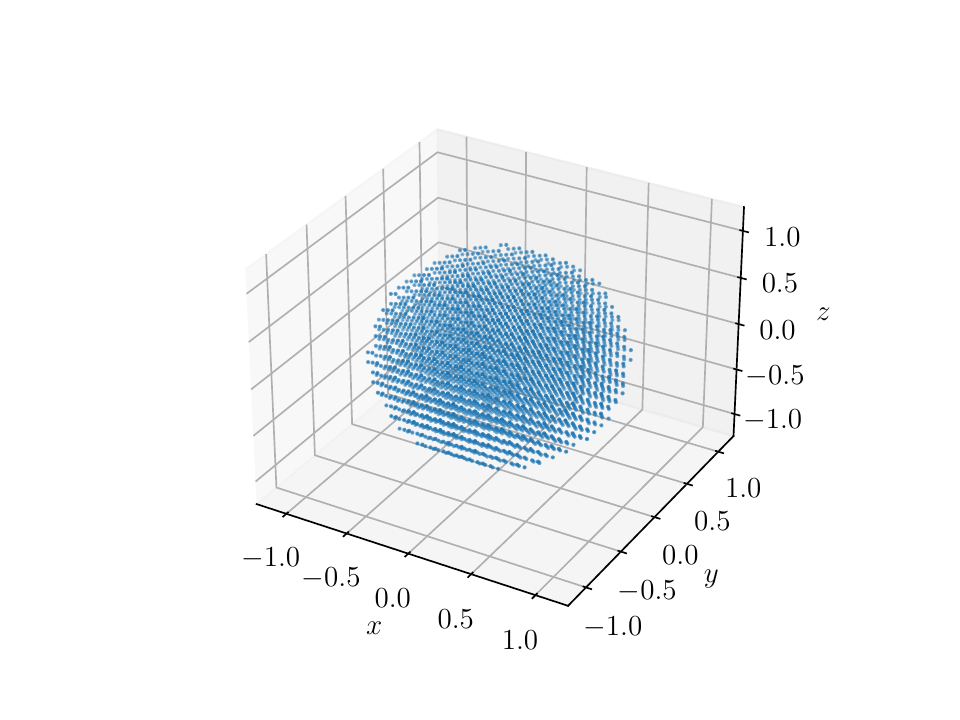}
        \caption{$p=0.01$}
    \end{subfigure}
    \begin{subfigure}[b]{.32\textwidth}
        \centering
        \includegraphics[trim=3.5cm 1cm 1.5cm 1.4cm, clip, scale=0.5]{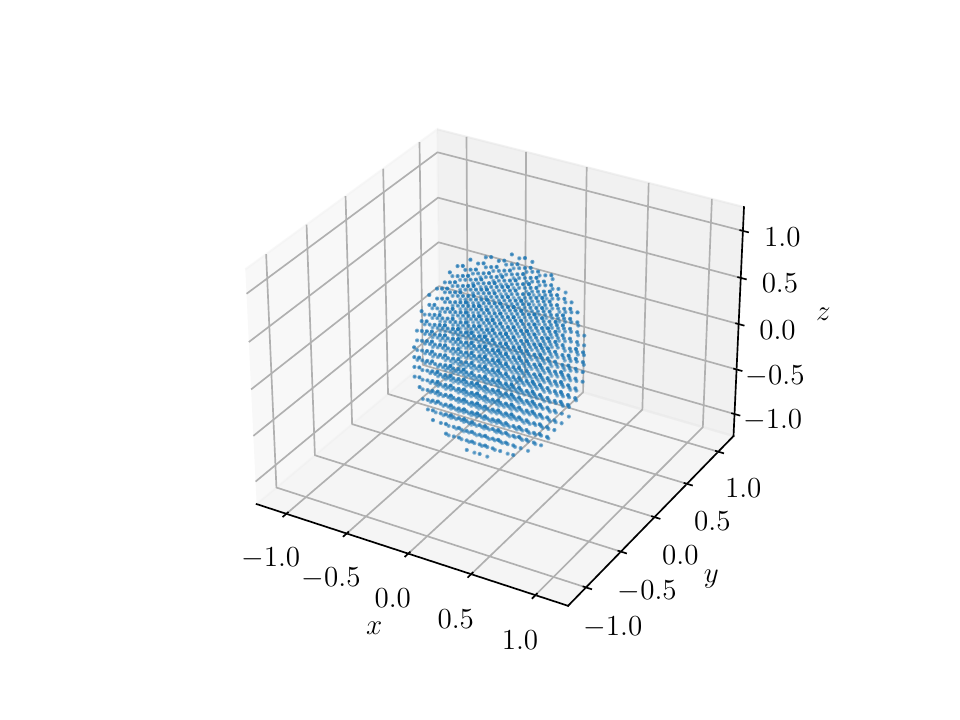}
        \caption{$p=0.1$}
    \end{subfigure}
    \begin{subfigure}[b]{.32\textwidth}
        \centering
        \includegraphics[trim=3.5cm 1cm 1.5cm 1.4cm, clip, scale=0.5]{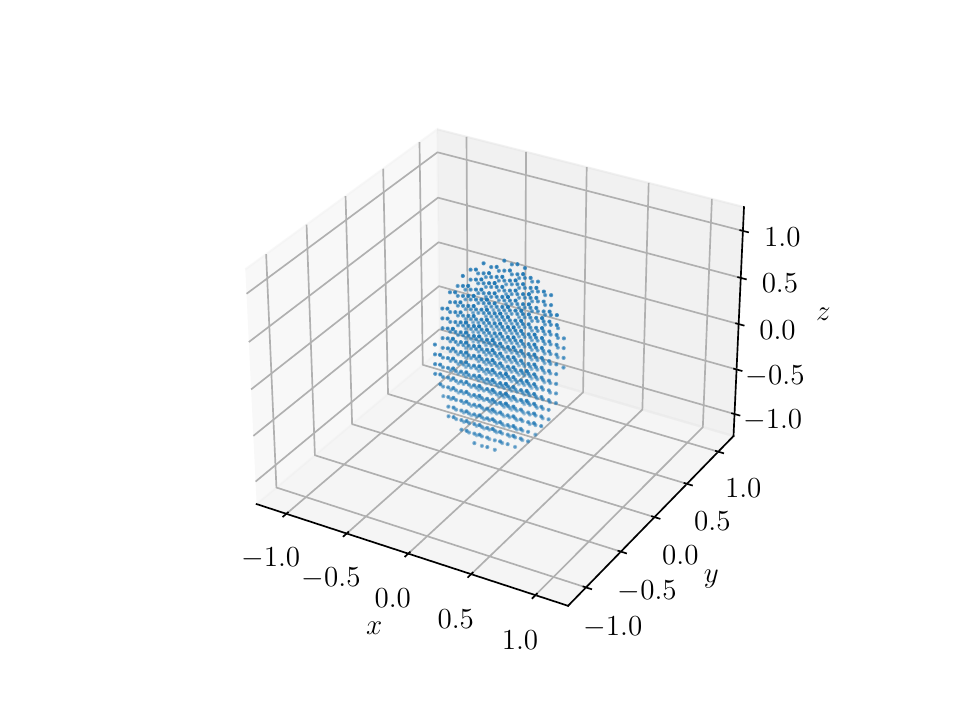}
        \caption{$p=0.3$}
    \end{subfigure}
    \caption{Regions of acceptable quality for the 4-design affected by the phase flip channel (see \equatref{eq:pflip}) for the model where noise is applied before the unitary operations, for different $p$.}
    \label{fig:regionsbeforepflip4}
\end{figure}

Regions of acceptable quality for the 2-design affected by the phase flip channel (see \equatref{eq:pflip}) are shown in \figref{fig:regionsbeforepflip2}.  For all $p$, the region of acceptable quality is an ellipsoid along the $z$-axis, that is, the region of acceptable quality is similar in shape and orientation to the region into which the Bloch sphere is deformed by the phase flip channel.  The shape and orientation of the region of acceptable quality can be attributed to the fact that states along the $z$-axis, which are closer to the eigenstates of the Pauli $Z$ operator, are less affected by phase flips since phase flips are performed by applying the Pauli $Z$ operator to a state.  For the 4-design, the regions of acceptable quality are again similar in shape and orientation, but smaller in size (see \figref{fig:regionsbeforepflip4}).

\subsubsection{Bit and phase flip channel}\label{append:bpflipnumregionsbefore} 

\begin{figure}
    \centering
    \begin{subfigure}[b]{.32\textwidth}
        \centering
        \includegraphics[trim=3.5cm 1cm 1.5cm 1.4cm, clip, scale=0.5]{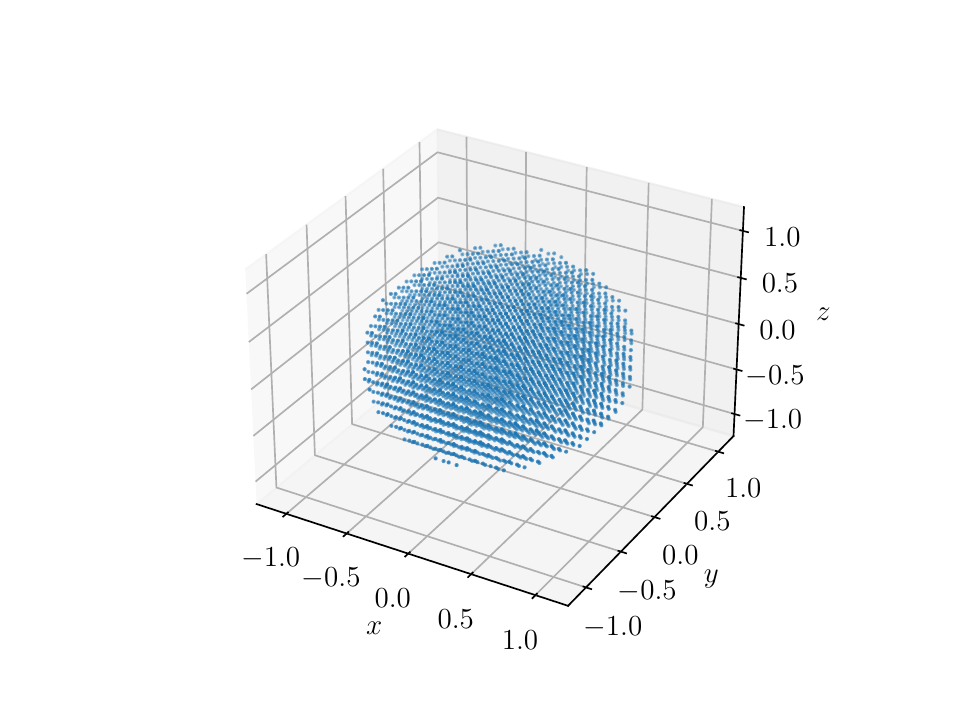}
        \caption{$p=0.01$}
    \end{subfigure}
    \begin{subfigure}[b]{.32\textwidth}
        \centering
        \includegraphics[trim=3.5cm 1cm 1.5cm 1.4cm, clip, scale=0.5]{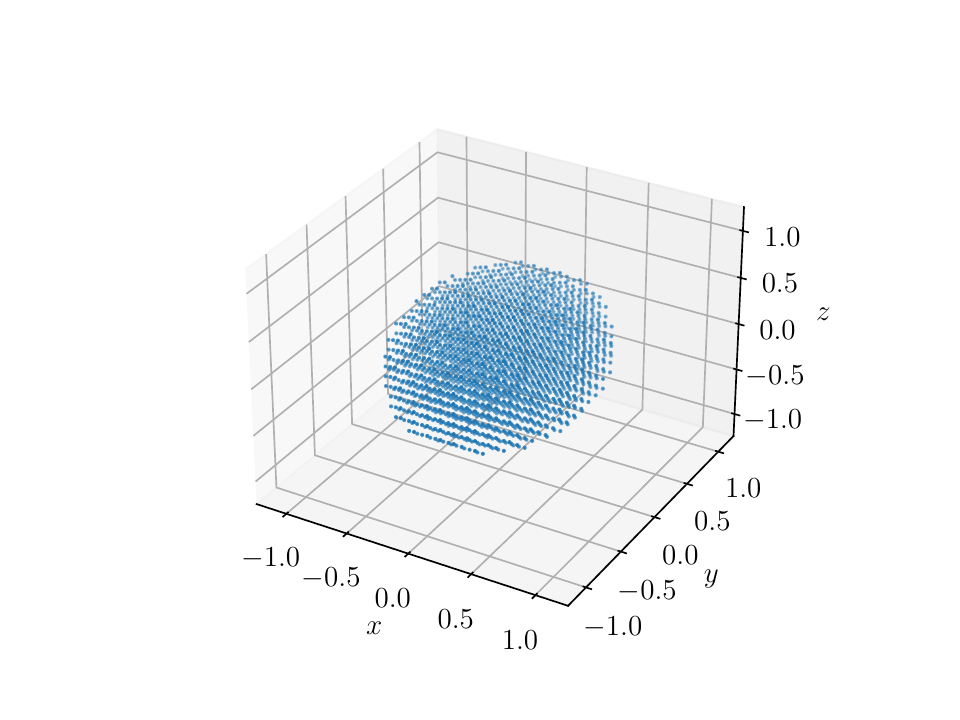}
        \caption{$p=0.1$}
    \end{subfigure}
    \begin{subfigure}[b]{.32\textwidth}
        \centering
        \includegraphics[trim=3.5cm 1cm 1.5cm 1.4cm, clip, scale=0.5]{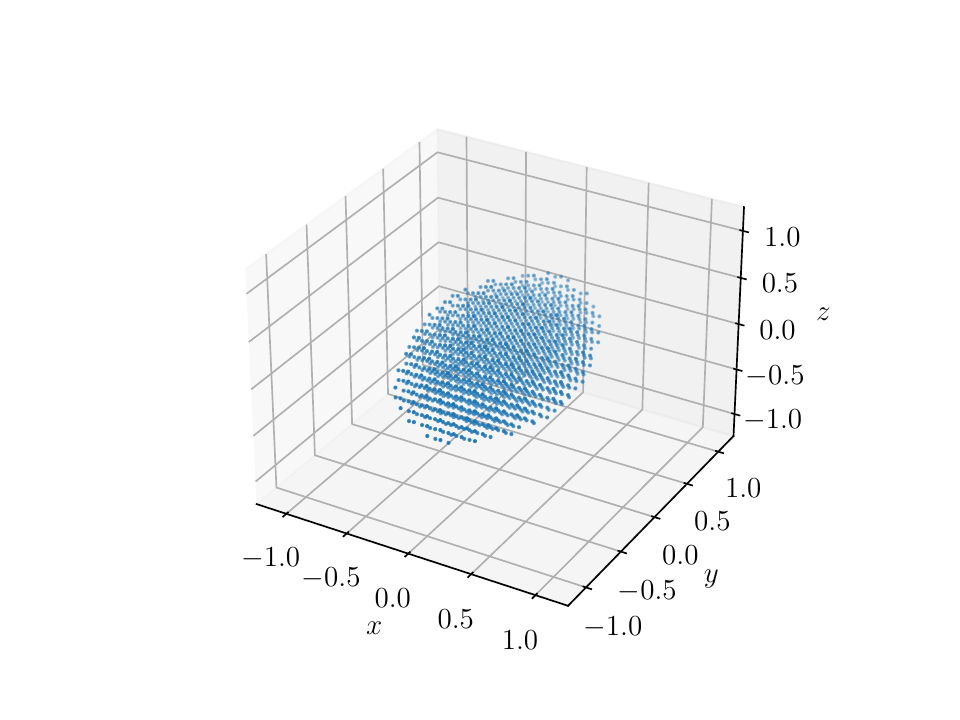}
        \caption{$p=0.3$}
    \end{subfigure}
    \caption{Regions of acceptable quality for the 2-design affected by the bit and phase flip channel (see \equatref{eq:bpflip}) for the model where noise is applied before the unitary operations, for different $p$.}
    \label{fig:regionsbeforebpflip2}
\end{figure}

\begin{figure}
    \centering
    \begin{subfigure}[b]{.32\textwidth}
        \centering
        \includegraphics[trim=3.5cm 1cm 1.5cm 1.4cm, clip, scale=0.5]{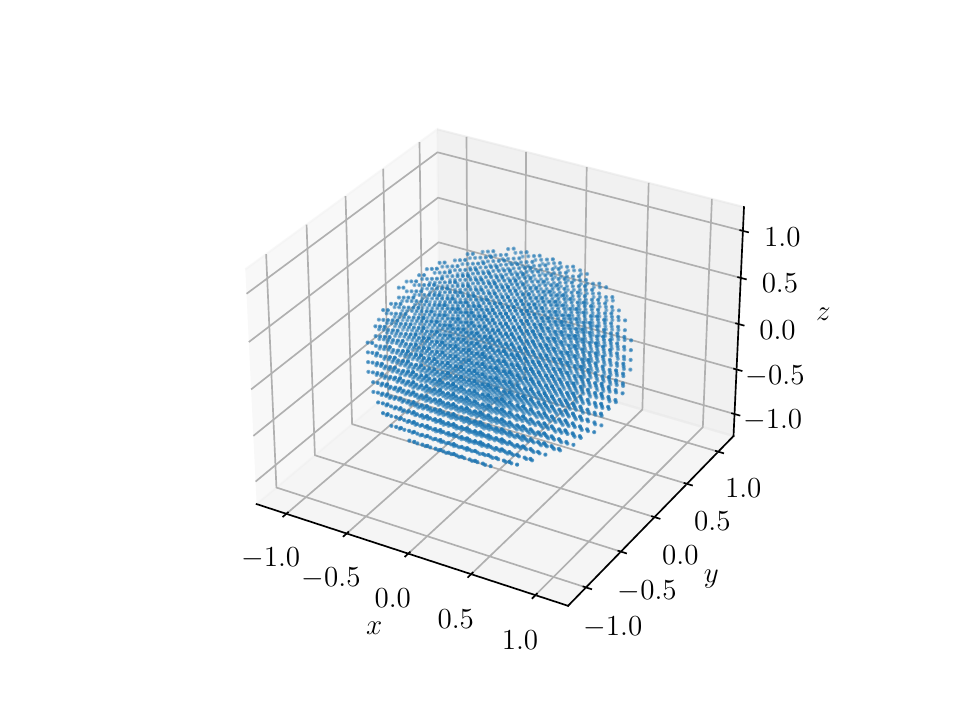}
        \caption{$p=0.01$}
    \end{subfigure}
    \begin{subfigure}[b]{.32\textwidth}
        \centering
        \includegraphics[trim=3.5cm 1cm 1.5cm 1.4cm, clip, scale=0.5]{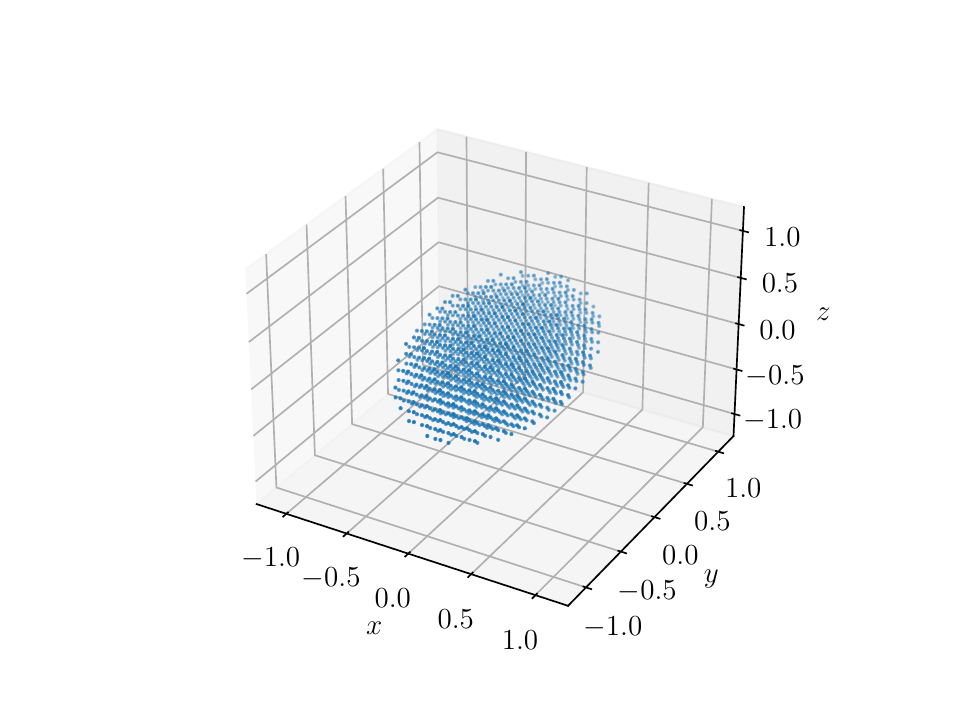}
        \caption{$p=0.1$}
    \end{subfigure}
    \begin{subfigure}[b]{.32\textwidth}
        \centering
        \includegraphics[trim=3.5cm 1cm 1.5cm 1.4cm, clip, scale=0.5]{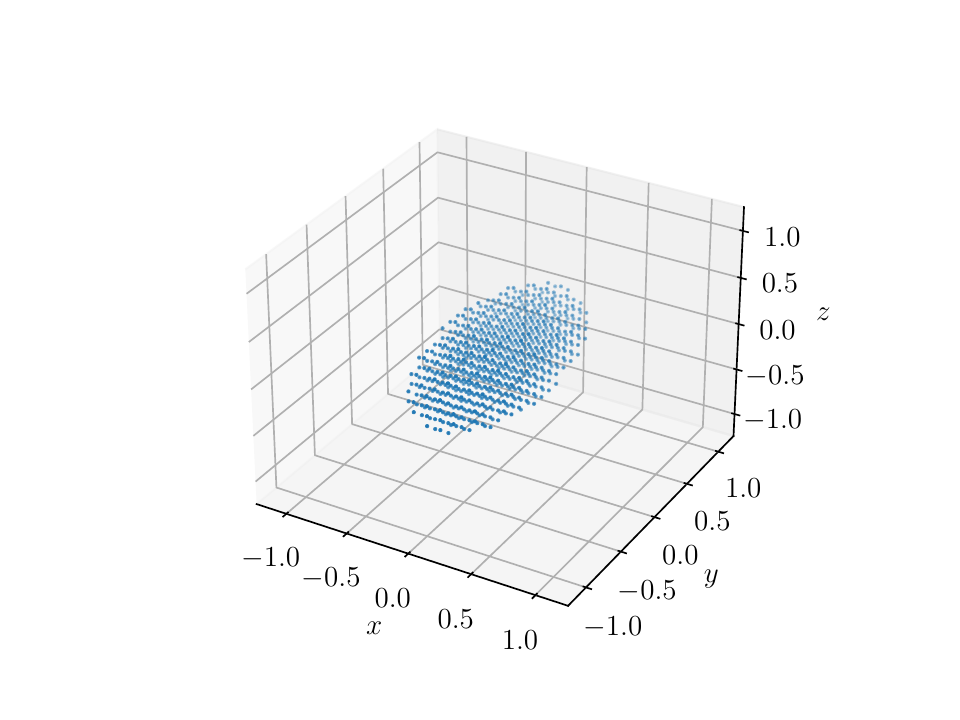}
        \caption{$p=0.3$}
    \end{subfigure}
    \caption{Regions of acceptable quality for the 4-design affected by the bit and phase flip channel (see \equatref{eq:bpflip}) for the model where noise is applied before the unitary operations, for different $p$.}
    \label{fig:regionsbeforebpflip4}
\end{figure}

As shown in Figs.~\ref{fig:regionsbeforebpflip2} and \ref{fig:regionsbeforebpflip4}, the regions of acceptable quality are similar in shape and orientation to the region into which the Bloch sphere is deformed by the bit and phase flip channel, that is, an ellipsoid along the $y$-axis.  Since bit and phase flips are performed by applying the Pauli $Y$ operator to a state, states along the $y$-axis, which are closer to the eigenstates of the Pauli $Y$ operator, are less affected by bit and phase flips.

\subsubsection{Phase damping channel}\label{append:phasenumregionsbefore} 

\begin{figure}
    \centering
    \begin{subfigure}[b]{.32\textwidth}
        \centering
        \includegraphics[trim=3.5cm 1cm 1.5cm 1.4cm, clip, scale=0.5]{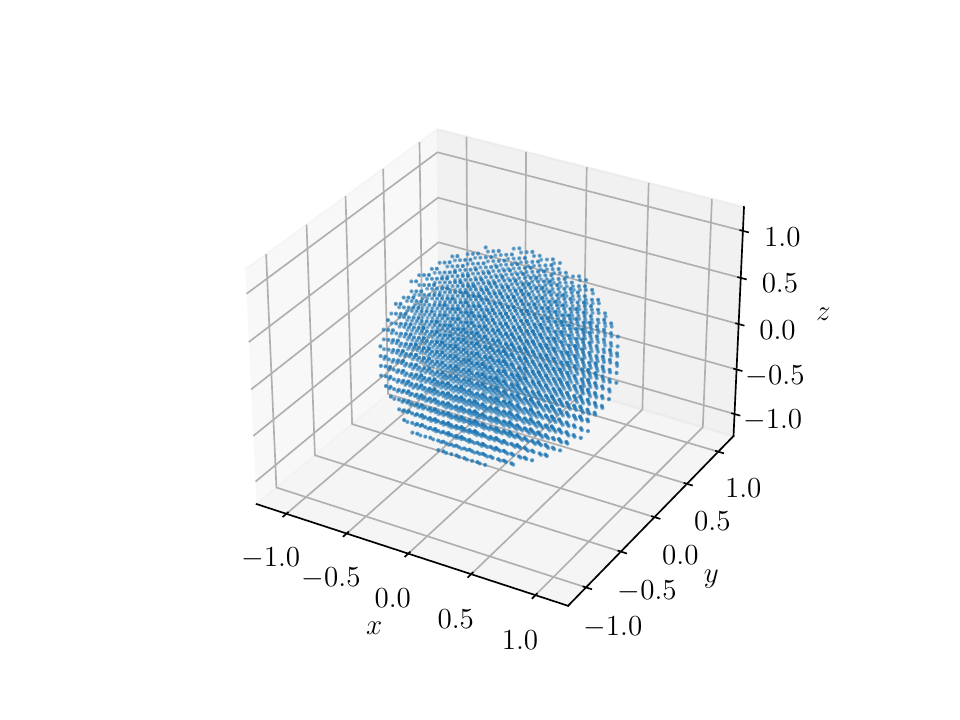}
        \caption{$\lambda=0.2$}
    \end{subfigure}
    \begin{subfigure}[b]{.32\textwidth}
        \centering
        \includegraphics[trim=3.5cm 1cm 1.5cm 1.4cm, clip, scale=0.5]{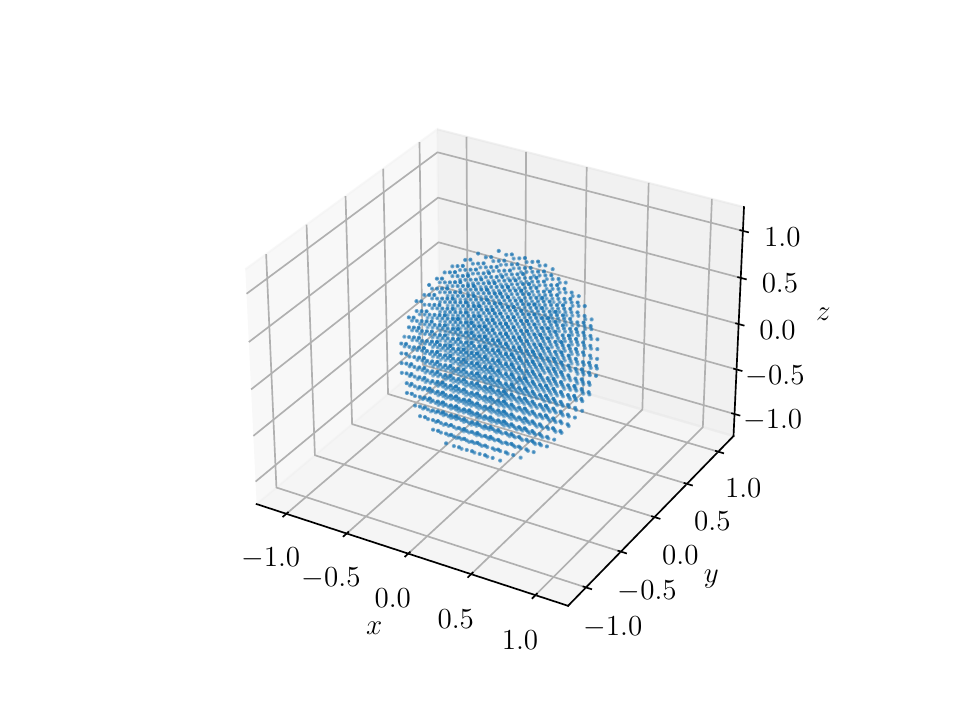}
        \caption{$\lambda=0.5$}
    \end{subfigure}
    \begin{subfigure}[b]{.32\textwidth}
        \centering
        \includegraphics[trim=3.5cm 1cm 1.5cm 1.4cm, clip, scale=0.5]{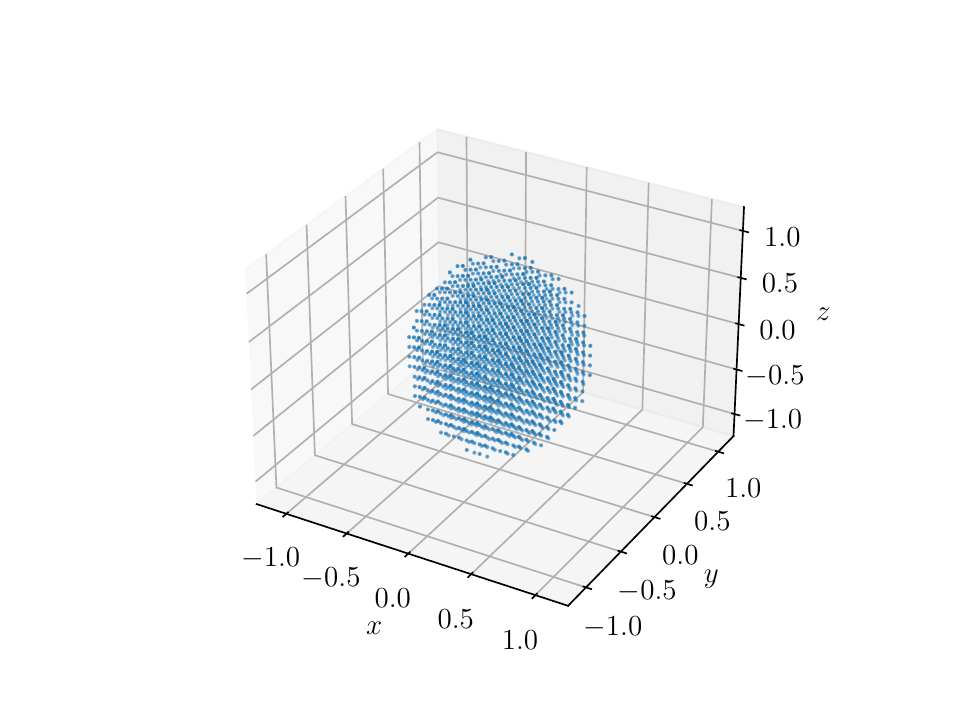}
        \caption{$\lambda=0.7$}
    \end{subfigure}
    \caption{Regions of acceptable quality for the 2-design affected by the phase damping channel (see \equatref{eq:phasedamp}) for the model where noise is applied before the unitary operations, for different $\lambda$.}
    \label{fig:regionsbeforephasedamp2}
\end{figure}

\begin{figure}
    \centering
    \begin{subfigure}[b]{.32\textwidth}
        \centering
        \includegraphics[trim=3.5cm 1cm 1.5cm 1.4cm, clip, scale=0.5]{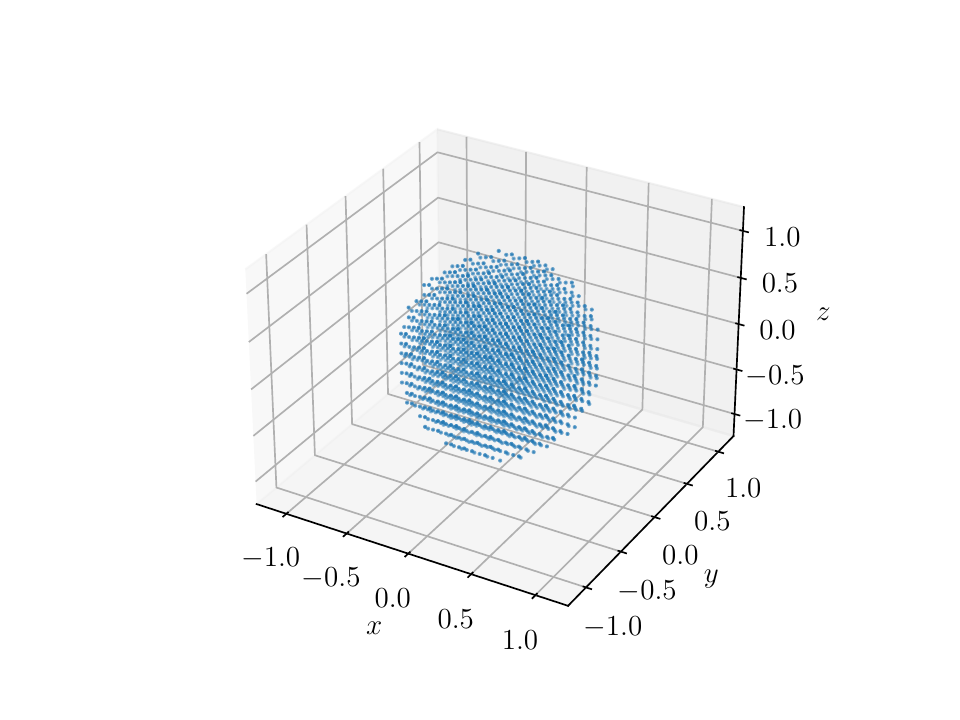}
        \caption{$\lambda=0.2$}
    \end{subfigure}
    \begin{subfigure}[b]{.32\textwidth}
        \centering
        \includegraphics[trim=3.5cm 1cm 1.5cm 1.4cm, clip, scale=0.5]{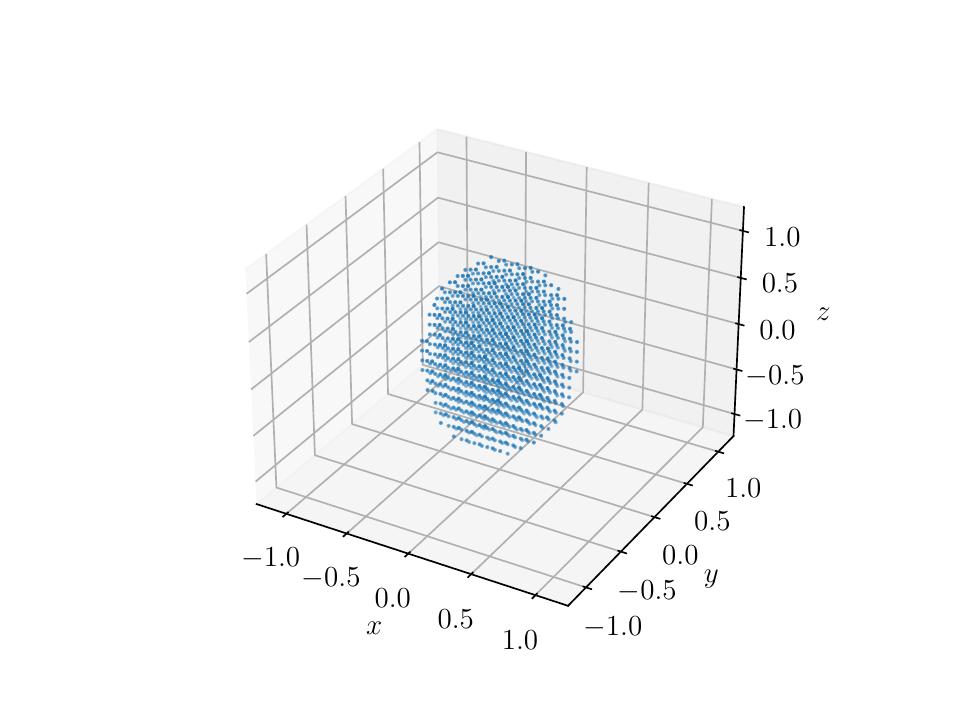}
        \caption{$\lambda=0.5$}
    \end{subfigure}
    \begin{subfigure}[b]{.32\textwidth}
        \centering
        \includegraphics[trim=3.5cm 1cm 1.5cm 1.4cm, clip, scale=0.5]{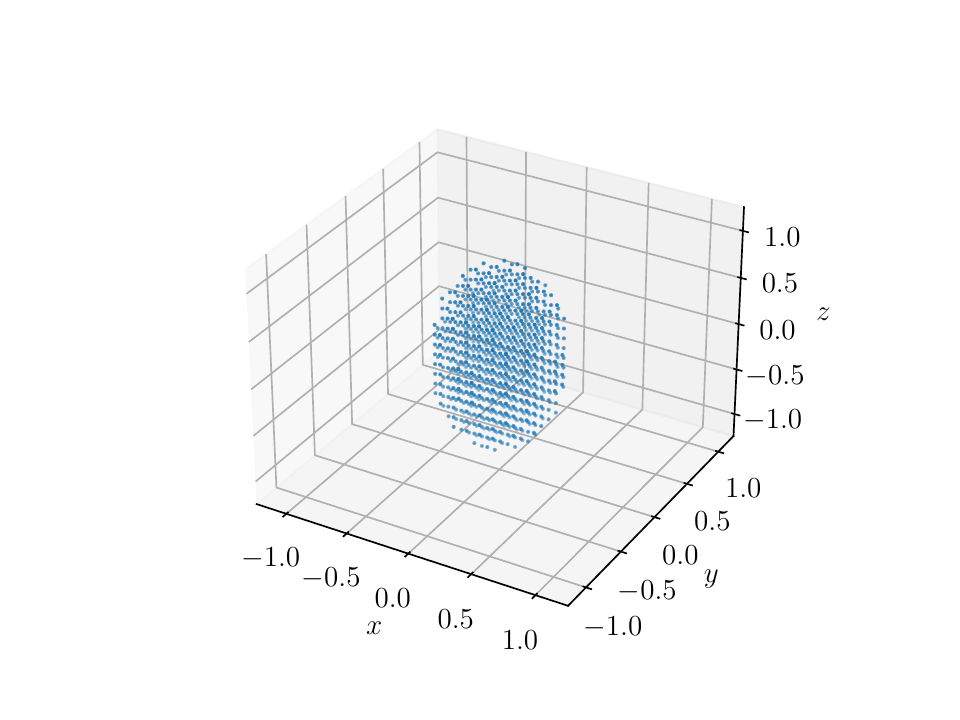}
        \caption{$\lambda=0.7$}
    \end{subfigure}
    \caption{Regions of acceptable quality for the 4-design affected by the phase damping channel (see \equatref{eq:phasedamp}) for the model where noise is applied before the unitary operations, for different $\lambda$.}
    \label{fig:regionsbeforephasedamp4}
\end{figure}

Regions of acceptable quality for the 2-design affected by the phase damping channel (see \equatref{eq:phasedamp}) are shown in \figref{fig:regionsbeforephasedamp2}, for different $\lambda$.  Just as for the phase flip channel, the regions of acceptable quality are ellipsoids along the $z$-axis.  This is to be expected, since the phase damping channel is equivalent to the phase flip channel, up to a reparameterisation (see \equatref{eq:damptoflip}).  For the 4-design, the regions of acceptable quality are again similar in shape, but smaller in size (see \figref{fig:regionsbeforephasedamp4}).

\subsubsection{Amplitude damping channel}\label{append:ampnumregionsbefore} 

\begin{figure}
    \centering
    \begin{subfigure}[b]{.32\textwidth}
        \centering
        \includegraphics[trim=3.5cm 1cm 1.5cm 1.4cm, clip, scale=0.5]{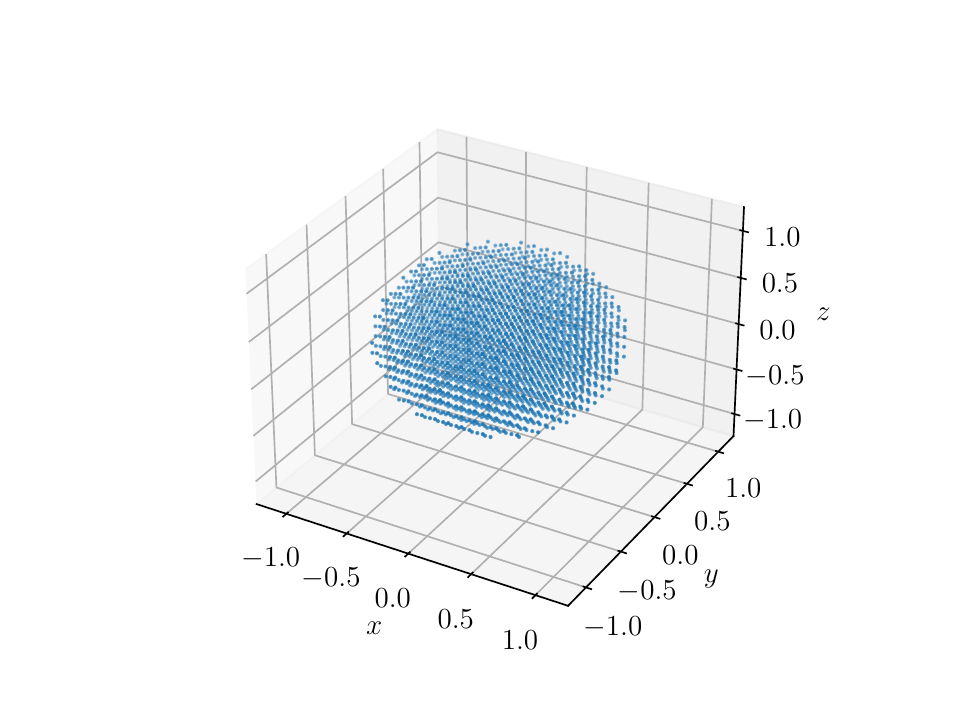}
        \caption{$\lambda=0.2$}
    \end{subfigure}
    \begin{subfigure}[b]{.32\textwidth}
        \centering
        \includegraphics[trim=3.5cm 1cm 1.5cm 1.4cm, clip, scale=0.5]{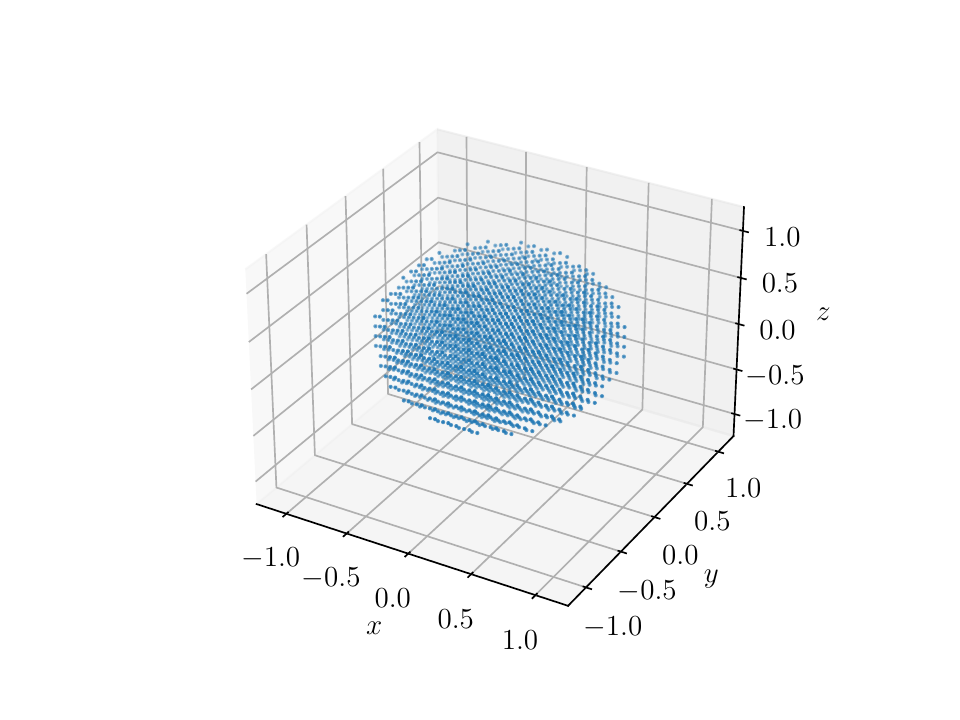}
        \caption{$\lambda=0.5$}
    \end{subfigure}
    \begin{subfigure}[b]{.32\textwidth}
        \centering
        \includegraphics[trim=3.5cm 1cm 1.5cm 1.4cm, clip, scale=0.5]{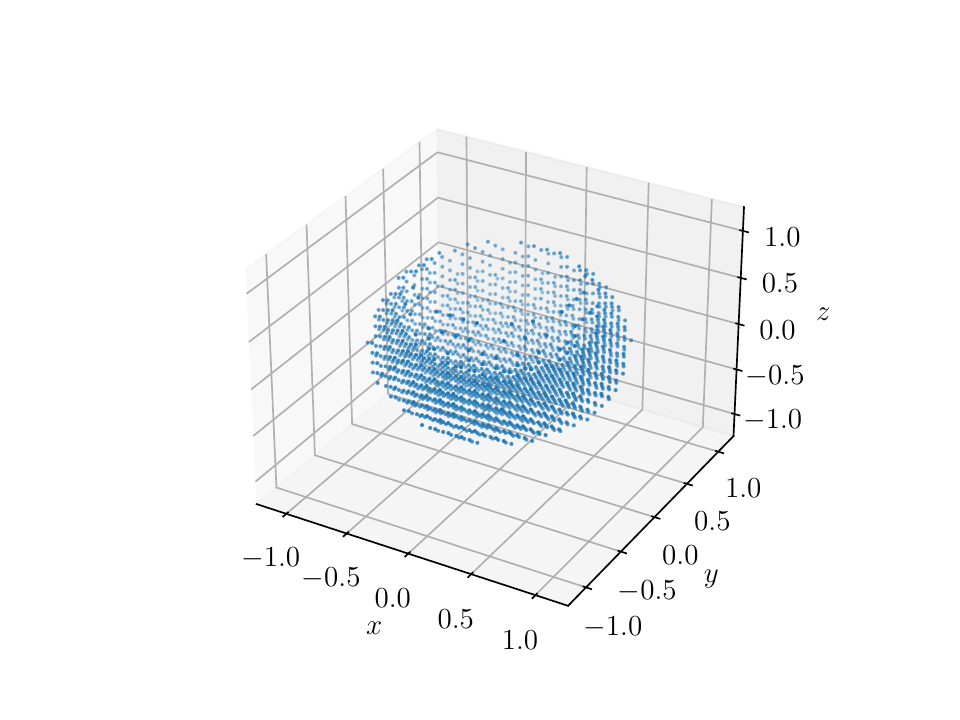}
        \caption{$\lambda=0.7$}
    \end{subfigure}
    \caption{Regions of acceptable quality for the 2-design affected by the amplitude damping channel (see \equatref{eq:ampdamp}) for the model where noise is applied before the unitary operations, for different $\lambda$.}
    \label{fig:regionsbeforeampdamp2}
\end{figure}

\begin{figure}
    \centering
    \begin{subfigure}[b]{.32\textwidth}
        \centering
        \includegraphics[trim=3.5cm 1cm 1.5cm 1.4cm, clip, scale=0.5]{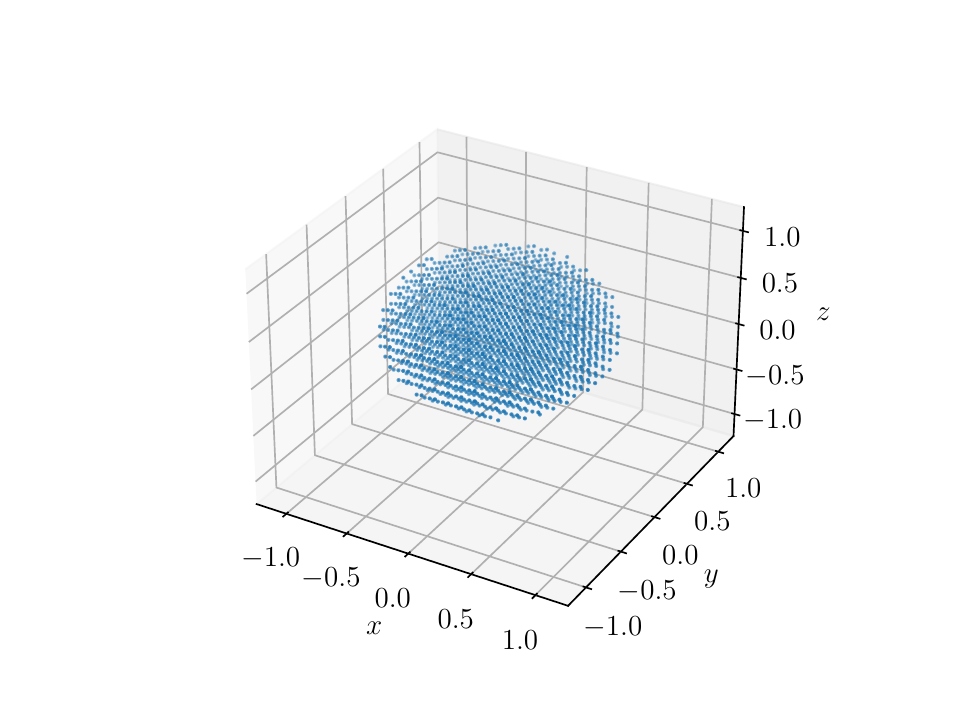}
        \caption{$\lambda=0.2$}
    \end{subfigure}
    \begin{subfigure}[b]{.32\textwidth}
        \centering
        \includegraphics[trim=3.5cm 1cm 1.5cm 1.4cm, clip, scale=0.5]{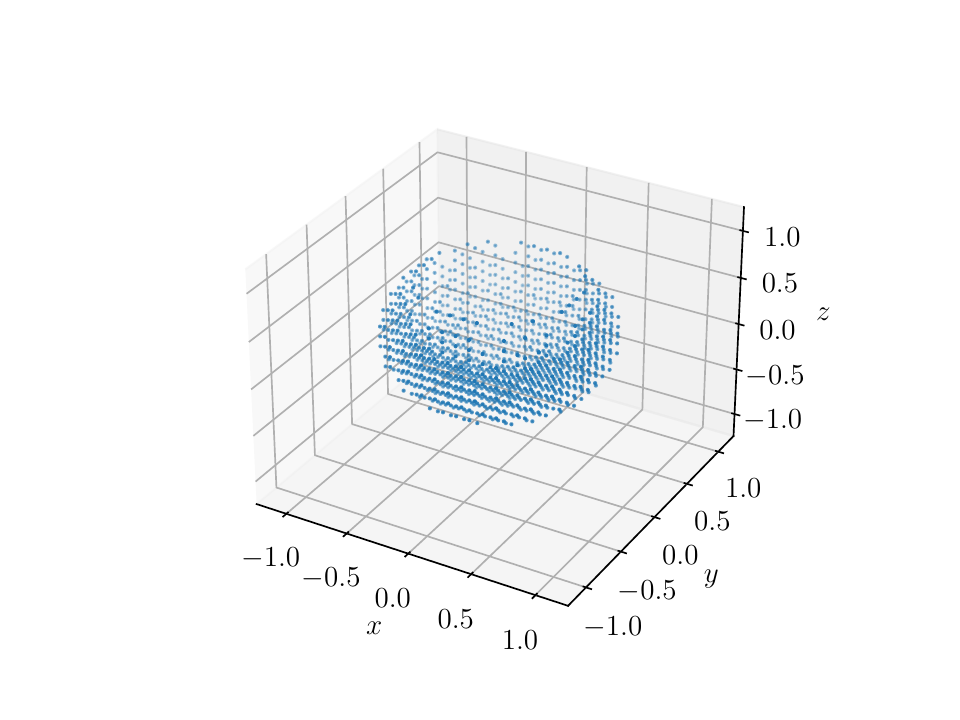}
        \caption{$\lambda=0.5$}
    \end{subfigure}
    \begin{subfigure}[b]{.32\textwidth}
        \centering
        \includegraphics[trim=3.5cm 1cm 1.5cm 1.4cm, clip, scale=0.5]{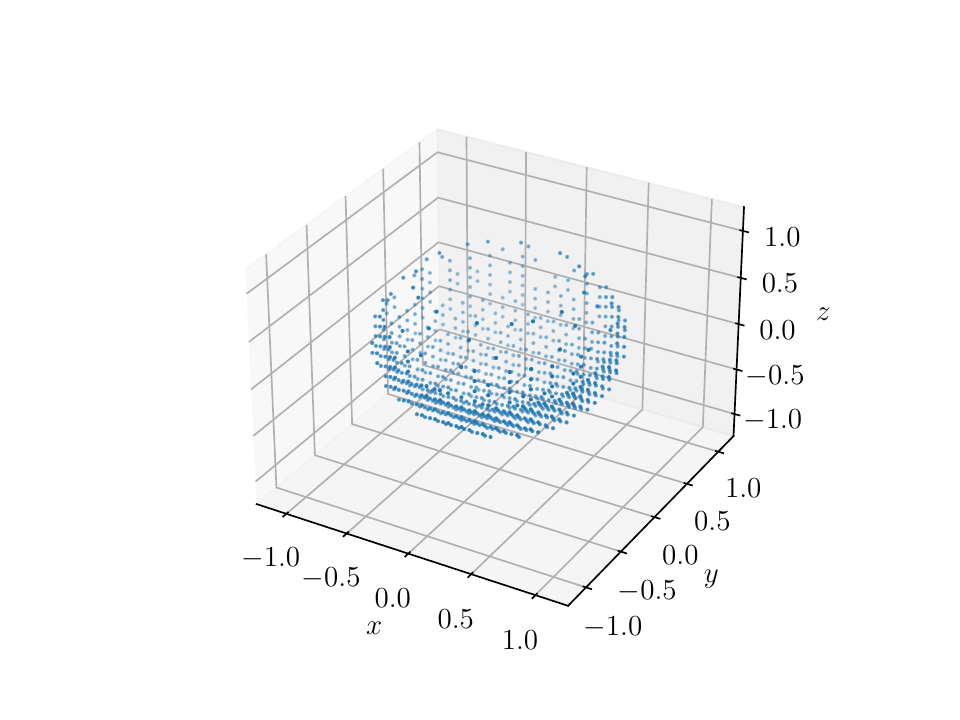}
        \caption{$\lambda=0.7$}
    \end{subfigure}
    \caption{Regions of acceptable quality for the 4-design affected by the amplitude damping channel (see \equatref{eq:ampdamp}) for the model where noise is applied before the unitary operations, for different $\lambda$.}
    \label{fig:regionsbeforeampdamp4}
\end{figure}

Regions of acceptable quality for the 2-design affected by the amplitude damping channel (see \equatref{eq:ampdamp}) are shown in \figref{fig:regionsbeforeampdamp2}, for various different $\lambda$.  For small $\lambda$, the region of acceptable quality is shifted up slightly towards the state \ket{0}, which seems to resemble the way in which the amplitude damping channel shrinks and shifts the Bloch sphere up towards the state \ket{0}.  However, the quality of the 2-design is not acceptable for states close to \ket{0} for large $\lambda$.  This is a result of setting the threshold for acceptable quality at $\epsilon=0.5$.  As discussed in \secref{sec:ampnumresultsbefore}, $\epsilon\to 1$ as $\lambda\to 1$ (see \figref{fig:beforeampdamp2}), so that states close to \ket{0} remain above the threshold of $\epsilon=0.5$ for large $\lambda$.  Just as for the other noise channels, regions of acceptable quality for the 3-design are identical to those obtained for the 2-design.  Regions of acceptable quality for the 4-design affected by the amplitude damping channel are shown in \figref{fig:regionsbeforeampdamp4}.  For small $\lambda$, the region of acceptable quality is shifted higher than for the 2-design, and for large $\lambda$, the region around \ket{0} for which the quality is not acceptable is much larger than for the 2-design.  Regions of acceptable quality for the 5-design affected by the amplitude damping channel are very similar to those obtained for the 4-design.

\subsubsection{Depolarising noise channel}\label{append:depnumregionsbefore}

\begin{figure}
    \centering
    \begin{subfigure}[b]{.32\textwidth}
        \centering
        \includegraphics[trim=3.5cm 1cm 1.5cm 1.4cm, clip, scale=0.5]{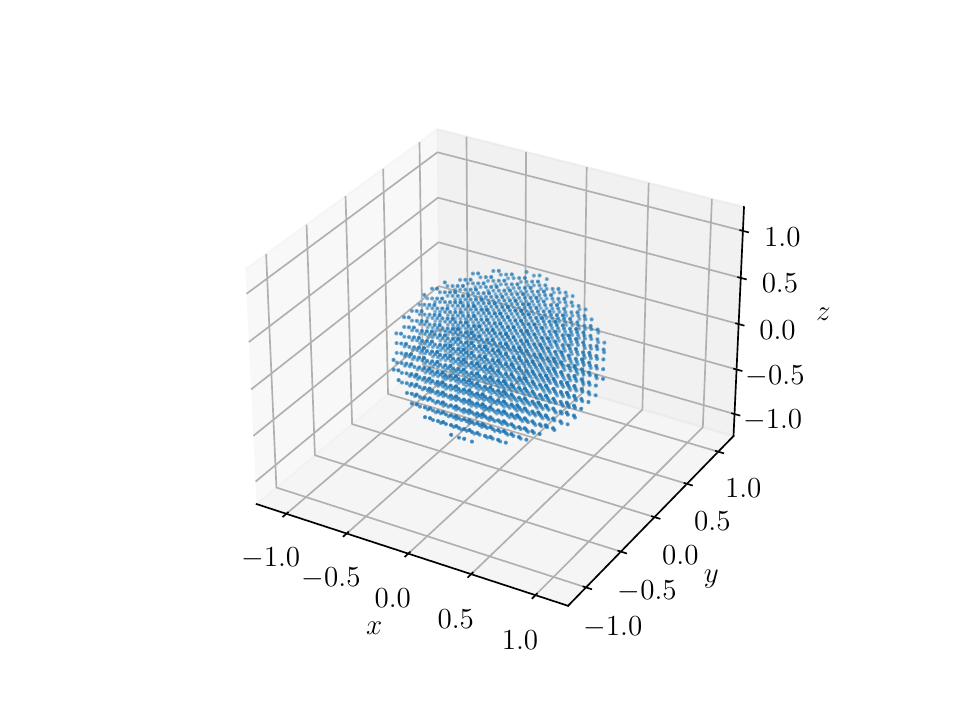}
        \caption{$p=0.2$}
    \end{subfigure}
    \begin{subfigure}[b]{.32\textwidth}
        \centering
        \includegraphics[trim=3.5cm 1cm 1.5cm 1.4cm, clip, scale=0.5]{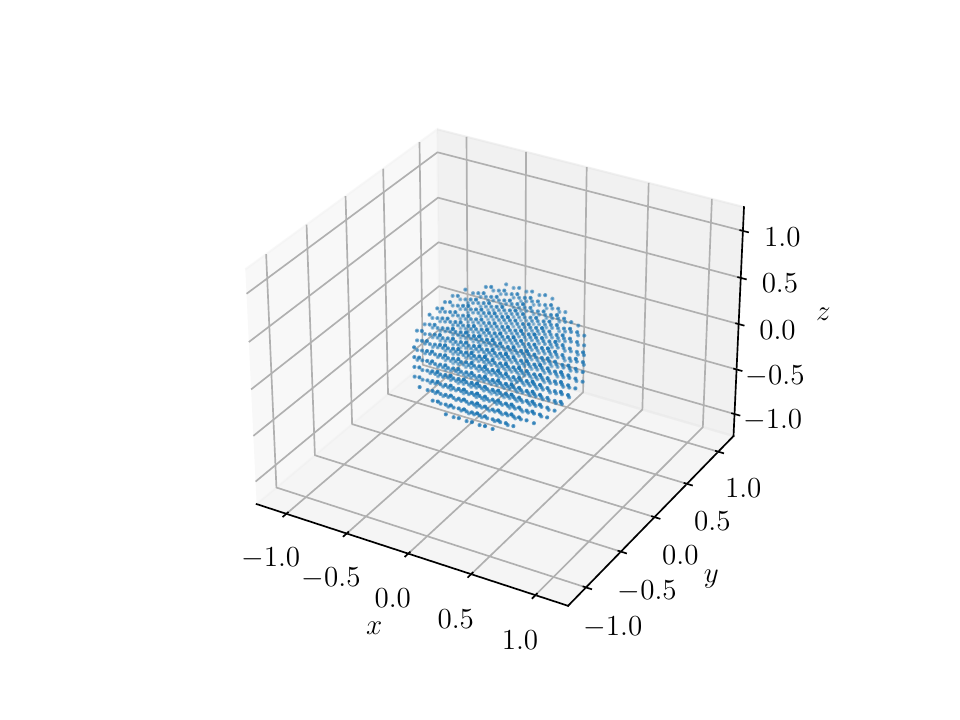}
        \caption{$p=0.5$}
    \end{subfigure}
    \begin{subfigure}[b]{.32\textwidth}
        \centering
        \includegraphics[trim=3.5cm 1cm 1.5cm 1.4cm, clip, scale=0.5]{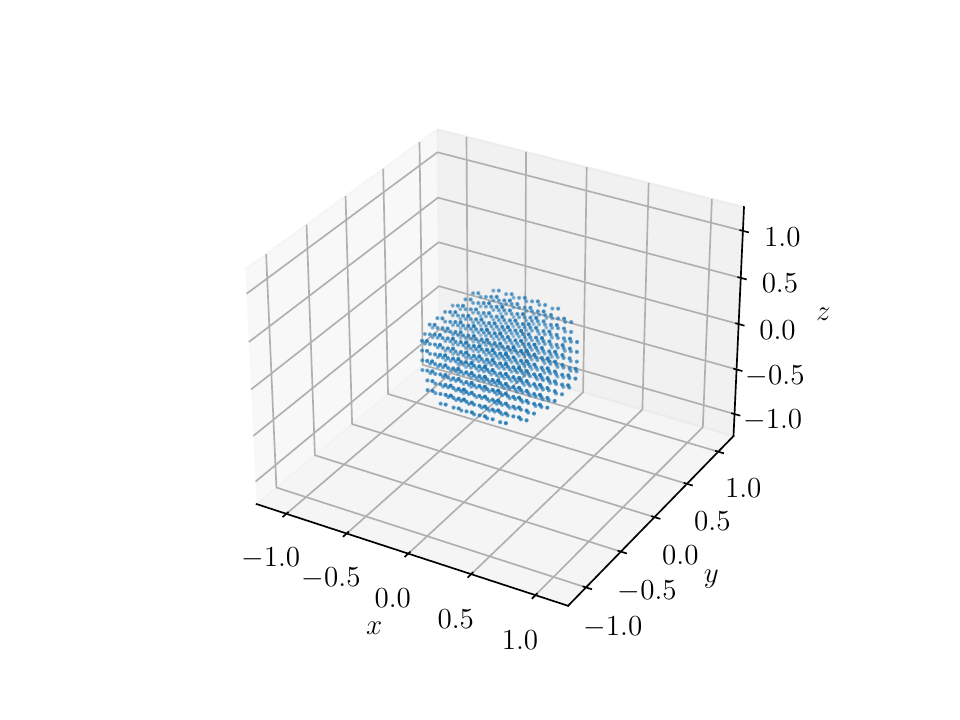}
        \caption{$p=0.7$}
    \end{subfigure}
    \caption{Regions of acceptable quality for the 2-design affected by the depolarising noise channel (see \equatref{eq:depnoise}) for the model where noise is applied before the unitary operations, for different $p$.}
    \label{fig:regionsbeforedepnoise2}
\end{figure}

\begin{figure}
    \centering
    \begin{subfigure}[b]{.32\textwidth}
        \centering
        \includegraphics[trim=3.5cm 1cm 1.5cm 1.4cm, clip, scale=0.5]{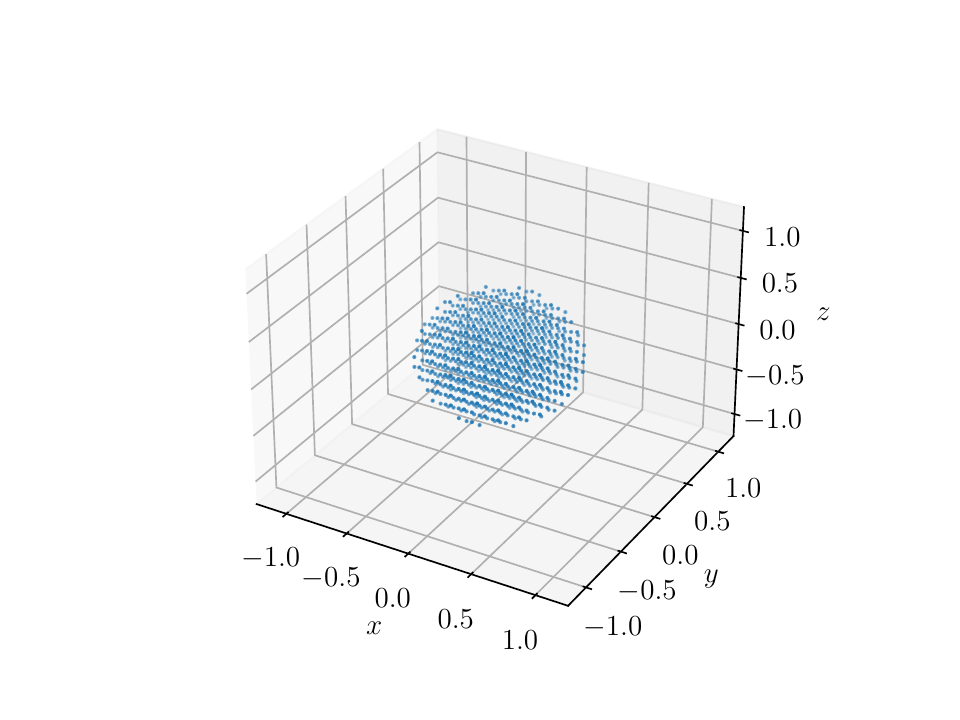}
        \caption{$p=0.2$}
    \end{subfigure}
    \begin{subfigure}[b]{.32\textwidth}
        \centering
        \includegraphics[trim=3.5cm 1cm 1.5cm 1.4cm, clip, scale=0.5]{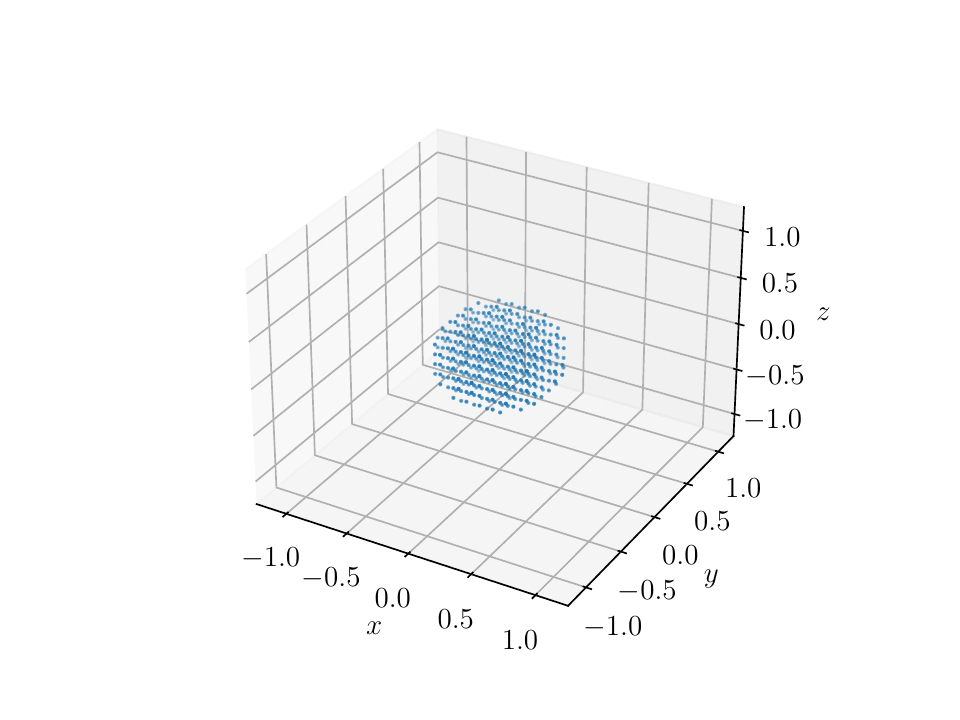}
        \caption{$p=0.5$}
    \end{subfigure}
    \begin{subfigure}[b]{.32\textwidth}
        \centering
        \includegraphics[trim=3.5cm 1cm 1.5cm 1.4cm, clip, scale=0.5]{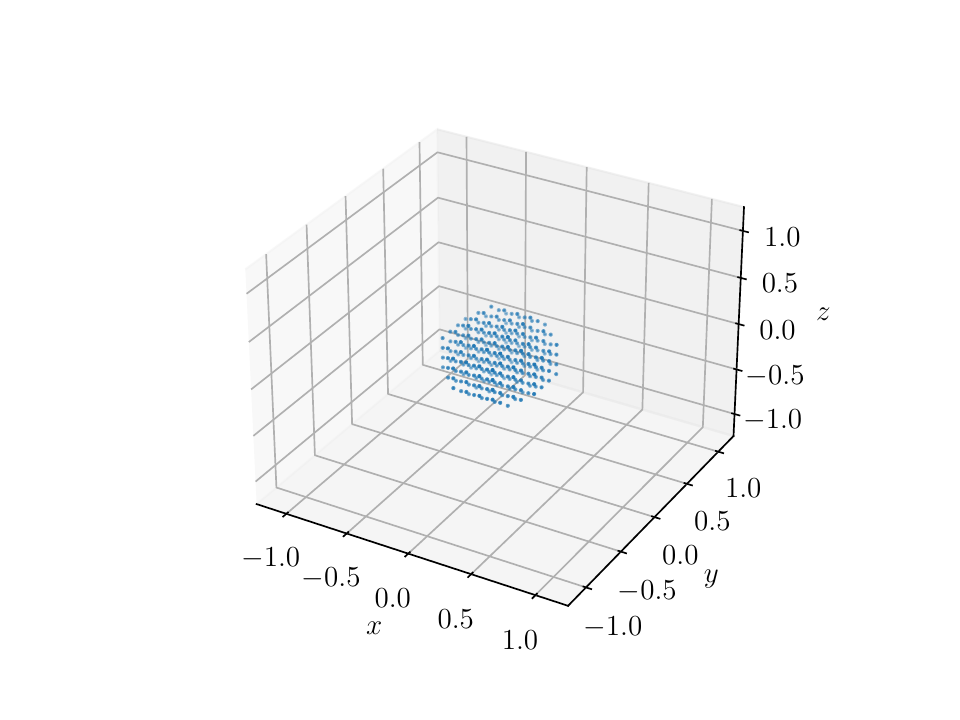}
        \caption{$p=0.7$}
    \end{subfigure}
    \caption{Regions of acceptable quality for the 4-design affected by the depolarising noise channel (see \equatref{eq:depnoise}) for the model where noise is applied before the unitary operations, for different $p$.}
    \label{fig:regionsbeforedepnoise4}
\end{figure}

Regions of acceptable quality for the 2-design affected by the depolarising noise channel (see \equatref{eq:depnoise}) are shown in \figref{fig:regionsbeforedepnoise2}.  The size of the region of acceptable quality decreases with increasing $p$, as expected.  For all $p$, the region of acceptable quality is a sphere centred at the origin, that is, the region of acceptable quality is similar in shape and orientation to the region into which the Bloch sphere is deformed by the depolarising channel.  Since the depolarising channel replaces a state by the maximally mixed state with probability $p$, states near the centre of the Bloch sphere, which are closer to the maximally mixed state, are least affected by depolarising noise.  Just as for the other noise channels considered, the region of acceptable quality for the 4-design is similar in shape to that of the 2-design, but smaller in size (see \figref{fig:regionsbeforedepnoise4}), once again confirming that the 4-design is more sensitive to noise than the 2-design.

\subsection{Numeric results for the model where noise is applied after the unitary operations}\label{append:numregionsafter} 

\subsubsection{Flip channels}\label{append:flipnumregionsafter}

\begin{figure}
    \centering
    \begin{subfigure}[b]{.32\textwidth}
        \centering
        \includegraphics[trim=3.5cm 1cm 1.5cm 1.4cm, clip, scale=0.5]{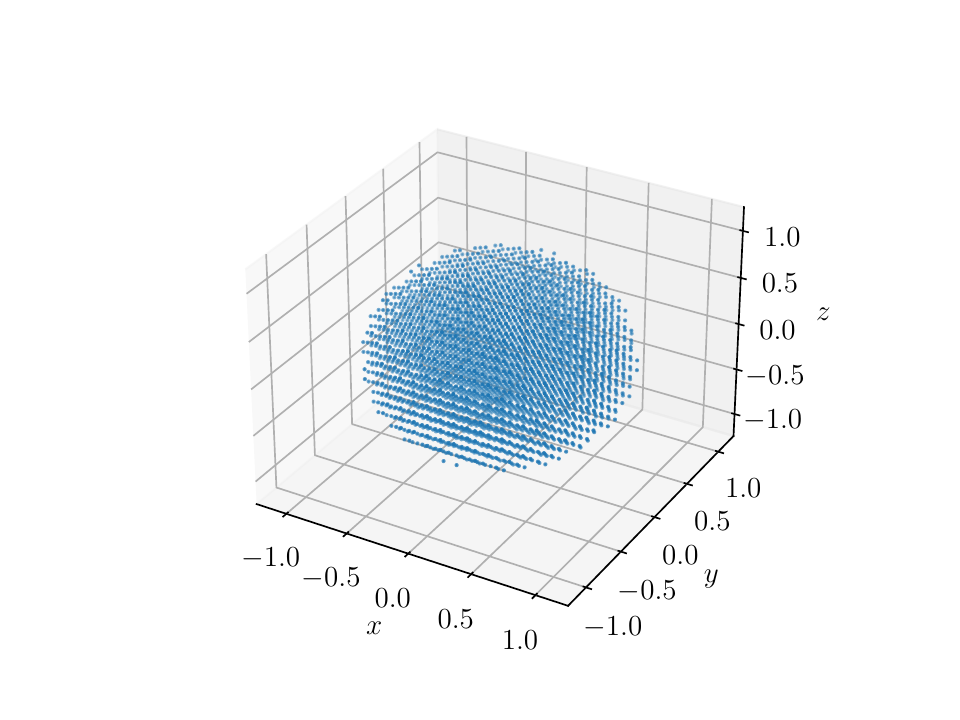}
        \caption{$p=0.01$}
    \end{subfigure}
    \begin{subfigure}[b]{.32\textwidth}
        \centering
        \includegraphics[trim=3.5cm 1cm 1.5cm 1.4cm, clip, scale=0.5]{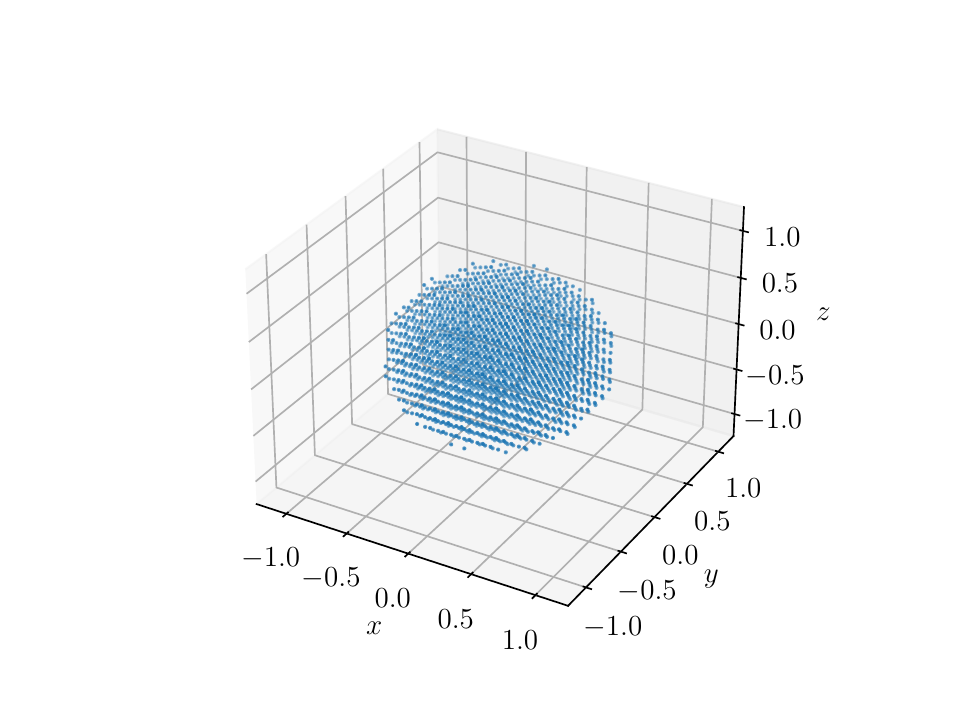}
        \caption{$p=0.1$}
    \end{subfigure}
    \begin{subfigure}[b]{.32\textwidth}
        \centering
        \includegraphics[trim=3.5cm 1cm 1.5cm 1.4cm, clip, scale=0.5]{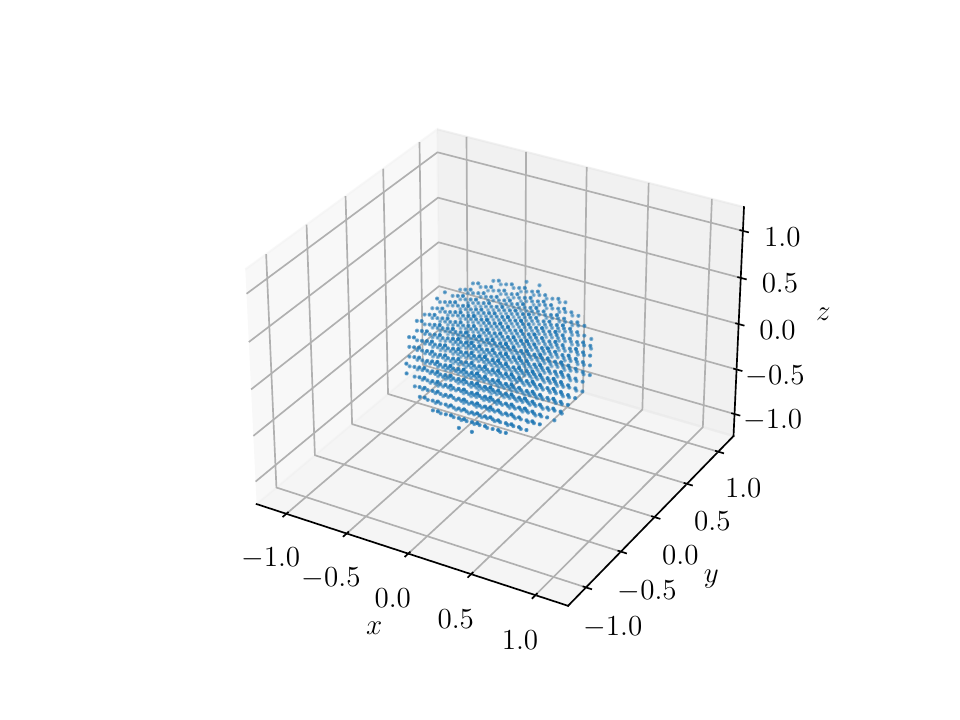}
        \caption{$p=0.3$}
    \end{subfigure}
    \caption{Regions of acceptable quality for the 2-design affected by the bit flip channel (see \equatref{eq:bflip}) for the model where noise is applied after the unitary operations, for different $p$.}
    \label{fig:regionsafterbflip2}
\end{figure}

Regions of acceptable quality for the 2-design affected by the bit flip channel (see \equatref{eq:bflip}) are shown in \figref{fig:regionsafterbflip2}.  For each $p$, the region of acceptable quality is a sphere centred at the origin, that is, the region of acceptable quality is similar in shape and orientation to the region of acceptable quality for the 2-design affected by the depolarising noise channel for the model where noise is applied before the unitary operations.  For the phase flip channel (see \equatref{eq:pflip}) and the bit and phase flip channel (see \equatref{eq:bpflip}), the regions of acceptable quality are identical to those for the bit flip channel (shown in \figref{fig:regionsafterbflip2}).  Thus we observe the transformation of each of the three flip channels into a depolarising channel when the flip channels are applied to states which have been randomised by unitary operators from the 2-design.  For all three flip channels, the region of acceptable quality for the 4-design is similar in shape and orientation to that of the 2-design, but smaller in size.

\subsubsection{Phase damping channel}\label{append:phasenumregionsafter} 

\begin{figure}
    \centering
    \begin{subfigure}[b]{.32\textwidth}
        \centering
        \includegraphics[trim=3.5cm 1cm 1.5cm 1.4cm, clip, scale=0.5]{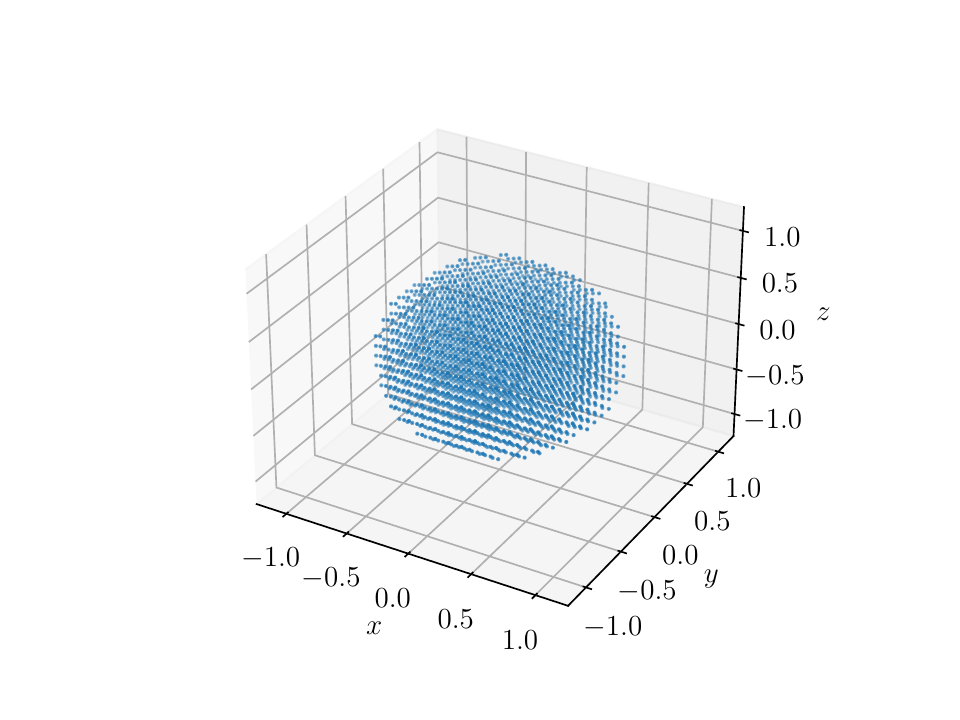}
        \caption{$\lambda=0.2$}
    \end{subfigure}
    \begin{subfigure}[b]{.32\textwidth}
        \centering
        \includegraphics[trim=3.5cm 1cm 1.5cm 1.4cm, clip, scale=0.5]{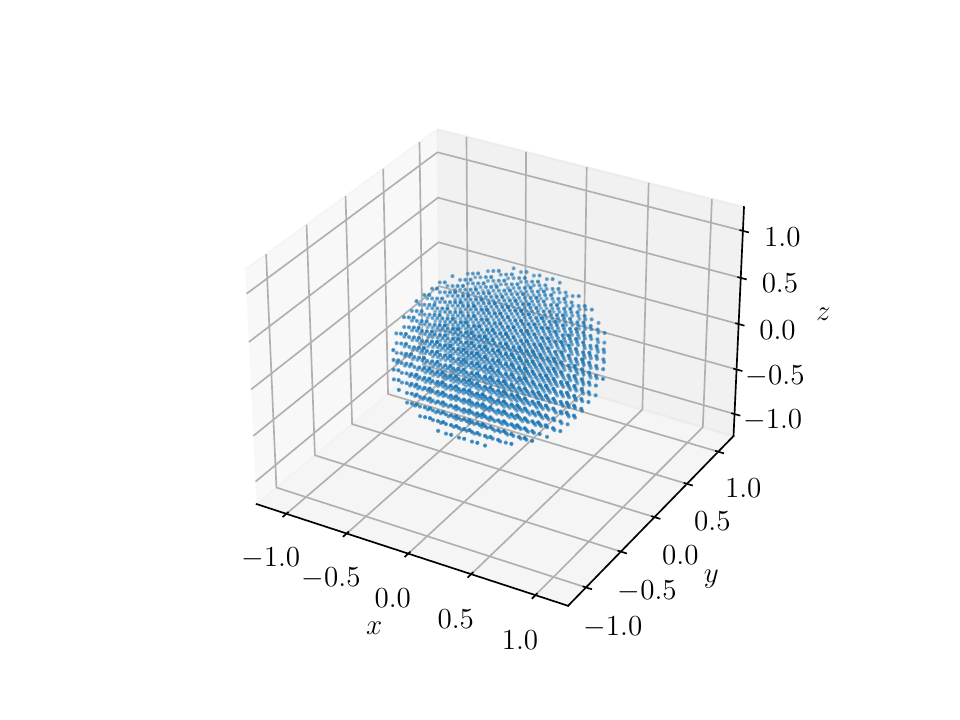}
        \caption{$\lambda=0.5$}
    \end{subfigure}
    \begin{subfigure}[b]{.32\textwidth}
        \centering
        \includegraphics[trim=3.5cm 1cm 1.5cm 1.4cm, clip, scale=0.5]{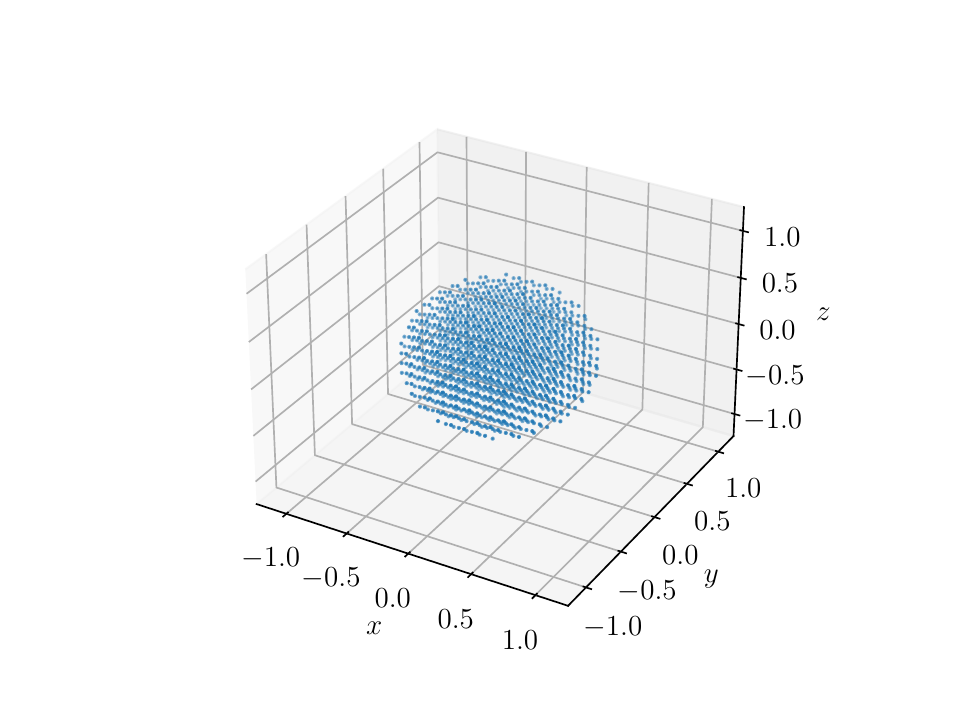}
        \caption{$\lambda=0.7$}
    \end{subfigure}
    \caption{Regions of acceptable quality for the 2-design affected by the phase damping channel (see \equatref{eq:phasedamp}) for the model where noise is applied after the unitary operations, for different $\lambda$.}
    \label{fig:regionsafterphasedamp2}
\end{figure}

For the phase damping channel (see \equatref{eq:phasedamp}), the regions of acceptable quality for both the 2-design (shown in \figref{fig:regionsafterphasedamp2}) and the 4-design are spheres centred at the origin, just as for the phase flip channel.

\subsubsection{Amplitude damping channel}\label{append:ampnumregionsafter} 

\begin{figure}
    \centering
    \begin{subfigure}[b]{.32\textwidth}
        \centering
        \includegraphics[trim=3.5cm 1cm 1.5cm 1.4cm, clip, scale=0.5]{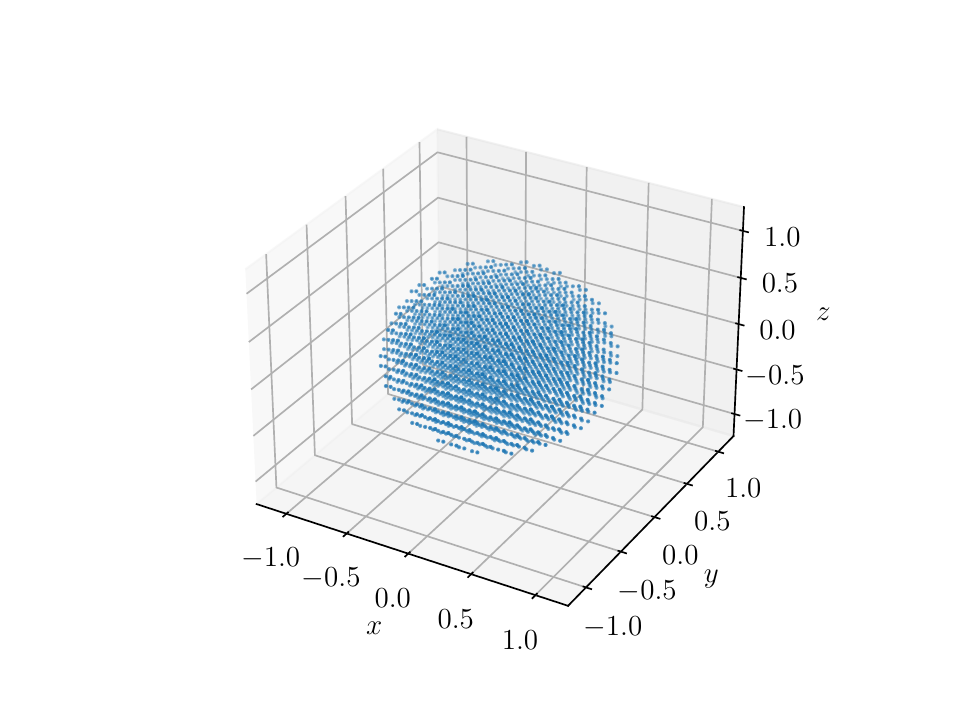}
        \caption{$\lambda=0.2$}
    \end{subfigure}
    \begin{subfigure}[b]{.32\textwidth}
        \centering
        \includegraphics[trim=3.5cm 1cm 1.5cm 1.4cm, clip, scale=0.5]{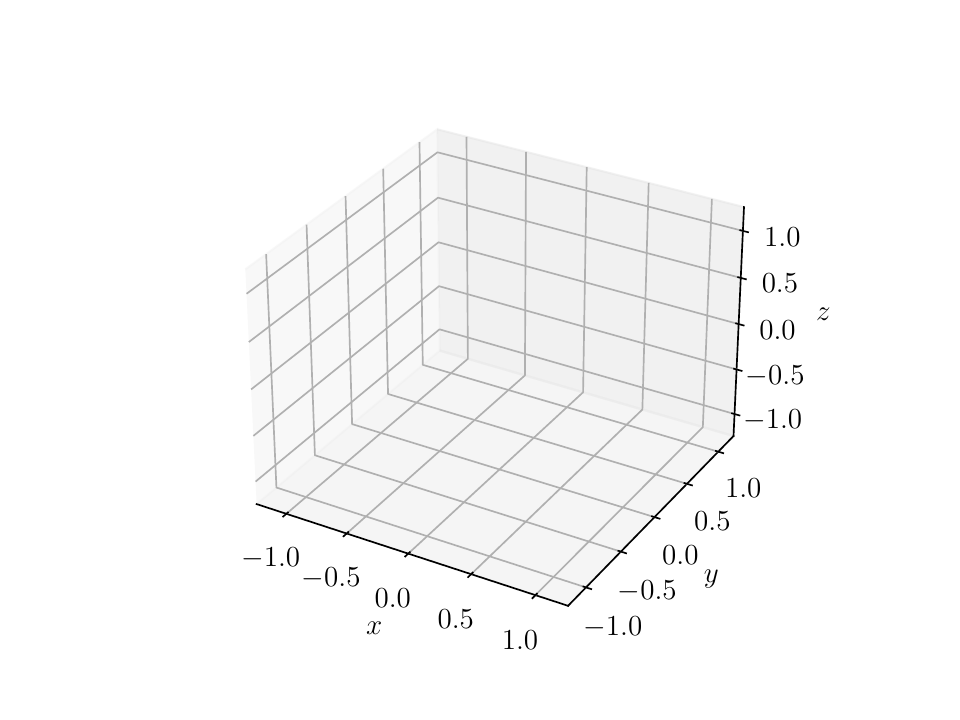}
        \caption{$\lambda=0.5$}
    \end{subfigure}
    \begin{subfigure}[b]{.32\textwidth}
        \centering
        \includegraphics[trim=3.5cm 1cm 1.5cm 1.4cm, clip, scale=0.5]{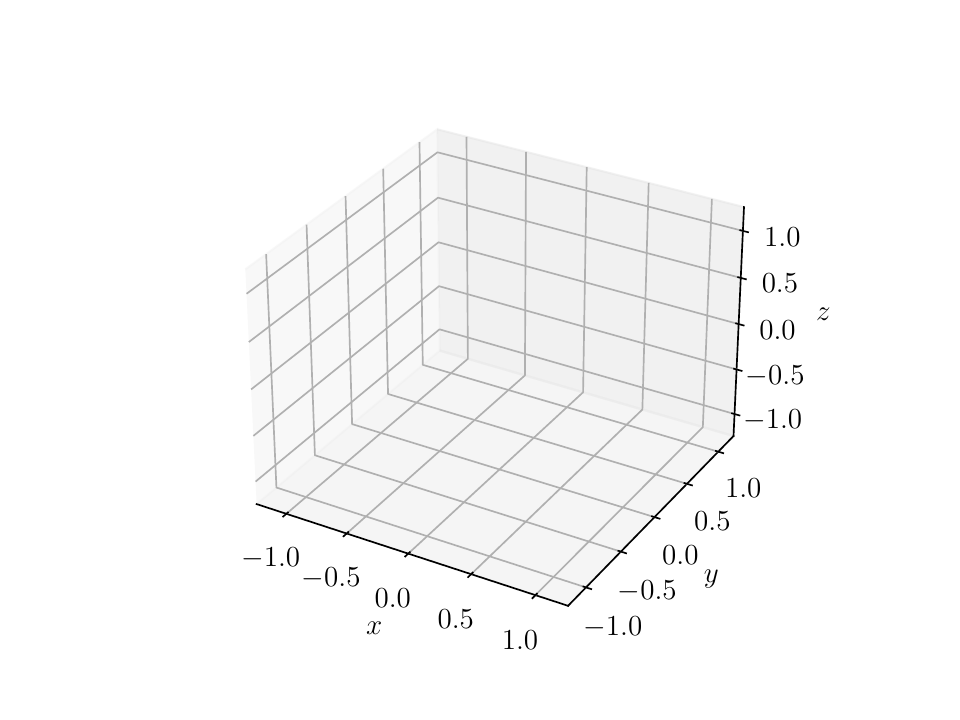}
        \caption{$\lambda=0.7$}
    \end{subfigure}
    \caption{Regions of acceptable quality for the 2-design affected by the amplitude damping channel (see \equatref{eq:ampdamp}) for the model where noise is applied after the unitary operations, for different $\lambda$.}
    \label{fig:regionsafterampdamp2}
\end{figure}

\begin{figure}
    \centering
    \begin{subfigure}[b]{.32\textwidth}
        \centering
        \includegraphics[trim=3.5cm 1cm 1.5cm 1.4cm, clip, scale=0.5]{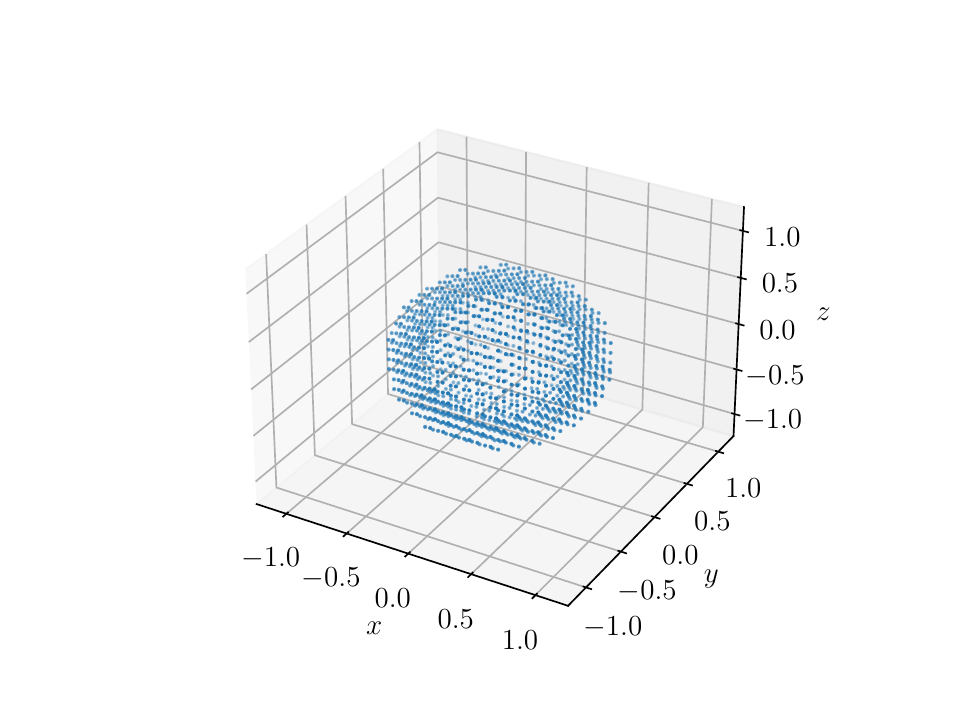}
        \caption{$\lambda=0.2$}
    \end{subfigure}
    \begin{subfigure}[b]{.32\textwidth}
        \centering
        \includegraphics[trim=3.5cm 1cm 1.5cm 1.4cm, clip, scale=0.5]{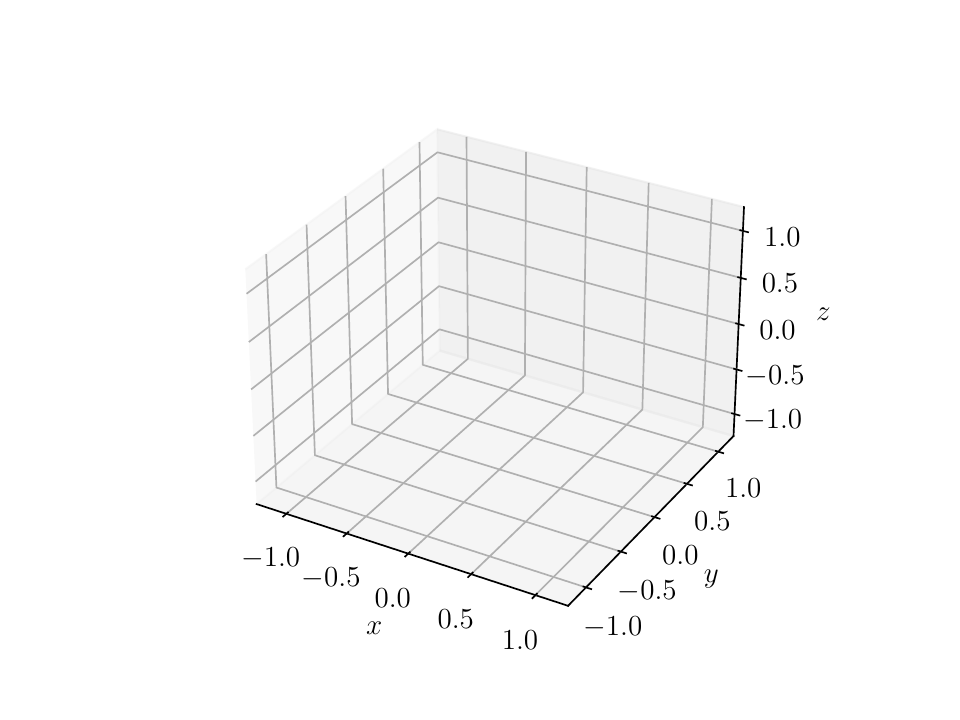}
        \caption{$\lambda=0.5$}
    \end{subfigure}
    \begin{subfigure}[b]{.32\textwidth}
        \centering
        \includegraphics[trim=3.5cm 1cm 1.5cm 1.4cm, clip, scale=0.5]{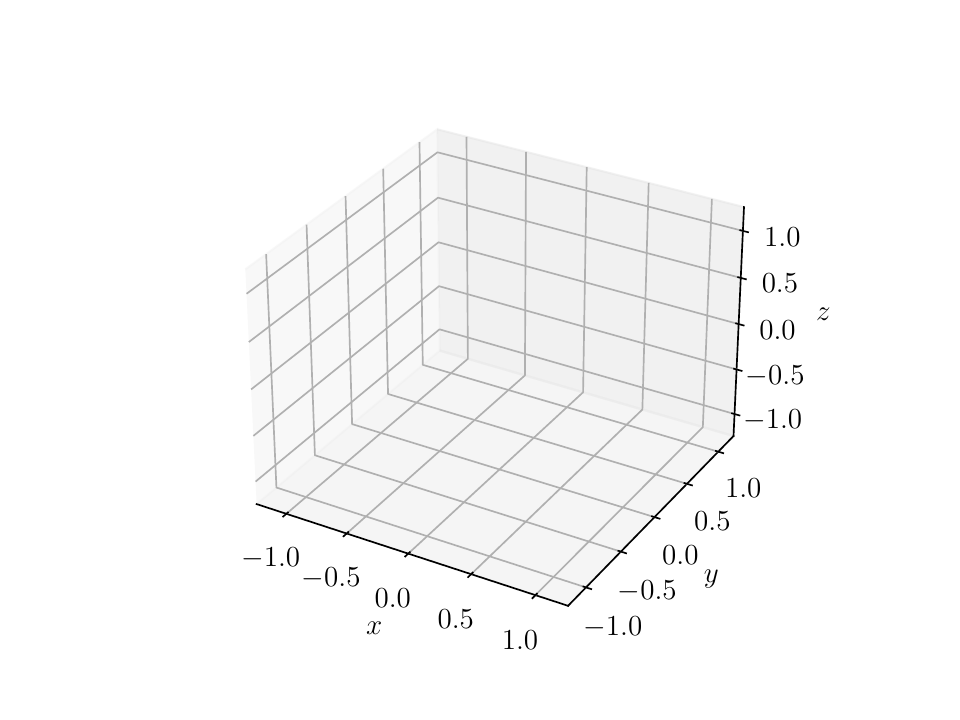}
        \caption{$\lambda=0.7$}
    \end{subfigure}
    \caption{Regions of acceptable quality for the 3-design affected by the amplitude damping channel (see \equatref{eq:ampdamp}) for the model where noise is applied after the unitary operations, for different $\lambda$.}
    \label{fig:regionsafterampdamp3}
\end{figure}

Regions of acceptable quality for the 2-design affected by the amplitude damping channel (see \equatref{eq:ampdamp}) are shown in \figref{fig:regionsafterampdamp2}, for different $\lambda$.  For small $\lambda$, the region of acceptable quality is once again a sphere centred at the origin, that is, the region of acceptable quality is once again similar to the region of acceptable quality for the 2-design affected by the depolarising noise channel for the model where noise is applied before the unitary operations.  For large $\lambda$, the region of acceptable quality disappears.  As discussed in \appendref{append:numresultsafter}, the $\epsilon$ attained for $\lambda=1$ is much larger for the model where noise is applied after the unitary operations (see \figref{fig:afterampdamp2}).  Hence, all states in the Bloch sphere remain above the threshold of $\epsilon=0.5$ for large $\lambda$.  Regions of acceptable quality for the 3-design affected by the amplitude damping channel are shown in \figref{fig:regionsafterampdamp3}.  For small $\lambda$, the region of acceptable quality is a hollow spherical shell centred at the origin.  Thus when amplitude damping is applied after the unitary operations, it is states close to the maximally mixed state, and not states close to \ket{0}, for which the quality is not acceptable.  For large $\lambda$, the region of acceptable quality once again disappears.  For the 4-design and the 5-design, the region of acceptable quality vanishes for $\lambda=0.2$, $\lambda=0.5$, and $\lambda=0.7$.

\subsubsection{Depolarising noise channel}\label{append:depnumregionsafter}

For the depolarising noise channel, we prove that the two noise models are equivalent (see \appendref{append:depolarising}).

\section{Numeric results for different truncations of the polar angle}\label{append:polar} 

We investigate the dependence of $\epsilon$ versus $p$ (or $\epsilon$ versus $\lambda$) on the truncation of the polar angle $\theta_t$.  In particular, we consider $\theta_t\in\left\{\frac{\pi}{6},\frac{\pi}{3},\frac{\pi}{2},\frac{2\pi}{3},\frac{5\pi}{6},\pi\right\}$ for a fixed truncation radius of $r_t=0.95$ (with $\phi_t=2\pi$) and obtain results numerically using the method described in \secref{sec:impresults}.

\subsection{Numeric results for the model where noise is applied before the unitary operations}\label{append:polarbefore} 

\subsubsection{Bit flip channel}\label{append:bflippolarbefore} 

For the bit flip channel (see \equatref{eq:bflip}), $\epsilon$ versus $p$ is independent of $\theta_t$.  This can be explained as follows.  For a fixed truncation radius, states along the positive $z$-axis are among the furthest from the eigenstates of the Pauli $X$ operator.  Hence these states are among the most sensitive to bit flips and therefore require the largest $\epsilon$ to satisfy inequality (\ref{eq:model}).  Since the states along the positive $z$-axis are included in the sample of density matrices for all $\theta_t$, the value of $\epsilon$ obtained for a given $p$ is simply this largest $\epsilon$ for all $\theta_t$, so that $\epsilon$ versus $p$ is the same for all $\theta_t$.

\subsubsection{Phase flip channel}\label{append:pflippolarbefore} 

For the phase flip channel (see \equatref{eq:pflip}), the maximum of $\epsilon$ versus $p$ increases as we increase $\theta_t$, up to $\theta_t=\frac{\pi}{2}$, after which the maximum remains more or less constant, for both the 2-design (shown in \figref{fig:polarbeforepflip2}) and the 4-design.  As $\theta_t$ increases from 0 to $\frac{\pi}{2}$, the sample of density matrices is expanded to include states which are further from the eigenstates of the Pauli $Z$ operator and therefore more sensitive to phase flips, which results in $\epsilon$ increasing.  When $\theta_t=\frac{\pi}{2}$, the states along the equator of the sphere, which are furthest from the eigenstates of the Pauli $Z$ operator and therefore most sensitive to phase flips, are included in the sample of density matrices and so further increasing $\theta_t$ does not further increase $\epsilon$.

\subsubsection{Bit and phase flip channel}\label{append:bpflippolarbefore} 

For the bit and phase flip channel (see \equatref{eq:bpflip}), $\epsilon$ versus $p$ is independent of $\theta_t$.  Just as for the bit flip channel, the states along the positive $z$-axis, which are included in the sample of density matrices for all $\theta_t$, require the largest $\epsilon$ to satisfy inequality (\ref{eq:model}), and so the value of $\epsilon$ obtained for a given $p$ is simply this largest $\epsilon$ for all $\theta_t$.  The states along the positive $z$-axis are among the furthest from the eigenstates of the Pauli $Y$ operator and are therefore among the most sensitive to bit and phase flips, which is why they require the largest $\epsilon$ to satisfy inequality (\ref{eq:model}).

\subsubsection{Phase damping channel}\label{append:phasepolarbefore}

\begin{figure*}
    \centering
    \begin{subfigure}[b]{.42\textwidth}
        \centering
        \includegraphics[scale=0.55]{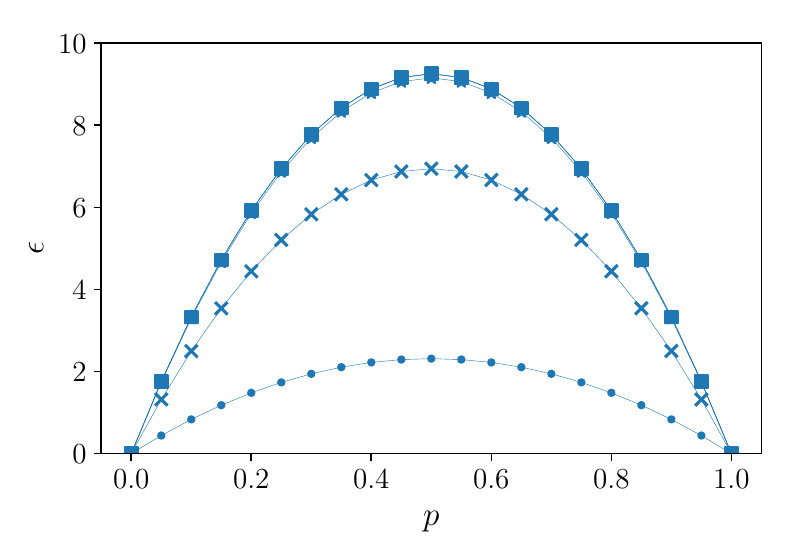}
        \caption{phase flip channel}
        \label{fig:polarbeforepflip2}
    \end{subfigure}
    \begin{subfigure}[b]{.42\textwidth}
        \centering
        \includegraphics[scale=0.55]{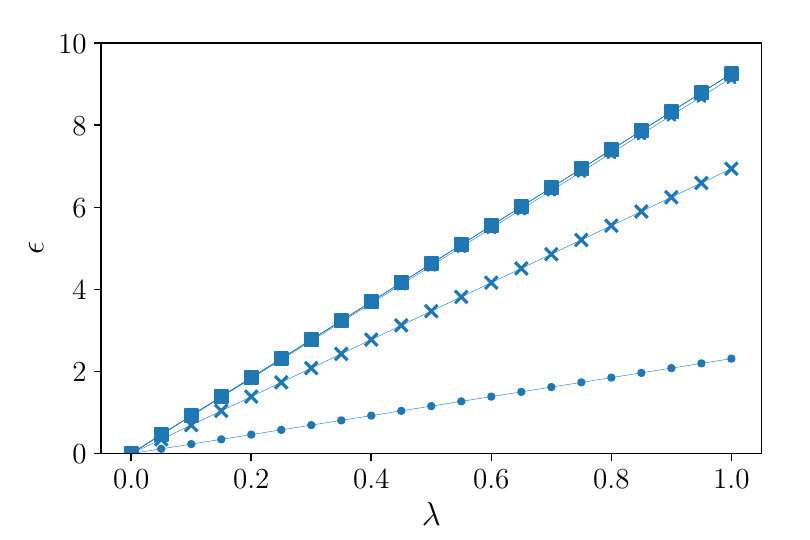}
        \caption{phase damping channel}
        \label{fig:polarbeforephasedamp2}
    \end{subfigure}
    \includegraphics[trim=0cm 3.8cm 0cm 7cm, clip, scale=0.52]{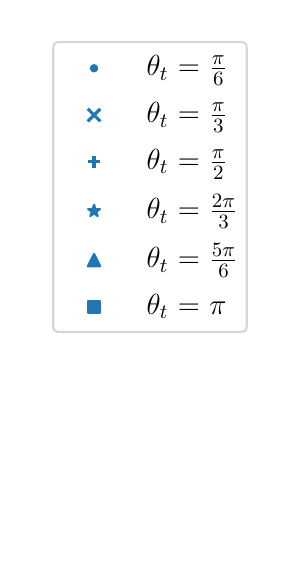}
    \caption{Effect of the (a) phase flip channel (see \equatref{eq:pflip}) and (b) phase damping channel (see \equatref{eq:phasedamp}) on the quality of the 2-design for the model where noise is applied before the unitary operations, for different truncations of the polar angle $\theta_t$, for a fixed truncation radius of $r_t=0.95$.}
\end{figure*}

For the phase damping channel (see \equatref{eq:phasedamp}), the gradient of $\epsilon$ versus $\lambda$ increases as we increase $\theta_t$, up to $\theta_t=\frac{\pi}{2}$, after which the gradient remains more or less constant, for both the 2-design (shown in \figref{fig:polarbeforephasedamp2}) and the 4-design.  As expected, the dependence of $\epsilon$ versus $\lambda$ on $\theta_t$ for the phase damping channel is similar to the dependence of $\epsilon$ versus $p$ on $\theta_t$ for the phase flip channel.

\subsubsection{Amplitude damping channel}\label{append:amppolarbefore} 

\begin{figure*}
    \centering
    \includegraphics[scale=0.55]{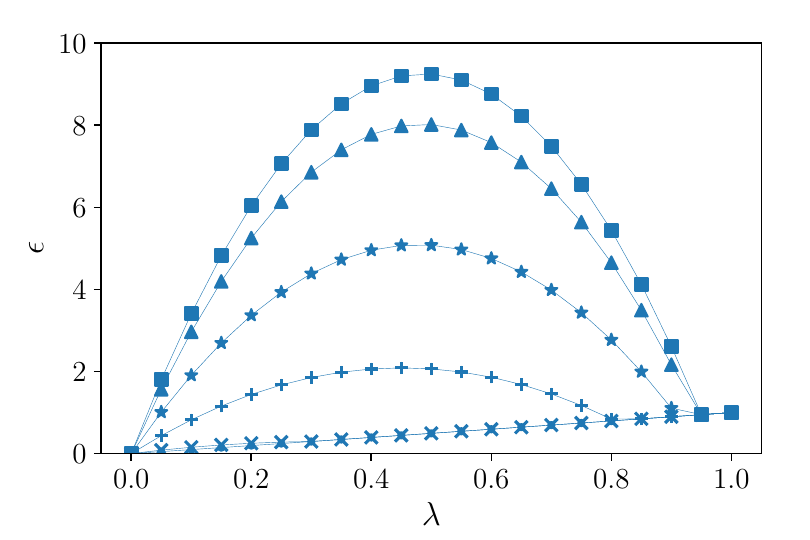}
    \includegraphics[scale=0.55]{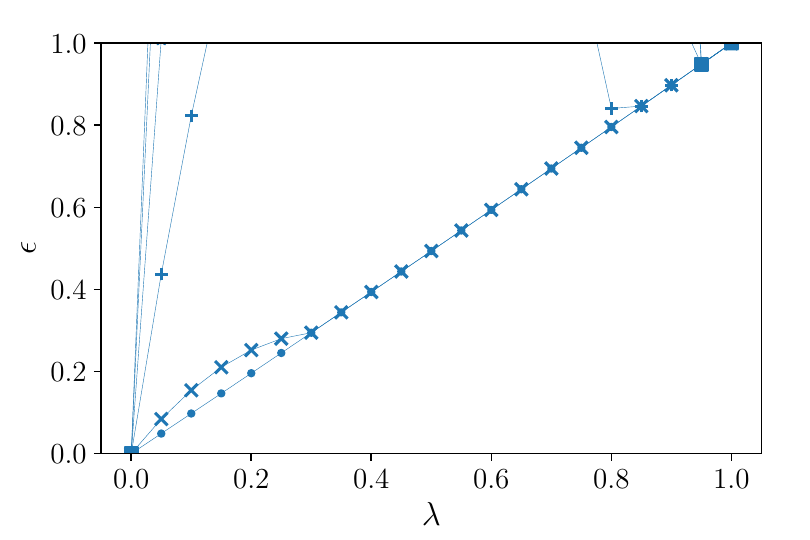}
    \includegraphics[trim=0cm 5cm 0cm 7cm, clip, scale=0.52]{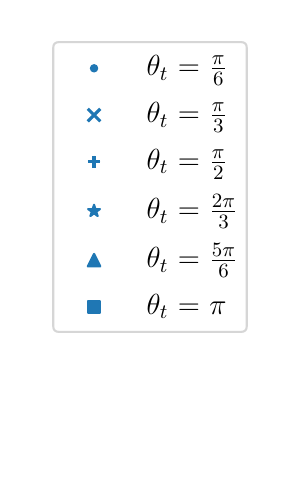}
    \caption{Effect of the amplitude damping channel (see \equatref{eq:ampdamp}) on the quality of the 2-design for the model where noise is applied before the unitary operations, for different truncations of the polar angle $\theta_t$, for a fixed truncation radius of $r_t=0.95$.  The full set of results is shown on the left and a zoomed-in region is shown enlarged on the right.}
    \label{fig:polarbeforeampdamp2}
\end{figure*}

For the amplitude damping channel (see \equatref{eq:ampdamp}), the maximum of $\epsilon$ versus $\lambda$ increases as we increase $\theta_t$, for the 2-design (shown in \figref{fig:polarbeforeampdamp2}), the 3-design, the 4-design, and the 5-design.  As $\theta_t$ increases, the sample of density matrices is expanded to include states which are further from the state \ket{0} and therefore more sensitive to amplitude damping, which results in $\epsilon$ increasing.

\subsubsection{Depolarising noise channel}\label{append:deppolarbefore}

For the depolarising noise channel (see \equatref{eq:depnoise}), $\epsilon$ versus $p$ is independent of $\theta_t$.  This is because all states at a given radial distance from the maximally mixed state satisfy inequality (\ref{eq:model}) with the same value of $\epsilon$, irrespective of the polar angles of the states.

\subsection{Numeric results for the model where noise is applied after the unitary operations}\label{append:polarafter} 

For the model where noise is applied after the unitary operations, numeric results are independent of $\theta_t$ for all six noise channels.  This reflects our observations from \appendref{append:regions}, that is, when the noise channel is applied after the states have been randomised by the unitary operations, all states at a given radial distance from the maximally mixed state satisfy inequality (\ref{eq:model}) with the same value of $\epsilon$, irrespective of the polar angles of the states.

\section{Numeric results for different truncations of the azimuthal angle}\label{append:azimuthal} 

We investigate the dependence of $\epsilon$ versus $p$ (or $\epsilon$ versus $\lambda$) on the truncation of the azimuthal angle $\phi_t$.  In particular, we consider $\phi_t\in\left\{\frac{\pi}{6},\frac{\pi}{3},\frac{\pi}{2},\frac{2\pi}{3},\frac{5\pi}{6},\pi,\frac{7\pi}{6},\frac{4\pi}{3},\frac{3\pi}{2},\frac{5\pi}{3},\frac{11\pi}{6},2\pi\right\}$ for a fixed truncation radius of $r_t=0.95$ (with $\theta_t=\pi$) and obtain results numerically using the method described in \secref{sec:impresults}.

\subsection{Numeric results for the model where noise is applied before the unitary operations}\label{append:azimuthalbefore} 

Numeric results are independent of $\phi_t$ for all six noise channels.  For the bit flip channel and the bit and phase flip channel, this can be attributed to the fact that states along the $z$-axis, which require the largest $\epsilon$ to satisfy inequality (\ref{eq:model}), are included in the sample of density matrices for all $\phi_t$, so that the value of $\epsilon$ obtained is simply this largest $\epsilon$ for all $\phi_t$.  For the phase flip channel, the phase damping channel, the amplitude damping channel, and the depolarising noise channel, this can be attributed to the fact that the smallest $\epsilon$ such that inequality (\ref{eq:model}) holds for a given state remains unchanged when that state is rotated about the $z$-axis, so that the value of $\epsilon$ obtained is completely independent of $\phi_t$.

\subsection{Numeric results for the model where noise is applied after the unitary operations}\label{append:azimuthalafter} 

For the model where noise is applied after the unitary operations, numeric results are also independent of $\phi_t$ for all six noise channels.

\end{document}